\documentclass[useAMS,usenatbib]{mn2e}
\usepackage{graphicx, float}
\usepackage{amsfonts}

\usepackage{epsfig}

\usepackage{color}

\usepackage{times}

\usepackage{amsmath}

\usepackage{amssymb}

\usepackage{xspace}

\usepackage{txfonts}

\usepackage{enumitem}

\usepackage{algorithm}

\usepackage{algorithmic}

\usepackage[outercaption]{sidecap}

\usepackage{subfig}

\usepackage{rotating}

\usepackage{hyperref}

\usepackage{bbm}

\def\Mpch{h^{-1} {\rm Mpc}}

\def\Mpch{h^{-1} {\rm Mpc}}
\def\Mpchk{h {\rm Mpc}^{-1}}

\newcommand{\Mspace}     {\mm{{\mathbb M}}}
\newcommand{\Rspace}     {\mm{{\mathbb R}}}

\newcommand{\Tspace}     {\mm{{\mathbb T}}}

\newcommand{\Exp}     {\mm{{\mathbb E}}}
\newcommand{\cH}  {{\mathcal H}}

\newcommand {\mm}[1] {\ifmmode{#1}\else{\mbox{\(#1\)}}\fi}

\def\LKC{Lipschitz-Killing curvature}
\def\LKCs{Lipschitz-Killing curvatures}

\begin{document}

\title[Topology and geometry  of Gaussian random fields II]{Topology and geometry  of Gaussian random fields II: on critical points, excursion sets, and persistent homology}

\author[Pratyush Pranav]
{Pratyush Pranav$^{1,2,3}$\thanks{pratyush.pranav@ens-lyon.fr}\\
	$^1$Univ Lyon, ENS de Lyon, Univ Lyon1, CNRS, Centre de Recherche Astrophysique de Lyon UMR5574, FV69007, Lyon, France\\
	$^2$Kapteyn Astronomical Institute, Univ. of Groningen, PO Box 800, 9700 AV Groningen, The Netherlands\\
	$^3$Technion -- Israel Institute of Technology, Haifa, 32000, Israel\\
}

\maketitle


\begin{abstract}
	
	This paper is second in the series, following Pranav et al. (2019), focused on the characterization of geometric and topological properties of 3D Gaussian random fields. We focus on the formalism of persistent homology, the mainstay of Topological Data Analysis (TDA), in the context of excursion set formalism. We also focus on the structure of critical points of stochastic fields, and their relationship with formation and evolution of structures in the universe.
	
	The topological background is accompanied by an investigation of Gaussian field simulations based on the LCDM spectrum, as well as power-law spectra with varying spectral indices. We present the statistical properties in terms of the intensity and difference maps constructed from the persistence diagrams, as well as their distribution functions. We demonstrate that the intensity maps encapsulate information about the distribution of power across the hierarchies of structures in more detailed than the Betti numbers or the Euler characteristic. In particular, the white noise ($n = 0$) case with flat spectrum stands out as the divide between models with positive and negative spectral index. It has the highest proportion of low significance features. This level of information is not available from the geometric Minkowski functionals or the topological Euler characteristic, or even the Betti numbers, and demonstrates the usefulness of hierarchical topological methods. Another important result is the observation that topological characteristics of Gaussian fields depend on the power spectrum, as opposed to the geometric measures that are insensitive to the power spectrum characteristics.

\end{abstract}

\begin{keywords}
 {large-scale structure of the universe, Gaussian random fields -- cosmology: theory; 
 Morse theory, Persistent homology, Betti numbers, Euler characteristic -- topology: theory;
 Topological Data Analysis -- data analysis: theory}
\end{keywords}


\section{Introduction}
\label{sec:intro}

The increased focus towards applications that geometry and topology have witnessed  in the recent decades, has produced state-of-the art novel methodologies that have the potential to reveal novel and crucial insights into the structure of cosmological data and fields. Within these recent developments,  \emph{Topological Data Analysis}  (TDA) is especially gaining prominence across disciplines as an applied tool in the computational setting. These developments on the computational side are crucially relevant given the recent thrust towards data driven investigation, ubiquitous across disciplines, leading to the demand for increasingly more sophisticated methods to detect patterns, and glean meaningful information in data. This paper presents an account of the foundational methodologies of TDA in the general framework of the excursion set formalism. On the application side, we present an investigation into the properties of Gaussian random fields using these methodologies. Gaussian random fields model a variety of natural phenomena, across a range of disciplines. In this paper, we present an investigation in the cosmological context. This paper follows the spirit of \cite{pranav2019a}, where we present a detailed characterization of the geometry as well as topology of Gaussian field models  in terms of Minkowski functionals and Betti numbers respectively. In terms of the analysis pipeline, this paper is also a follow-up to \cite{feldbrugge2019}, where we present a semi-analytical case-study of the persistent homology characteristics of Gaussian and local-type non-Gaussian random fields.

\subsection{Gaussian random fields}

Given a parameter space $\mathcal{S}$, a \emph{Gaussian random field} is a stochastic process, $\mathcal{X}$, such that the
vector $(\mathcal{X}(s_1),\dots,\mathcal{X}(s_k))$ has a $k$-dimensional, multivariate normal distribution for any collection of points $(s_1,\dots,s_k)$ in $\mathcal{S}$. 


Gaussian random fields serve as the baseline reference model for a variety of astrophysical phenomena. At the largest cosmological scales, Gaussian random fields serve as the null hypothesis model for the primordial perturbation field. A substantial body of theoretical and observational evidence supports this consensus. The primary theoretical argument for the primordial field to be Gaussian emanates from the standard inflationary scenarios. According to this fundamental cosmological paradigm, the early universe undergoes a phase
transition at around $t \approx 10^{-35}$ sec after the Big Bang, leading to a 
rapid exponential expansion of space over at least 60 e-folding \citep{guth1981,linde1981,kolb1990,liddle2000}. The inflationary expansion freezes the quantum fluctuations
in the driving inflaton (field) and leads to the generation of cosmic density and velocity
fluctuations. The distribution of this stochastic fluctuation field is an adiabatic and a homogeneous Gaussian
random field, with a near scale-free Harrison-Zeldovich spectrum,
$P(k) \propto k^1$ \citep{harrison1970,zeldovich1972,mukhanov1981,guthpi1982,starobinsky1982,bardeen1983}. Theoretical considerations emanating from the Central Limit Theorem (CLT) also advocate for the primordial field to have Gaussian characteristics. In terms of observations, the most important evidence for the Gaussian premise is the 
the temperature fluctuations in the Cosmic Microwave Background (CMB) radiation. These directly reflect
the density and velocity perturbations on the surface of last scattering, and thus the mass distribution at the
recombination epoch 379,000 years after the Big Bang, at a redshift of $z \approx 1090$ \citep[see e.g.][]{Pee80,jones2017,pranav2019a}.
The general consensus from the measurements by the COBE, WMAP and Planck satellites is that the CMB temperature fluctuation field is characterized by a homogeneous and isotropic Gaussian random field \citep{smoot1992,bennett2003,spergel2007,komatsu2011,planck2016,bfs17,planck2015cosmoparams}. At later epochs, the large scale structure of the Universe emerges from the primordial perturbation field. As a result, Gaussian random fields also model the large scale structure of the Universe \citep{weygaertlecturenotesi,weygaertlecturenotesii}, including HI signals from structures at the epoch of reionization \citep{mondal2015}. At yet smaller scales, Gaussian random fields serve as models for galactic scale magnetic fields \citep{irinaMagnet,makarenko2018}. 

Due to its central role in describing a multitude of fields of 
interest arising in cosmology and astrophysics, the 
characterization of Gaussian random fields has been a focused endeavor 
\citep{doroshkevich1970,bbks,gdm86,bertschinger1987,scaramella1991,schmalzing1997,matsubara2010}. Since a Gaussian field is completely characterized by the correlation function or the power spectrum, determination of these quantities has been a key focus in the investigation of theoretical models as well as observational probes. Approaches based on higher order correlation functions have been employed to characterize possible deviations from the Gaussian premise \citep{VWHK2000,naselsky2005,HCGM08,bartolo2010,matsubara2010,xia2011}.

\subsection{Characterization of the properties of random fields}

Approaches towards characterizing the stochastic random fields arising in the  cosmological context chiefly focus on either their properties in the frequency space by means of signal decomposition in the Fourier domain, or in the real space by means of examining the topo-geometrical properties. Each of these broad frameworks has yielded novel methodologies, as well as extremely insightful views into the structural properties of stochastic random fields. 

Methodologies in the former category chiefly focus on spectral analysis, through the determination of power spectrum, as well as the higher order bi-spectrum and tri-spectrum \citep{verde2001,sefusatti2007,fergusson2009,matsubara2010}. However, the higher order spectral moments turn out to be extremely resource intensive from the computational point of view. Methodologies in the latter category focus on the properties in the real space chiefly via topo-geometrical characterization. The geometrical aspects involve the notion of volume, area and so on, via the \emph{Minkowski functionals}, or the \emph{Lifshitz-Killing curvatures} \citep{crofton1868,hadwiger1957,adler1981}, while the topological characterization involves the notion of \emph{critical points} \citep{mil63,edelsbrunner2010}, and topological properties associated with them, such as \emph{topological cycles} or equivalently the \emph{topological holes}, finding their basis in \emph{homology theory} \citep{munkres1984,edelsbrunner2010,weygaert2011a,pranavthesis,pranav2017}. The notion of \emph{Euler characteristic} \citep{euler1758,gdm86,gott1989,ppc13,appleby2018,appleby2020} provides a bridge between purely topological and purely geometrical concepts, as while being a purely topological quantity, it can be expressed in a  purely integral geometric setting, as established by the \emph{Theorema Egrerium} due to Gauss \citep{gauss1900}. 

There are a number of ways to study the topo-geometrical complexity of random fields. Common to all of them is examining the sample paths of the field that are tractable from both topo-geometrical as well as probabilistic viewpoints \citep{adler2011}. Given a random field $f$ on a support manifold $\Mspace$, a natural choice is to examine the image of the parameter space $\Mspace$, or subsets of $\Mspace$, with the mapping $f : \Mspace \to \Rspace^d$. If $f$ and $\Mspace$ are intrinsically smooth, so is $f(\Mspace)$, in which case it is natural to express the structure of $f(\Mspace)$ as a combination of the  topological structure of $\Mspace$, the probabilistic structure of $f$, as well as the ambient dimension of $\Mspace$, $dim (\Mspace)$. Despite such obviousness, there are no known results in this direction yet \citep{adl10}. 

An alternative is to examine the inverse problem, where we examine sets in the parameter space $\Mspace$, where the random field exhibits specific properties. These sets are known as \emph{excursion sets} \citep{adler1981,bbks,gdm86,mecke94,park2001,shethwey2004,adl10}, and studying them has yielded extremely insightful results about the structure of stochastic random fields. Specifically, the endeavor to study the excursion sets of random fields has yielded close-form expressions about their geometric characteristics, culminating in the \emph{Gaussian Kinematic Formula}, which gives the expressions for all the Minkowski functionals for Gaussian and Gaussian-related processes in a single compact expression \citep{taylor2006,adl10}. 

While there exist closed-form expressions for the geometric characteristics, this is not the case for purely topological measures, the exception being the Euler characteristic. Beyond the Euler characteristic, it has been extremely hard, if not impossible, to obtain closed-form expressions for purely topological measures such as the distribution of critical points, which are local in nature, or measures such as the homology properties, that are, by definition non-local in nature. This is chiefly due to the fact that the geometric quantities can be expressed in an integral geometric setting, depending solely on the local properties of the manifold, while this is not the case even for local topological measures. Recently though, \cite{cheng2015} have developed methods that yield numerically solvable integrable equations for the expectation and height distribution of all critical points of a smooth isotropic Gaussian random field, restricted to the case of $\mathbb{S}^2$, and up to 3D in the Euclidean setting. Theoretical limitations aside, there have been developments in the recent past on the computational side in topology, that have paved ways for exact computations of topological quantities, both local and non-local in nature, such as the information on critical points, as well as the information on homology properties \citep{munkres1984}, in a hierarchical setting \citep{elz02,edelsbrunner2010}, which is of crucial relevance  in the cosmological scenario, given the hierarchical nature of structure formation in the cosmos.



\subsection{Topology and geometry in cosmology}

Topology and geometry, being fundamental properties of space,  have a long history of interaction with astrophysics and cosmology. At the largest scales, the global topology and geometry of the Universe we live in is an open question, and a wide body of literature exists concerning these studies \citep{Luminet1999,SteinerHyperbolic,roukema2004,SteinerMultiConnected,divalentino2019}. Topological and geometric measures have also been employed for many decades to characterize the mass distribution in the Universe. This is because topology and geometry study properties of space such as shape and connectivity. Therefore, topo-geometrical studies have important repercussions for understanding the organization and connectivity of structures in the cosmos. As an example, the formation and evolution of the large scale structure in the Universe is expected to follow different paths in different structure formation scenarios. In top-down models like Zeldovich's \emph{pancake formalism} smaller structures form through hierarchical fragmentation of larger structures \citep{zeldovich1972}. The opposite is the hierarchical structure formation scenario, where progressively larger structures form by aggregation of smaller objects \citep{Pee80,bond1996,weygaertlecturenotesi}. Understanding the connectivity characteristics of the large scale Universe through topological measures has the potential to reveal deeper information about the path structure formation in the Universe follows. 

The earliest topological methods employed in cosmology chiefly consisted of a characterization of the cosmic mass distribution through the Euler characteristic, or equivalently, the \emph{genus} \citep{bbks,gott1989,sahni1991,rhoads1994,matsubara1996,hikage2002,gott2008,gott2009,eecestimate}. Gott and collaborators pioneered the usage of the Euler characteristic in a series of studies to investigate the large scale structure of the Universe. In a seminal paper, they studied the structure of the matter distribution from the CfA galaxy catalog \citep{gdm86}. A major finding was that the Euler characteristic indicated that the topology of the galaxy distribution at median thresholds was \emph{sponge-like}, meaning that the structure is a single connected object, indented by tunnels and loops. Later, a richer characterization of the topo-geometrical structure of cosmological fields was introduced via the Minkowski functionals \citep{mecke1991,mecke94,schmalzing1996,schmalzinggorski,sahni1998}.

There are ample motivations for extending the topo-geometrical description as supplied by the Minkowski functionals and the Euler characteristic. While the Minkowski functionals are predominantly geometric in nature, the Euler characteristic supplies limited topological information. The Euler-Poincar\'{e} formula states that the Euler characteristic is the alternating sum of another topological invariant called the \emph{Betti numbers} \citep{Bet71,edelsbrunner2010,pranav2017}. Like the Minkowski functionals, there are $d+1$ such topological quantifiers for a $d$-dimensional space, and they are the ranks of the \emph{homology groups} of this space \citep{edelsbrunner2010}. Homology studies the connectivity properties of a space by identifying and characterizing the \emph{topological holes} in the space. There can be zero up to $d$-dimensional topological holes in a $d$-dimensional space. In three spatial dimensions, the topological holes have an intuitive meaning. A $0$-dimensional hole is a \emph{gap} between two connected objects. A $1$-dimensional hole is a \emph{tunnel}, and a $2$-dimensional hole is a \emph{cavity} fully enclosed by a surface \citep{pranavthesis,pranav2017,pranav2019a}. The $p$-dimensional Betti numbers count the number of independent $p$-dimensional holes of a $d$-dimensional space ($p = 0, \dots, d$). 

That the Betti numbers provide more topological information compared to the Euler characteristic is clear from theoretical arguments and practical examples. Theoretically, as the Euler-Poincar\'{e} formula states that the Euler characteristic of a space is the alternating sum of its Betti numbers, two spaces with the same Euler characteristic may turn out to have different topologies when characterized in terms of the Betti numbers. In a practical example, \cite{pranav2019a} show that the Betti characterization of Gaussian field models reveals that the topology is not strictly sponge-like at median density thresholds even for this baseline reference model. Instead, it is composed of many disconnected pieces, each of which is indented by a set of loops and tunnels. This information is not available through an examination of the Euler characteristic. Assuming Gaussian initial conditions for the structure formation scenario in the Universe, it is clear that the large scale structure of the Universe at the later epoch will exhibit a more complex topology.

Another compelling reason for adopting a richer topological language for describing the structures in the Universe is the observation that they are proposed to form hierarchically. A topological formalism that is inherently hierarchical in nature presents a powerful extension to the available methods. The hierarchical extension of homology is \emph{persistent homology}. Introduced by Edelsbrunner and collaborators \citep{elz02,edelsbrunner2010}, there is vast ongoing research in the field in terms of theoretical developments  \citep[see e.g.][]{carlsson2005,carlsson2009a,carlzom09,bobrowski2012,kahle2014,bubenik2015} and applications \citep[see e.g.][]{braintrip,neuronTopology,kannan2019,moraleda2019,pranavReview2021}. It is deeply connected to \emph{Morse theory} \citep{mil63,edelsbrunner2010,pranav2017}, which is the study of properties of scalar functions via the topological structure they induce on the manifold. A Morse function $f$ defined on a compact manifold has a finite number of well separated critical points $\nu_i$, with distinct critical values $f(\nu_i)$. Arranging these critical points in a monotonically decreasing sequence, the manifolds corresponding to the superlevel sets defined by these critical points form a nested hierarchy called a \emph{filtration}. The $p$-dimensional \emph{persistence diagram} is a summary of birth and death of all $p$-dimensional classes across the filtration. In 3D, the persistence diagrams of different dimensions record the creation and merger, or filling up, of  components, tunnels and voids across the filtration. Studying the topological properties of cosmic fields in such a hierarchical setting has the potential to reveal deep information about the hierarchical structure and connectivity of the field. As an example, \cite{pranav2017} proposed that the $0$-dimensional persistence diagrams possess similar information to the familiar merger trees, as both indicate the creation and merger of isolated objects. On these lines, \cite{busch2020} have presented a characterization of the nested hierarchy of density peaks in cosmological simulations using their persistence characteristics. The $1$-dimensional persistence diagrams is representative of the percolation properties of the field. In general, the characterization of giant cycles of all ambient dimensions that permeate through the manifold also characterizes the percolation properties of the manifold. A study of such giant cycles has recently been conducted by \cite{eulerPercolation}. Finally, the $2$-dimensional persistence diagrams maybe regarded as topological  representatives of the void characteristics and distribution of the field, and have the potential to be a useful tool in studies concentrating on the characteristics and distribution of cosmic voids.

\subsection{Topological data analysis}

The recent developments in computational topology have ushered a new era in data analysis, giving rise to \emph{Topological Data Analysis} (TDA) as a field in its own right. The mainstay of TDA has been the development of computational persistent homology, of which the persistence diagrams  are the central pillar. While TDA rests on strong mathematical foundations from topology, in applications, it has faced challenges due to difficulties in handling issues of statistical reliability and robustness, often leading to an inability to make scientific claims with verifiable levels of statistical confidence.  Recently though, there have been developments in the field that begin to address this issue, with the introduction of various summary statistics that condense the information contained in the persistence diagrams in a meaningful way.  There are many statistical approaches to the analysis of a persistence diagram. A detailed account of these methods can be found in the excellent review by \cite{wasserman2016}; also see \cite{summariesPD} for a more recent account. Notable among them are the \emph{persistence silhouette} \cite{chazal2014a}, the \emph{persistence landscape} \citep{bubenik2015},  the \emph{accumulated persistence function} (APF) \citep{APF}, the \emph{persistence entropy function} \citep{persistenceEntropy}, the \emph{Replicating Statistical Topology} (RST) method \citep{RST,Adler2019}, the persistence \emph{intensity function} \citep{edelsbrunner2012,pranavthesis,pranav2017} and the related \emph{persistence images} \citep{adams2017}. The end product of some of these  methods is a summary statistic that yields a single number. While they are of great convenience when model comparison is the objective, the results can be hard to interpret. The landscape method is based on converting the diagrams into a set of 1D functions \cite{bubenik2015}, and the results are hard to interpret as well \citep{chenintensitypd}. The RST technique, in a related yet different approach, develops a method that enables modeling and replication of persistence diagrams. In some real world scenarios, it is the case that only a single realization of data is available, as example the CMB, as well as the large scale Galaxy distribution. In such cases, in the absence of a reliable model that can give rise to multiple realizations, it is a hard task to assign levels of confidence to the results. RST method addresses this issue by treating the persistence diagram as a point cloud, and using Monte-carlo methods to model and replicate the persistence diagram. This results in a suite of diagrams which are constructed and drawn from a distribution whose expected mean, variance and possibly more parameters are the ones that are measured or estimated from the the original diagram. While all the methods enumerated above are powerful, as \cite{chenintensitypd} point out, the persistent intensity function \citep{edelsbrunner2012,pranavthesis,pranav2017}, and the related persistence images \citep{adams2017} are appealing methods in their own right. This is because the end product of these approaches is a \emph{vectorization} of the persistence diagrams, which can then be subjected to rigorous statistical methods, as well as the state-of-the-art machine learning techniques \cite{adams2017,pun2018}. 

In view of the above observations, this paper presents a numerical investigation of the topological properties of an ensemble of
Gaussian random field simulations through persistent homology. The paper is second in a series, aimed at characterizing the topological and geometric properties of a class of cosmologically relevant Gaussian random field models. In the first part of the series \citep{pranav2019a}, we presented an account of the geometric characteristics of Gaussian field through Minkowski functionals. In the process we also provided a brief description of the GKF in the cosmological context. \cite{pranav2019a} also presents the topological characteristics of Gaussian fields through homology, as quantified by the Betti numbers. 

The topo-geometrical analyses of Gaussian fields follow the broad program laid out in \cite{isvd10} and \cite{pranavthesis}, aimed at introducing topological data analysis in the cosmological context. Following this, in \cite{pranav2017} we described in formal detail the mathematical foundations and computational aspects of the framework. This is keeping in mind the fact that these methods carry with them the power of an elaborate and solid mathematical framework behind them, and complement the existing methods for gleaning meaningful information out of the ever-growing cosmological datasets. That these methodologies are useful is evident from a recent proliferation of their use in the astronomical and cosmological disciplines, in a variety of contexts including structure detection and identification \cite{sousbie1,sousbie2, shivashankar2015,xu2019}, statistical characterization of ISM \cite{makarenko2018,utreras2020}, statistical characterization of cosmological density fields \cite{ppc13,pranavde,chen2015a,codis2018,cole2018,pranav2019a,feldbrugge2019,kono2020,biagetti2020,wilding2020}, as well as the topological characterization of the temperature fluctuations in the cosmic microwave background \citep{pranav2019b}.


Section~\ref{sec:topology_ch2} presents a brief description of 
the topological background, which also includes a brief discussion on the excursion set formalism. Section~\ref{sec:grf} lists the essential concepts and definitions related to Gaussian random field, and also presents a brief description of the models analyzed in this paper. We begin our results with an analysis of the persistent homology characteristics of power-law Gaussian field models in Section~\ref{sec:intensity_map_result_plaw}, followed by the investigation of the characteristics of simulations based on the Lambda Cold Dark Matter (LCDM) model. Section~\ref{sec:intensity_map_stats}  presents an analysis of the statistical properties of persistence diagrams of the models, by means of investigating the marginal distributions. These sections present the main results of the paper. We present a brief discussion and concluding remarks in Section~\ref{sec:discussion_ch2}.

\label{sec:morseLemma}
\begin{figure*}
	\centering
	\subfloat[]{\includegraphics[width=0.4\textwidth]{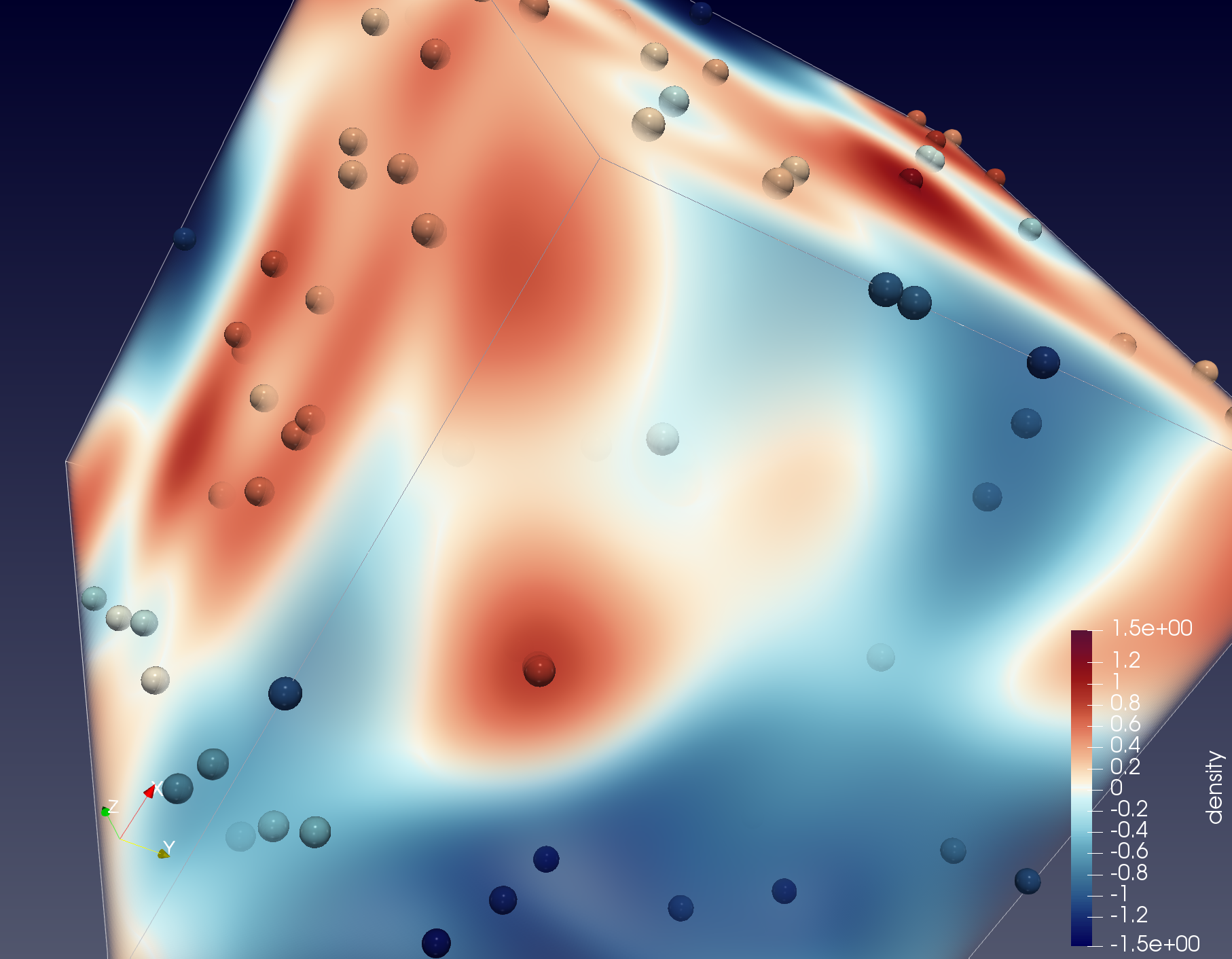}}
	\subfloat[]{ \includegraphics[width=0.4\textwidth]{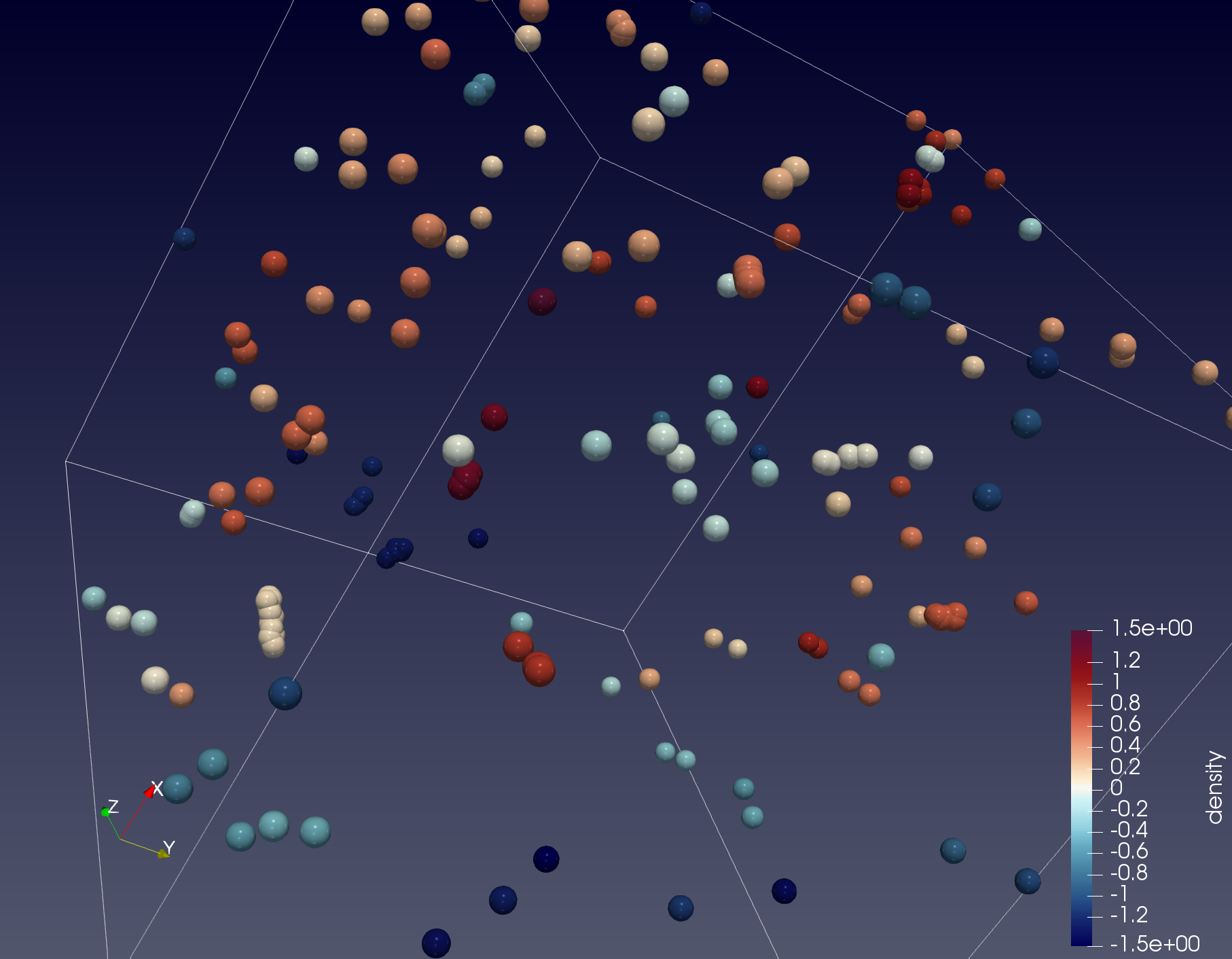}}\\
	
	\caption{ Visualization of the density field and critical points of a simulation of smooth Gaussian random field. Left: Volume rendering of the density field. The critical points of the field are marked with balls color coded by density value. Right: A  visualization of the critical points only.}
	\label{fig:volren_CP}
\end{figure*}

\begin{figure}
	\centering
	\subfloat[]{\includegraphics[width=0.2\textwidth]{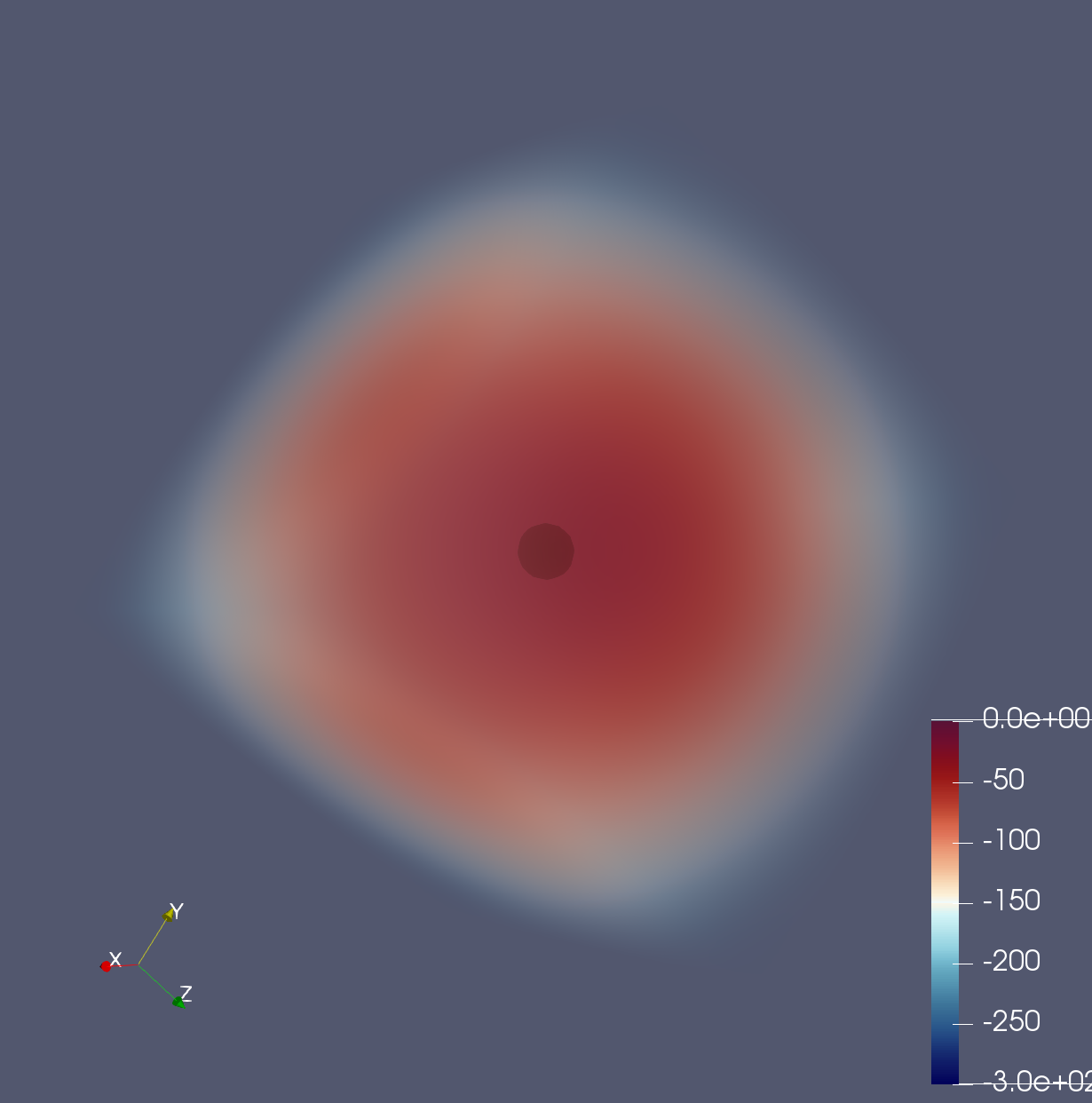}}
	\hspace{0.001\textwidth}
	\subfloat[]{ \includegraphics[width=0.2\textwidth]{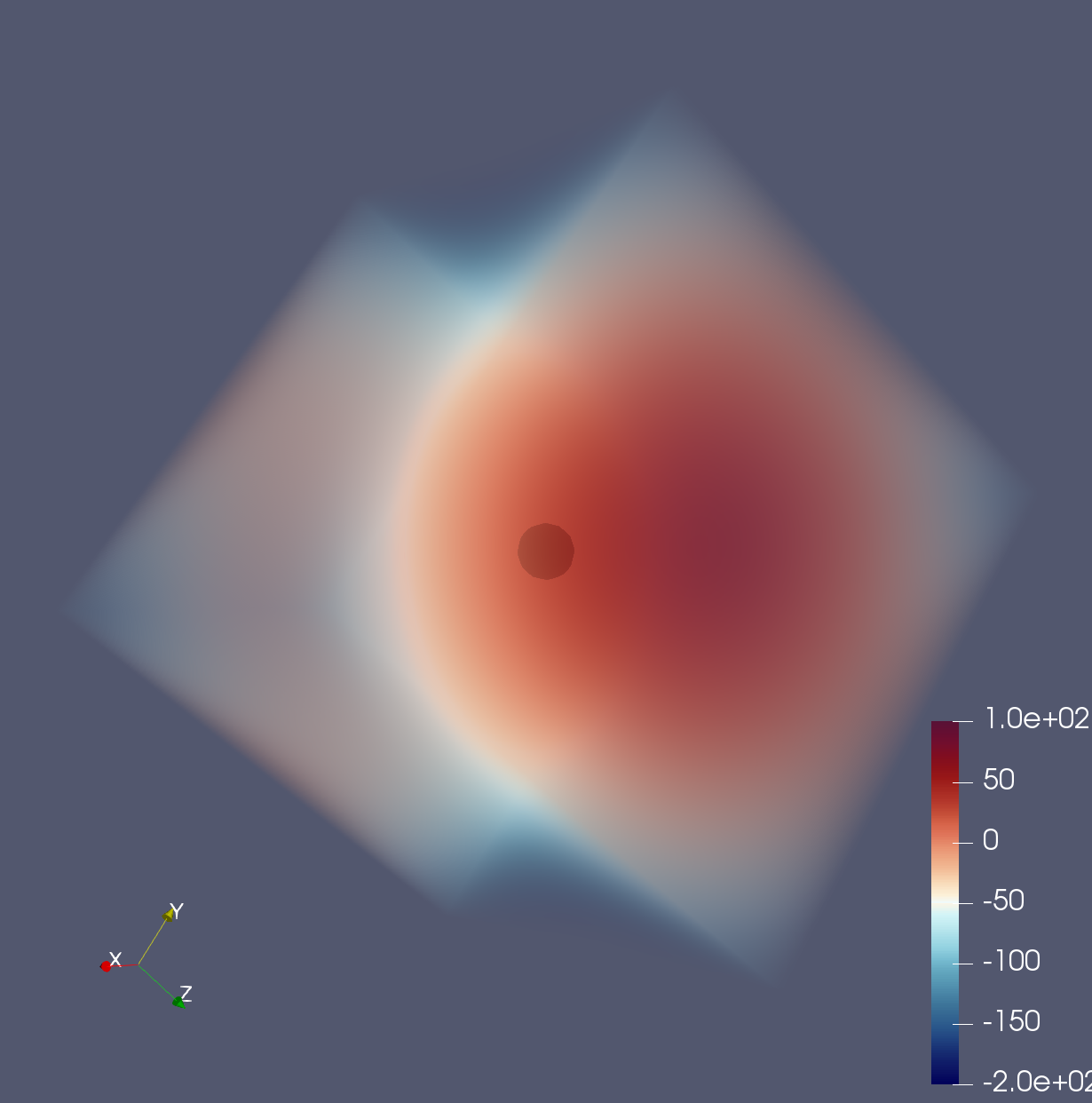}}\\
	
	\subfloat[]{\includegraphics[width=0.2\textwidth]{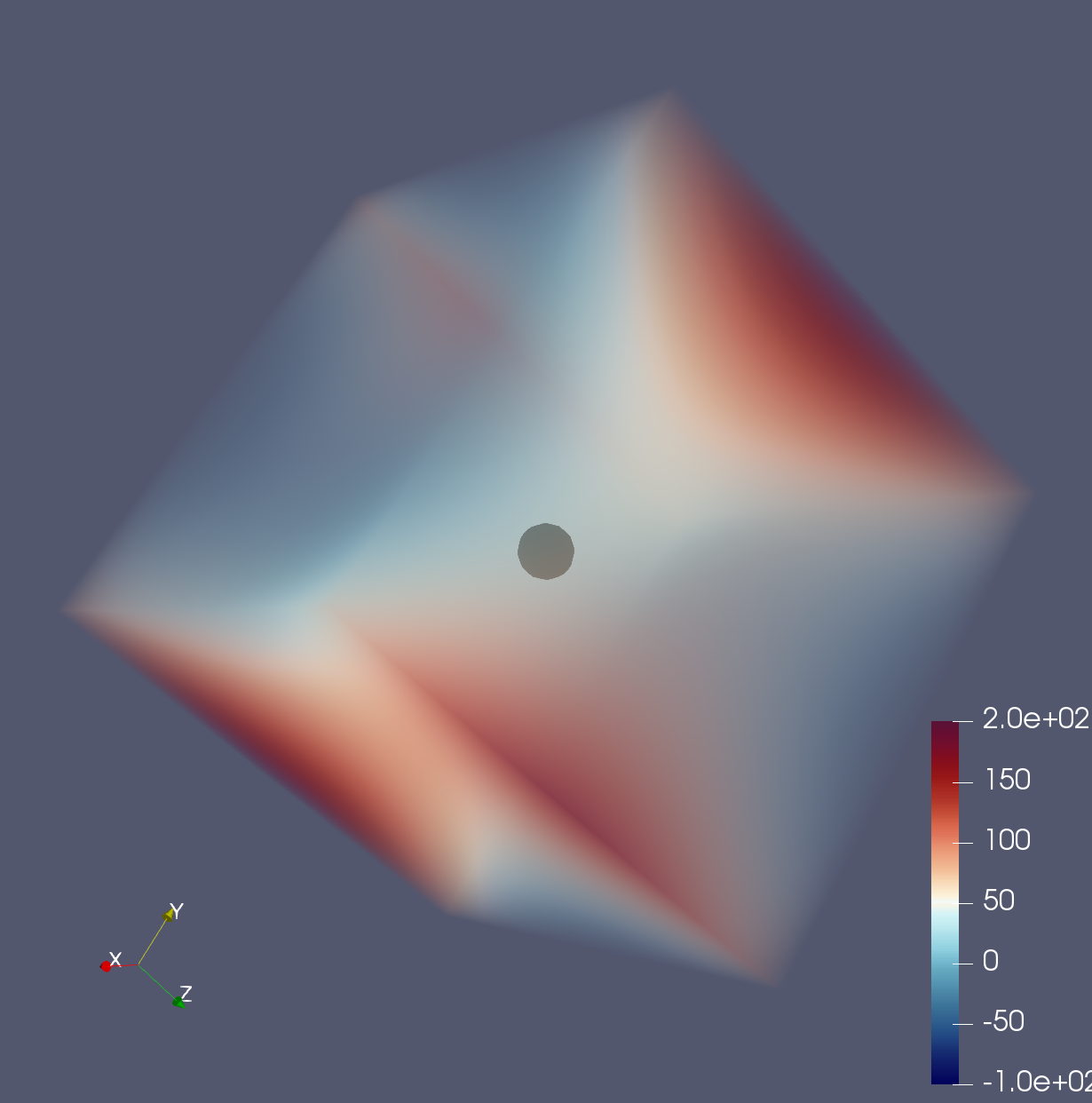}}
	\hspace{0.001\textwidth}
	\subfloat[]{ \includegraphics[width=0.2\textwidth]{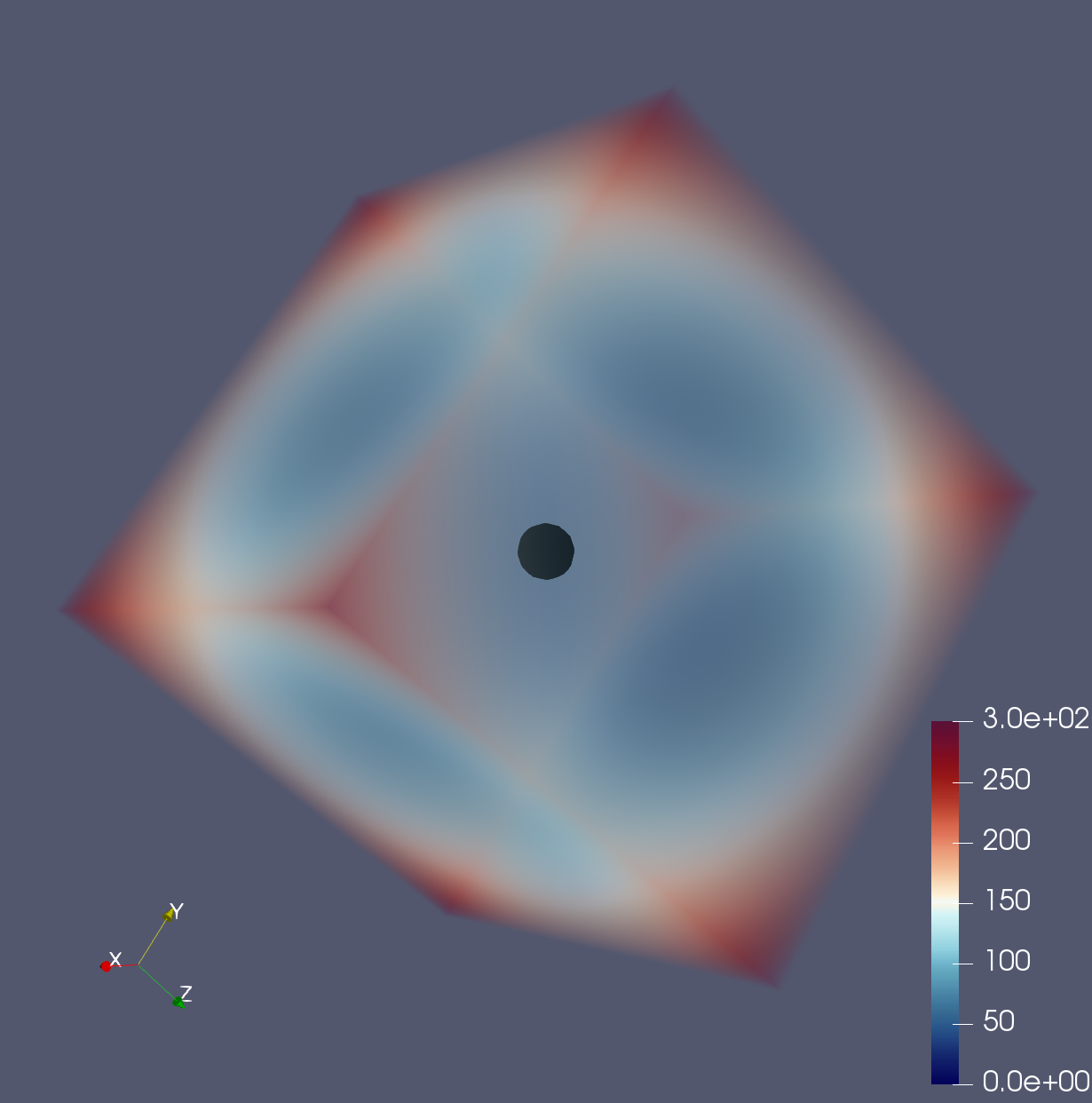}}\\
	
	\caption{ Volume rendering of the density field in the neighborhood of the four kinds of singularities in 3D. The critical point is indicated by a solid ball. From top-left, clockwise:  maximum, $2$-saddle, $1$-saddle and minimum.}
	\label{fig:singularities}
\end{figure}

\begin{figure}
	\centering
	\subfloat[]{\includegraphics[width=0.2\textwidth]{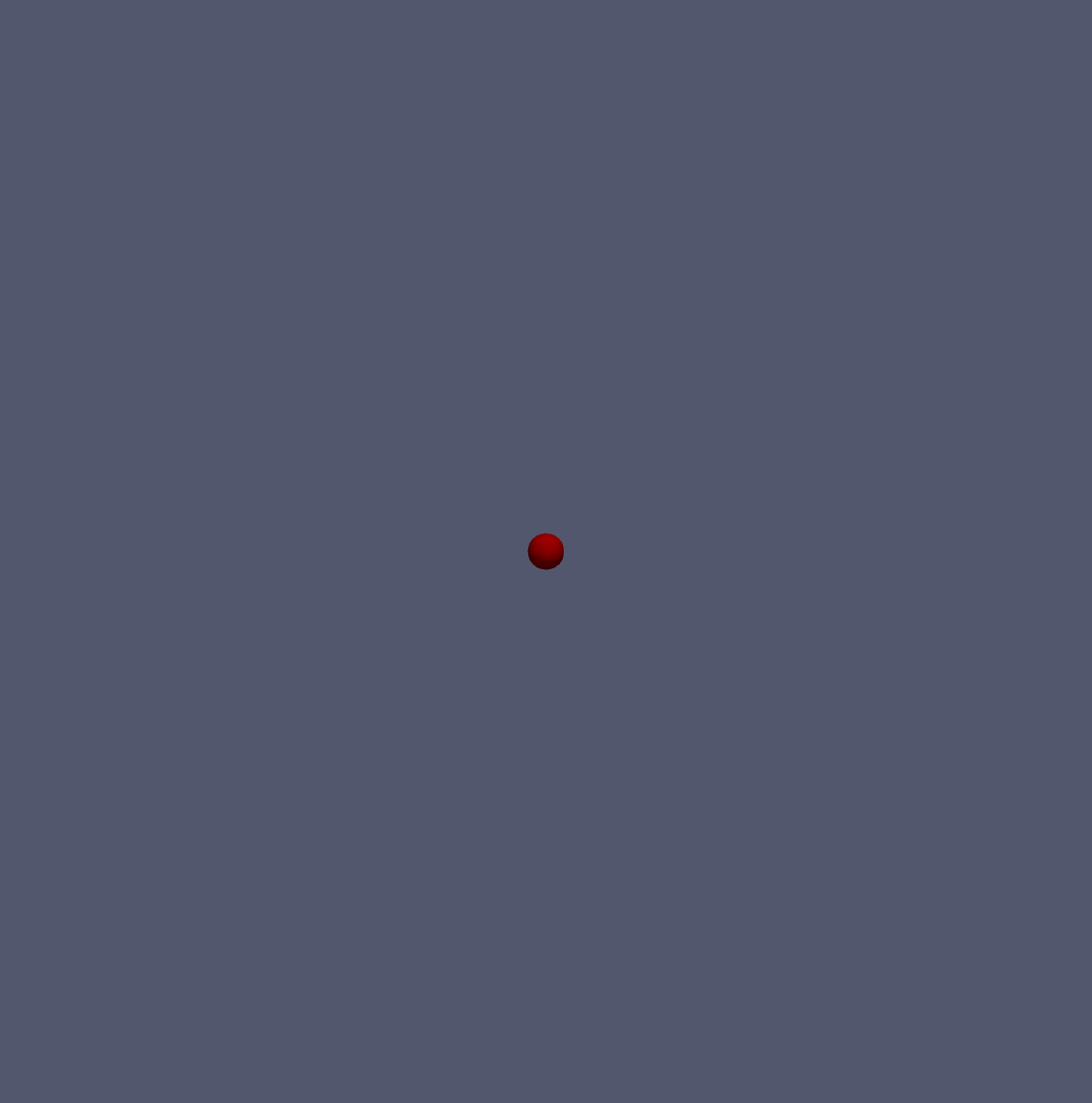}}
	\hspace{0.001\textwidth}
	\subfloat[]{ \includegraphics[width=0.2\textwidth]{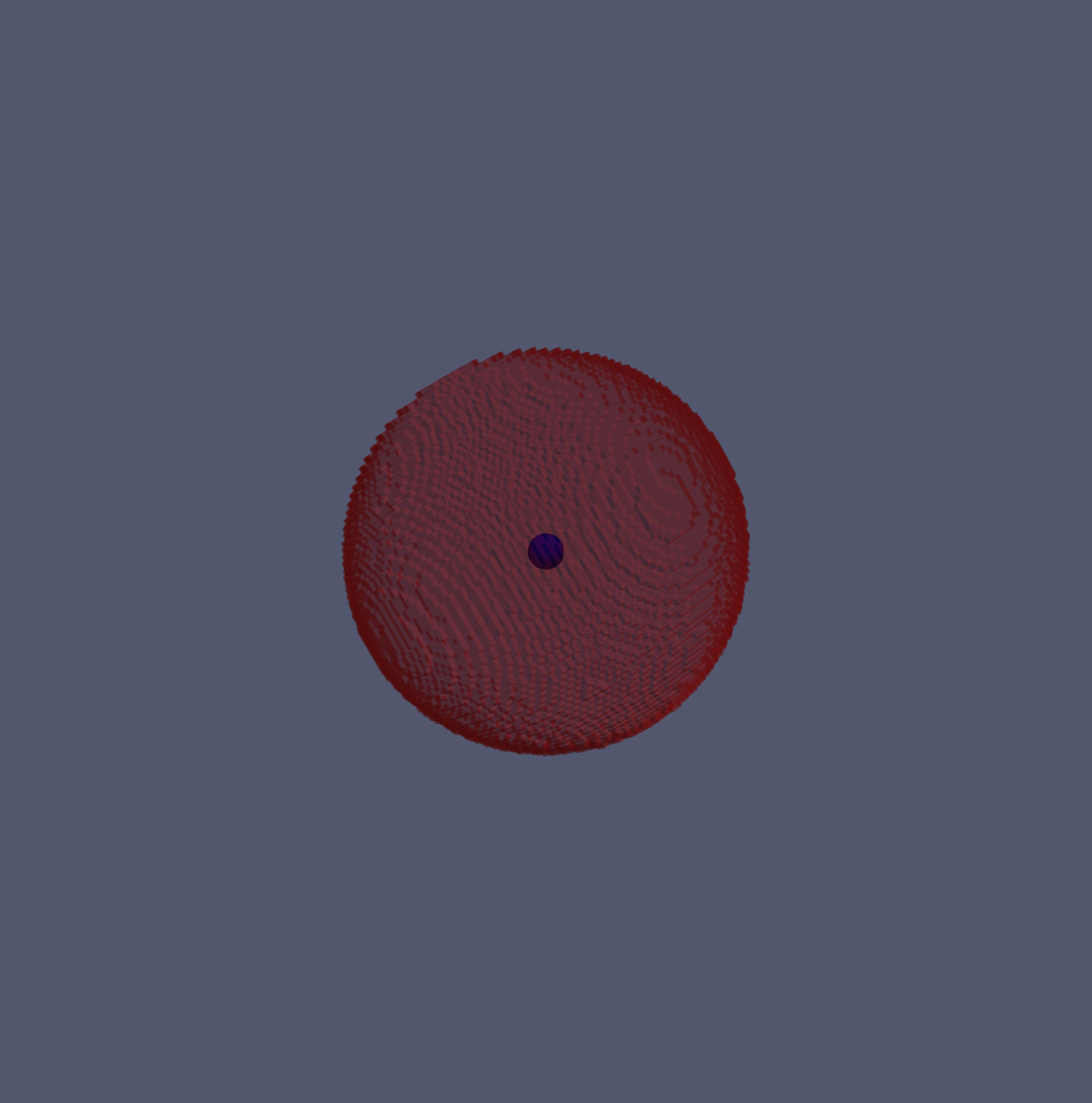}}\\
	
	\subfloat[]{\includegraphics[width=0.2\textwidth]{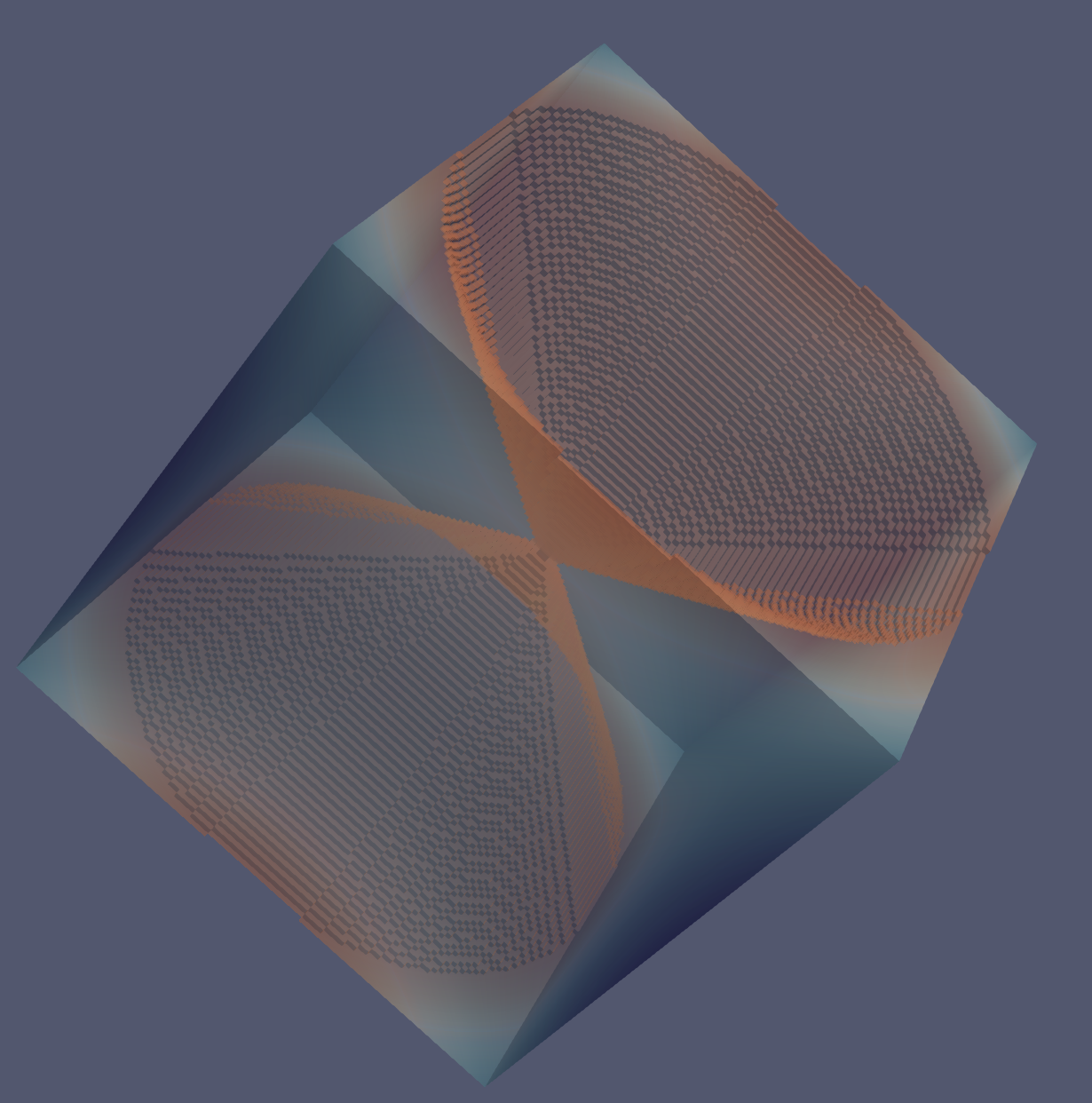}}
	\hspace{0.001\textwidth}
	\subfloat[]{ \includegraphics[width=0.2\textwidth]{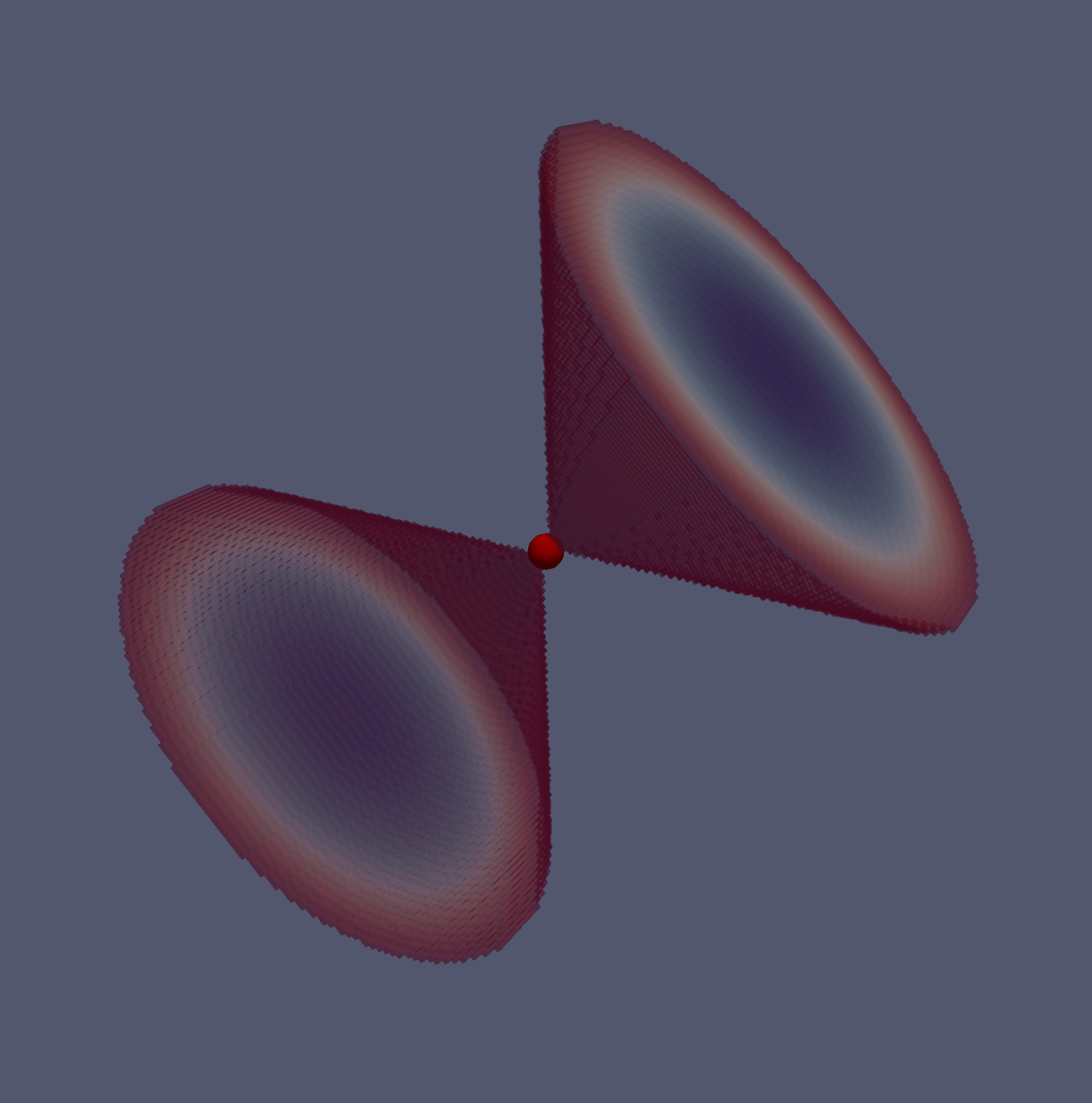}}\\
	
	\caption{ Visualization of the geometry of the ascending and descending manifolds in the neighborhood of the four kinds of singularities in 3D. The ascending manifold of a maximum is empty, as shown in panel (a). Its descending manifold is an open ball, as depicted in panel (b). The situation is reversed for a minimum: its ascending manifold is an open ball, and its descending manifold is empty. The ascending manifold of a 1-saddle is a two sheet hyperboloid meeting at the saddle, as depicted in panel (c). The descending manifold of a 1-saddle is a one sheet hyperboloid (an hourglass figure), that approaches along a circle of directions till it joins at the saddle. The situation is reversed for a 2-saddle, whose its ascending manifold is a one sheet hyperboloid, and the descending manifold is a two sheet hyperboloid.}
	\label{fig:ascDescManifold}
\end{figure}

\begin{figure*}
	\centering
	\subfloat[]{\includegraphics[width=0.4\textwidth]{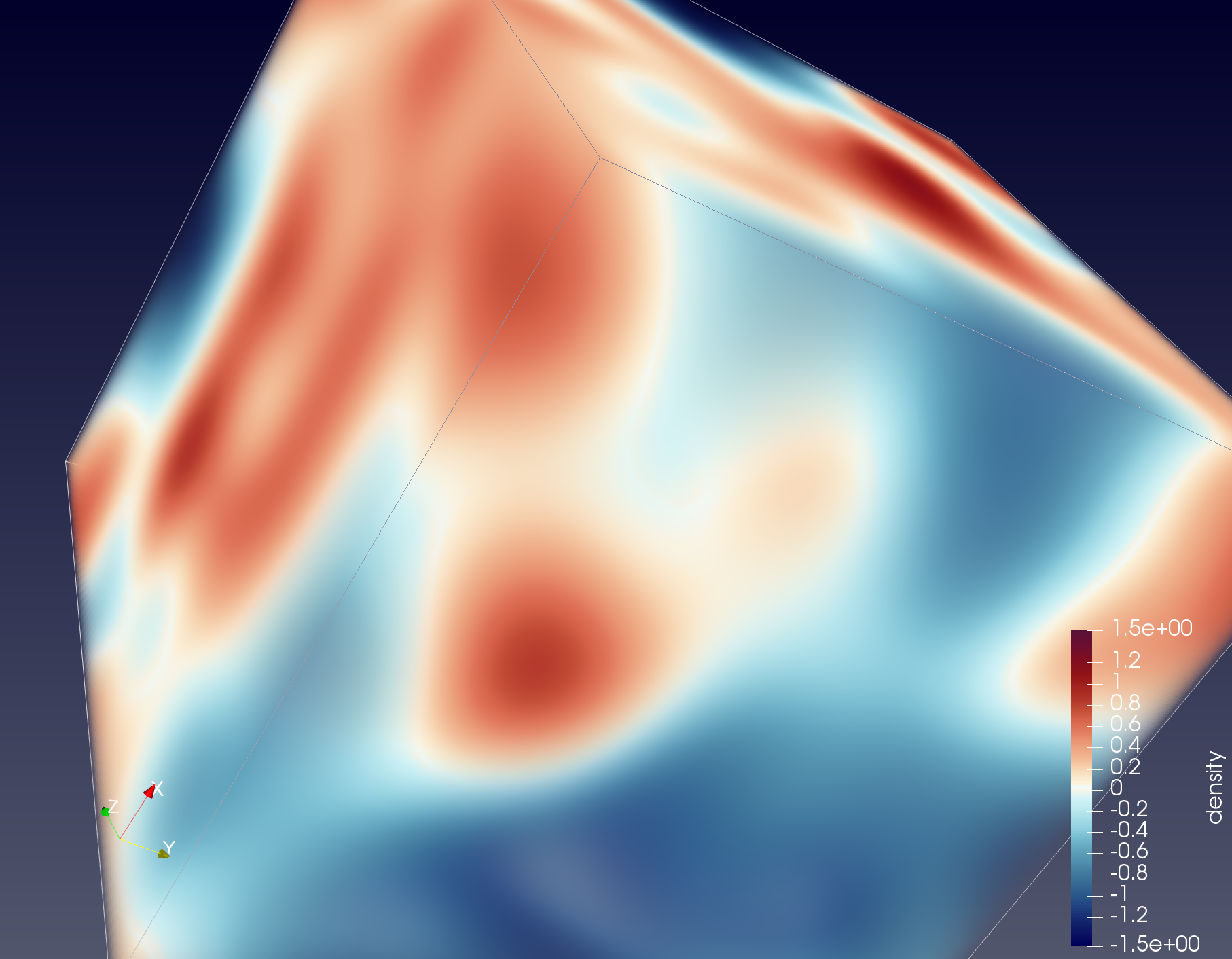}}
	\hspace{0.001\textwidth}
	\subfloat[]{\includegraphics[width=0.4\textwidth]{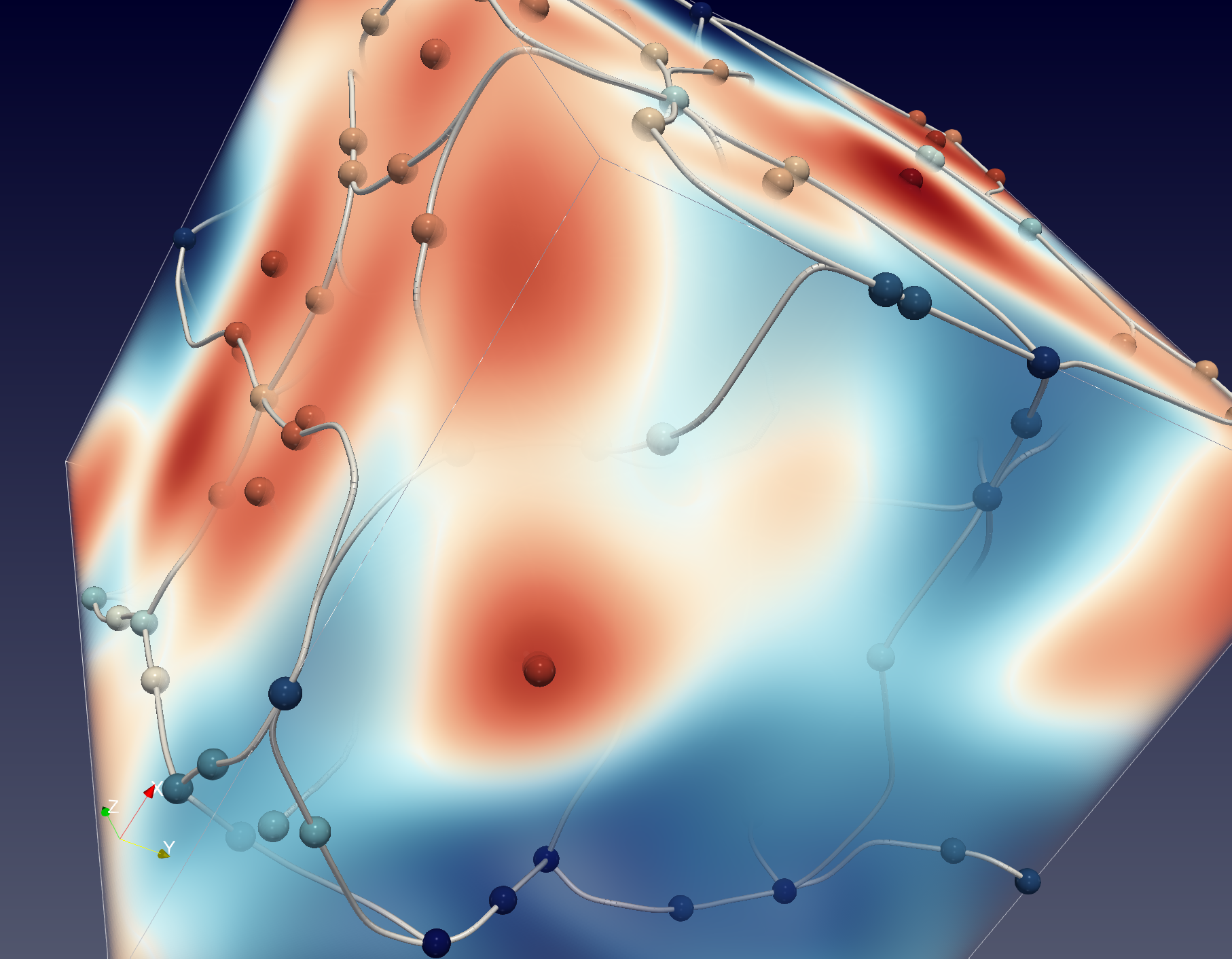}}\\
	\subfloat[]{\includegraphics[width=0.4\textwidth]{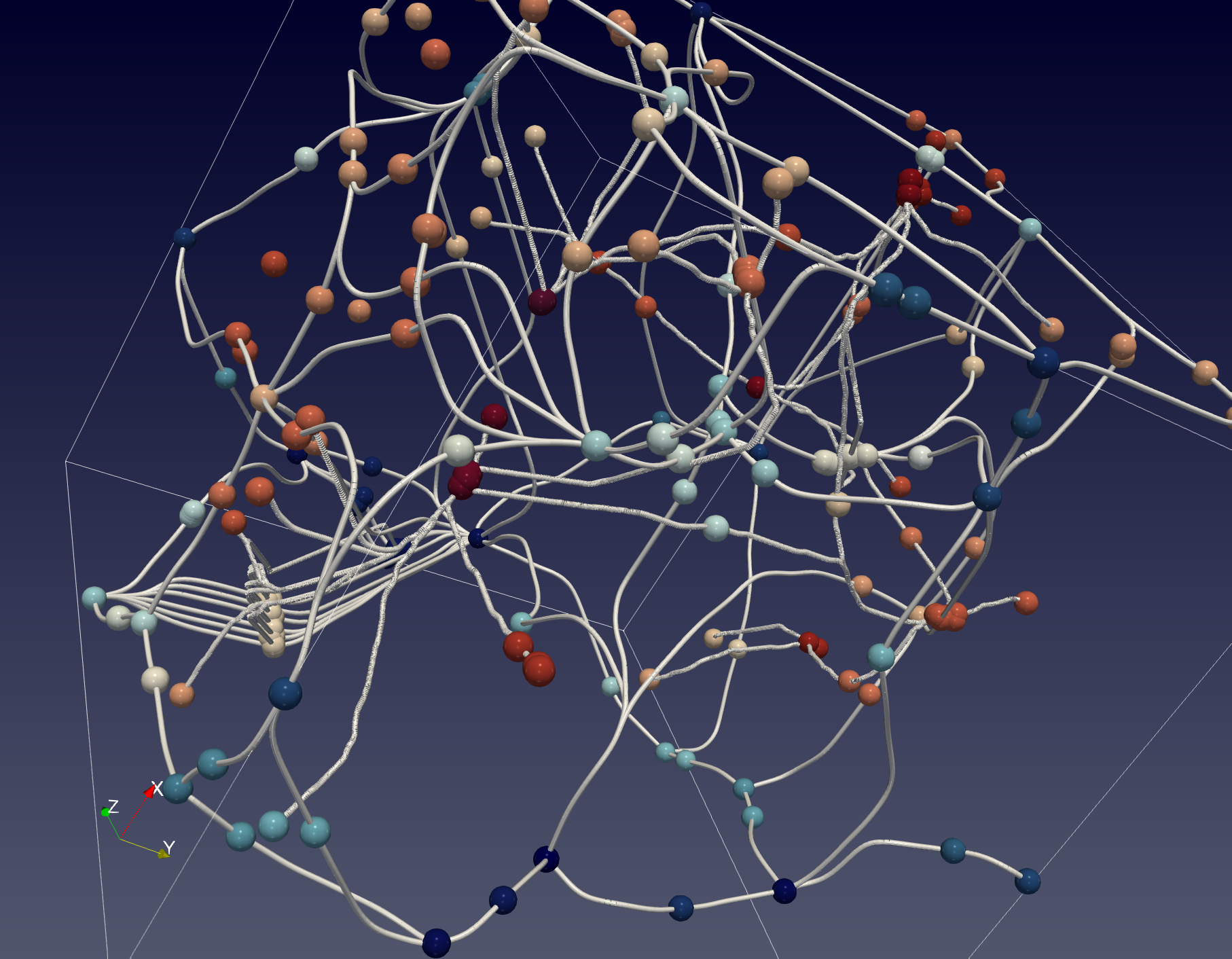}}
	\hspace{0.001\textwidth}
	\subfloat[]{\includegraphics[width=0.4\textwidth]{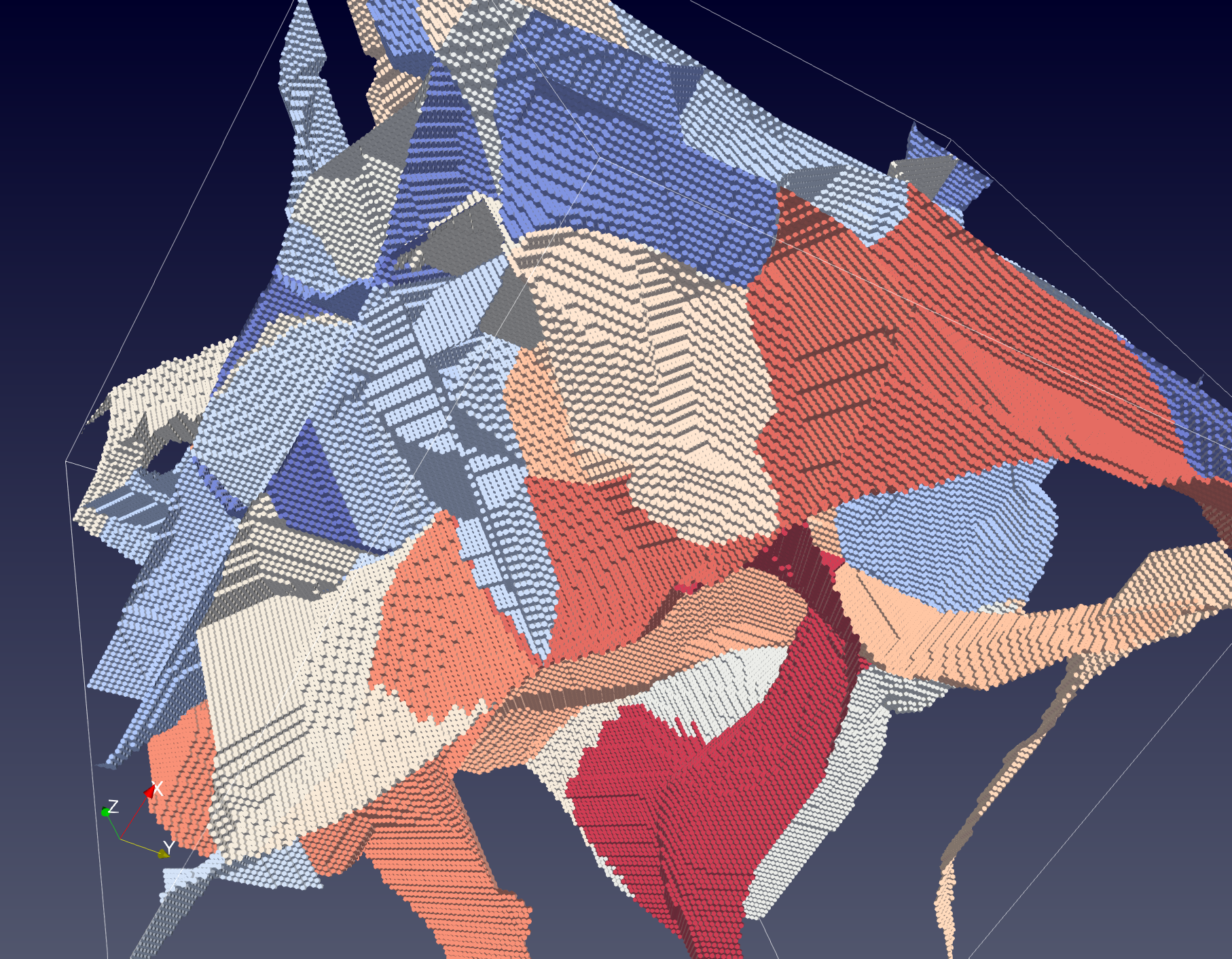}}\\
	
	\caption{ Visualization of the density field and the corresponding Morse-Smale complex. Panel (a) illustrates the volume rendering of the field. The Morse-Smale complex of the field is presented in panel (b). The opacity of the 3D volume rendering decreases while moving away from the center, in order to facilitate a visualization of the interior. The \emph{nodes} (0D), \emph(arcs) (1D), \emph{sheets} (2D) and the \emph{basins} (3D) together form the various components of the Morse-Smale complex. The nodes together with the arcs form the skeleton of the Morse-Smale  complex. This is referred to as the \emph{combinatorial representation} of the Morse-Smale complex, and illustrated in panel (c). Panel (d) illustrates the 2 dimensional sheets that form the boundary of the 3D voids that terminate at minima.}
	\label{fig:MScomplex}
\end{figure*}

\section{Stochastic random fields on manifolds: topology and geometry}
\label{sec:topology_ch2}

A large fraction of the generalized field of astronomy and cosmology, as well as other disciplines, is the study of stochastic functions, and fields \footnote{When the parameter space has dimension $D > 1$, the function is referred to as a \emph{field}.} on manifolds. Stochastic fields, cosmologically relevant examples of which are the 3D matter density distribution field, or the temperature fluctuation field of the cosmic microwave background, have been studied through a variety of techniques\footnote{The parameter space for the spatial mass distribution is the 3D Euclidean space, while the parameter space for the CMB is $\mathbb{S}^2$.}. In this section, we present a discussion on various inter-related notions, that form the essential background to understanding topo-geometrical properties of stochastic fields, through the deformation characteristics they induce on their support manifolds. A major part of our understanding of the topo-geometrical characteristics of stochastic functions arises through studying the \emph{excursion sets}, and we devote a part of the theoretical background to the \emph{excursion set formalism}. Throughout, we employ visualizations based on simulations of Gaussian random fields to augment the understanding of the reader. We propose a set of standard references for this section. On the side of pure geometric and topological discourse, we refer the reader to \cite{rice1944}, \cite{kac1943}, \cite{adler1981}, \cite{adl10} and \cite{edelsbrunner2010}.  For the development of excursion set formalism in the cosmological setting essential references are \cite{ps74,peacock1990,bond1991,sheth1998,kerscher2001b,sheth2002,shethwey2004}; also see the review by \cite{zentner2007}, and the references there in. 
For a concise and technical, yet user-friendly, description on the topo-geometrical background, as well as computational intricacies, we refer the reader to \cite{pranav2017} and \cite{pranav2019a}.

\subsection{Singularities and the structure of manifold in their neighborhood: critical points and Morse-Smale complex}
\label{sec:morse_ch2}

 When considering spatial stochastic processes, a common approach is to investigate their multi-scale properties via Fourier transform analysis, or equivalently the (higher)-order correlation functions. Alternative approaches aim at understanding the properties of these fields via characterizing the properties of critical point distribution, and geometric properties of the manifold in the neighborhood of critical points. Such investigations find their root in topo-geometrical methods. 

One such methodology for investigating stochastic fields finds its basis in \emph{Morse theory} \cite{mil63}, if they satisfy three conditions: that they are continuous, smooth and generic. Smoothness implies the function is at least twice differentiable, and genericity implies that the \emph{critical points} are non-degenerate. As a consequence a Morse function defined on a compact manifold has a finite number of critical points and critical values. Assume a smooth function $f$ defined on a 3D compact manifold $\Mspace$. A critical point of $f$ is the location $\vec{x}$ where the gradient vanishes
\begin{equation}
	\nabla f |_{\vec{x}} = 0.
\end{equation}
\noindent The function value at the critical point is called the \emph{critical value}. All locations in $\Mspace$ that are not critical are called \emph{regular points}. The function takes \emph{regular values} at regular points. Figure~\ref{fig:volren_CP} presents a visualization of a smooth Gaussian random field, with the critical points drawn in solid balls and colored according to the density value. The \emph{Hessian} of $f$ at a location $\vec{x}$ is the matrix of the second derivatives $H(\vec{x})$. A critical point is \emph{non-degenerate} if the Hessian is non-singular, i.e., the determinant is non-zero. The number of negative eigen values of the Hessian is independent of the coordinates and is called the \emph{index} of the critical point. There are $d+1$ kinds of critical points  for a $d$-dimensional manifold. In 3D, the four kind of critical points are the \emph{maximum} (index-3), \emph{2-saddle} (index-2), \emph{1-saddle} (index-1) and the \emph{minimum} (index-0). Figure~\ref{fig:singularities} presents the volume rendering of the density function in the local neighborhood of various kinds of critical points in 3D. Starting from top-left, we present the density field around a maximum, $2$-saddle, $1$-saddle and minimum respectively.

Given a Morse function $f$, the gradient flow can be used to decompose the manifold according to where the flow originates and terminates \citep{edelsbrunner2010}. An \emph{integral line} in $\Mspace$ is a curve whose tangent is aligned with the gradient of $f$ at every point $\vec{x}$ in $\Mspace$. The function value along the integral line is monotonic, and the integral line originates as well as terminates at critical points of $f$. Integral lines originating at a critical point $p$, along with $p$, constitute the \emph{ascending manifold} of $p$. The \emph{descending manifold} of the critical point $p$ is the set of integral lines that terminate at $p$, along with $p$. Figure~\ref{fig:ascDescManifold} illustrates the geometry of the ascending and descending manifolds of the different kinds of critical points. The ascending manifold of a maximum is empty, as shown in panel (a). Its descending manifold is an open ball, as depicted in panel (b). The situation is reversed for a minimum: its ascending manifold is an open ball, and its descending manifold is empty. The ascending manifold of a 1-saddle is a two sheet hyperboloid meeting at the saddle, as depicted in panel (c). The descending manifold of a 1-saddle is a one sheet hyperboloid (an hourglass figure), that approaches along a circle of directions till it joins at the saddle \citep{edelsbrunner2010}. The situation is reversed for a 2-saddle, whose its ascending manifold is a one sheet hyperboloid, and the descending manifold is a two sheet hyperboloid.

One can decompose $\Mspace$ into collections of ascending and descending manifolds that share common origin and destination, based on the integral lines of $f$. If additionally, the ascending and the descending manifolds intersect only \emph{transversally}, the given function is a  \emph{Morse-Smale function}, and this decomposition forms the \emph{Morse-Smale complex}. Transversality implies that if the ascending and descending manifolds intersect, the dimension of the intersection is the same as the difference in the indices of the critical points. Figure~\ref{fig:MScomplex} illustrates the Morse-Smale complex decomposition of a typical smooth Gaussian random field. Panel (a) illustrates the volume rendering of the field. The Morse-Smale complex of the field is presented in panel (b). The opacity of the 3D volume rendering decreases while moving away from the center, in order to facilitate a visualization of the interior. The \emph{nodes} (0D), \emph(arcs) (1D), \emph{sheets} (2D) and the \emph{basins} (3D) together form the various components of the Morse-Smale complex. The nodes together with the arcs form the skeleton of the Morse-Smale  complex. This is referred to as the \emph{combinatorial representation} of the Morse-Smale complex, and illustrated in panel (c). Panel (d) illustrates the 2 dimensional sheets that form the boundary of the 3D voids that terminate at minima.

\subsection{Excursion set formalism}
\label{sec:excursion}

There are a number of ways to study the topo-geometrical complexity of random fields. Common to all of them is examining the sample paths of the field that are tractable from both topo-geometrical as well as probabilistic viewpoints. One of the ways would be to examine the image of the parameter space $\Mspace$, or subsets of $\Mspace$, with the mapping $f : \Mspace \to \Rspace^d$. If $f$ and $\Mspace$ are intrinsically smooth, so is $f(\Mspace)$, in which case it would be natural to express the structure of $f(\Mspace)$ as a combination of the  topological structure of $\Mspace$, the probabilistic structure of $f$, as well as the ambient dimension of $\Mspace$, $dim (\Mspace)$. Despite such obviousness, there are no known results in this direction yet. 

An alternative is to examine the inverse problem, where we examine sets in the parameter space $\Mspace$, where the random field exhibits specific properties. These sets are known as \emph{excursion sets}, which we define as 

\begin{equation}
\label{rob:eq:AH}
A_\cH \ \equiv \ A_\cH\left(\vec f,\Mspace\right) \ = \ \left\{x\in \Mspace: \vec f(x)\in \cH\right\},
\end{equation}

\noindent where,  $\cH\subset\mathbb R^d$, is called a \emph{hitting set}. When $d=1$, so that $f$ is real-valued, and $\cH$ is the set $[\nu,\infty)$, we are looking at \emph{super level sets} of $f$, so that 
\begin{equation}
\label{intro:Au:equn}
A_\nu \ \equiv\  A_u\left(\vec f,\Mspace\right)\ =\  \{x\in \Mspace: f(x)\geq \nu\}.
\end{equation}

Investigating the excursion sets has yielded extremely important insights about the structure of random fields, resulting in close-form expressions of topo-geometrical characteristics in specific cases. These include some classical results such as the number of up-crossings of a random function, as well as computations of closed form expressions of \LKCs of excursion sets of Gaussian and Gaussian related random fields in a wide variety of settings, via the \emph{Gaussian Kinematic formula} \cite{adler1981,adl10,pranav2019a}. The GKF yields close form expression for geometrical characteristics of the random functions, yet through celebrated \emph{Gauss-Bonnet-Chern} theorem, it also establishes a link to the purely topological Euler characteristic. Despite the remarkable success of excursion set formalism, some crucially important problems, predominantly purely topological in nature, remain intractable as of yet in terms of closed form expressions. Prominent examples include the height distribution of critical points, the exception being the alternating sum of critical points, which is another representation of the Euler characteristic. More complex topological measures, such as the homology characteristics of excursion sets of random fields, which is the main theme of this paper, also remain intractable as of now. The most prominent hurdle in the case of topological measures is their non-local nature, which is a hindrance towards expressing them in integral-geometric settings. The only, and the most celebrated exception is the Euler characteristic

\subsubsection{Kac-Rice formula}

The beginning of excursion set formalism is traced back to the Kac-Rice theorem, derived by \cite{rice1944} and \cite{kac1943}, in independent settings, which gives the number of up-crossings of a random function in 1D. We sketch it here, for its ease of understanding.

Let $f$ be a real-valued random process on the positive half-real line in the interval $[0,T]$, and $u \in \Rspace$. Let $N^+_u (0,T)$ denote the number of \emph{up-crossings} of $f$ at level $u$, such that

\begin{equation}
N^+_u (0,T) \equiv \#\{t \in [0, T] : f(t) = u, f'(t) > 0\}.
\end{equation}

\noindent Assuming $N^+_u (0,T)$ is well defined and finite, the technical conditions of which are given in detail in \cite{adl10}, we compute the expectation $\Exp \{N^+_u (0,T)\}$. A crucial requirement is that the \emph{up-crossing} points of $f$, meaning the points $t \in [0, T]$ at which $f(t) = u$ and $f'(t) > 0$, are isolated. This condition implies there is a small interval $I$ in which there are no other up-crossings and that $f'(t) > 0$ throughout. Under technical conditions defined in \cite{adl10},  invoking the properties of the \emph{Dirac-delta} function $\delta_u (x)$, and treating it as a smooth function, a change of variable  gives

\begin{equation}
1 = \int_{\mathbb R} \delta_u (y)\,dy = \int_I \delta_u (f(t)) \cdot f'(t)\,dt.
\end{equation}

\noindent Joining all such intervals $I$, with the observation that there is no contribution to the following integral outside of them, we have 

\begin{equation}
N^+_u (0,T) = \int_{0}^{T} \delta_u (f(t)) \cdot \mathbbm{1}_{[0, \infty]} (f'(t)) \cdot f'(t)\,dt,
\end{equation}

\noindent where, $ \mathbbm{1}_{[0, \infty]}$, is the \emph{indicator function} that selects intervals with positive values of $f'(t)$. Assuming that the pairs of random variables $(f(t), f'(t))$ have joint probability densities $p_t$, we take the expectation, exchanging orders of integration, and we get 

\begin{eqnarray}
\Exp \{N^+_u (0,T)\} &=& \int_{0}^{T} dt \int_{-\infty}^{\infty} dx \int_{0}^{\infty} dy\, y\delta_u(x) p_t(x,y)\\ \nonumber
&=& \int_{0}^{T}  \int_{0}^{\infty} y\,p_t(u,y)\,dy\,dt.
\end{eqnarray}

The above is the \emph{Rice-formula} in its most basic form, valid for all processes on $\Rspace$ for which the above operations are justifiable \citep{adl10}. However,  it is extremely hard to compute, unless $f$ is a Gaussian or a Gaussian-related process. Assuming $f$ is Gaussian, with zero mean and unit variance, and noting that $f(t)$ and $f'(t)$ are independent of each $t$, we denote the variance of $f'(t)$ by $\lambda_t$, so that 

\begin{eqnarray}
\Exp \{N^+_u (0,T)\} &=&\frac{e^{-u^2/2}}{2\pi}  \int_{0}^{T}  \int_{0}^{\infty} y \frac{e^{-y^2/2\lambda_t}}{\sqrt{\lambda_t}}\,dy\,dt\\ \nonumber
&=& \frac{e^{-u^2/2}}{2\pi}  \int_{0}^{T} \lambda_t^{1/2}\, dt.
\end{eqnarray}

For a stationary process, $\lambda_t = \lambda_2$, the second spectral moment, so that the formula simplifies further to 

\begin{equation}
\Exp \{N^+_u (0,T)\} = \frac{\lambda_2^{1/2}T}{2\pi} e^{-u^2/2},
\end{equation}

\noindent giving the classical Rice formula.

\subsubsection{Expectation meta-theorem in higher dimensions}

The generalization of Kac-Rice theorem in higher dimensions leads to the \emph{Expectation meta-theorem}, where the basic idea is to arrive at expressions where a vector-valued random field takes values in a particular set. This is especially relevant for cosmological applications, where we hardly encounter examples in 1D. We describe the general setting below, followed by a commentary on its usage in particular settings. 

Let $f = (f^1,\ldots,f^N)$ and $g = (g^1,\ldots,g^N)$ be $\Rspace^N$ - and $\Rspace^K$-valued random fields, for $N, K \ge 1$. Let $T \subset \Rspace^N$ be a compact parameter set, and $B \subset R^K$ be an open set, with additional technical conditions on their boundary given in \cite{adl10}. Since $f$ is defined on $\Rspace^N$, its gradiant $\nabla f$ is an $N \times N$ matrix of the first order partial derivatives of $f$, given by 

\begin{equation}
	(\nabla f) (t) \equiv \nabla f (t) \equiv (f_j^i (t))_{i,j=1,\ldots,N} \equiv \left(\frac{\partial f^i (t)}{\partial t_j}\right)_{i,j=1,\ldots,N}.
\end{equation}

\noindent Given all of the above, if 

\begin{equation}
N_u \equiv N_u (T) \equiv N_u (f, g: T, B)
\end{equation}

\noindent denotes the number of points in $T$ where
\begin{equation}
f (t) = u \in R^N
\end{equation}
\noindent and
\begin{equation}
g(t) \in B \subset R^K, 
\end{equation}

\noindent and $p_t (x, \nabla y, v)$ denotes the joint densities of $(f_t, \nabla f_t, g_t)$, then under technical conditions detailed in \cite{adl10}, which have mostly to do with adequate regularity conditions of continuity and boundedness, and most of which can be lifted if both $f$ and $g$ are Gaussian, we have with $D = N(N+1)/2 + K$

\begin{equation}
\Exp \{N_u\} = \int_T \int_{\Rspace^D} \Exp \{|\text{det}\, \nabla f  (t)| \mathbbm{1}_B g(t)| f (t) = u\} p_t (u)\, dt 
\end{equation}

\noindent which may also be written as 

\begin{equation}
	\Exp \{N_u\} = \int_T \Exp \{| \text{det}\, \nabla f (t)| \mathbbm{1}_B g(t)| f (t) = u\} p_t (u)\, dt,
\end{equation}
\noindent where $p_t$ is the density of $f(t)$, and $\mathbbm{1}$ is the indicator function defined in the previous subsection \footnote{This setting is applicable in the cosmological scenario, where we are interested in, say the height distribution of critical points of the cosmological density field. In such a case, let $h$ denote the scalar field, in which case $f$ is the derivative of the scalar field, and $\nabla f$ the second derivative or the \emph{Hessian} of the scalar field. While the gradient of the scalar field $f$ is used to determine the "zeros" or the critical points of the scalar field, $\nabla f$, further determines the characteristics of the critical point. The function $g$ acts as a mapping of $\nabla f$ to correctly determine and include the points with the right characteristics in the counting exercise.}. 

While the expectation meta-theorem  has paved the way for qualitatively assessing certain mean characteristics of the excursion set,  it is typically very difficult, if not impossible, to obtain closed-form expression for them, even for the most basic scenarios, as for example the expected number of critical points, $\Exp {\mu_k}$, where $\mu_k$ is the number of critical points of index-$k$.  What has been tractable though, is the expected value of the height distribution of the alternating sum of critical points, given by

\begin{equation}
	\Exp \left\{ \sum_{k = 0}^{N} (-1)^k \mu_k \right\} = \frac{(-1)^N\,|T||\Lambda|^{1/2}}{2\pi^{(N+1)/2}\sigma^N}\, H_{N-1} \left(\frac{u}{\sigma}\right)\, e^{-u^2/2\sigma^2}, 
\end{equation}

\noindent where $H_N$ are the hermite polynomials of order $N$, $\Lambda$ is the variance co-variance matrix of the gradient of $f$, and $\sigma^2$ is the the variance of $f$. However, recent developments in this direction by \cite{cheng2015} has resulted in integral expressions that can be numerically evaluated, and give expressions for critical points of all indices. These expressions, however, are restricted to the case of isotropic and homogeneous Gaussian random fields, defined either on $\mathbb{s}^2$ or up to 3D in the Euclidean setting. 

The above equation,  in fact, is the equation for the Euler characteristic of excursion sets of $f$, computed in detail in \cite{adl10} and \cite{pranav2019a} in the setting of the Gaussian Kinematic formula. An important point to note here is that the expression is regularly used as an asymptotic approximation for maxima at high levels. 

\subsubsection{Excursion set formalism in cosmology}

Even though finding its roots in the methodologies pioneered by \cite{kac1943} and \cite{rice1944} in 1D, and generalization in higher dimensions by \cite{adler1981}, excursion set formalism in cosmology has independent development,  keeping in focus, and adapting to the requirements of the problems in the cosmological setting, specifically directed towards predicting and matching the observational reality of galaxy number distribution and their mass function. Motivated by specific cosmological problems, the excursion set formalism in the cosmological setting has yielded exact expressions, often though under simplified assumptions. Most prominent among such results is the \emph{halo mass-function}, which gives the expression for the number distribution of halos as a function of their mass. In the cosmological setting, the standard model envisages galaxy formation and evolution in the setting where the gravitationally interacting dark matter acts as the womb in which galaxy formation and evolution proceeds. Subsequently, most theories act  in the exclusive setting of gravitational physics of dark matter. As halos form at the seat of local maxima of the density field, these number distributions are strongly related to the topological characteristics of the density fluctuation field.  

At the most basic level, the formalism attempts to solve the \emph{barrier-crossing problem}, which is related to determining the number of up-crossings of a random function at a given barrier height. The earliest deployment of excursion set formalism in cosmology can be traced back to the seminal \emph{Press-Schechter formalism} \citep{ps74}, which predicts the number density of collapsed objects as a function of scale based on heuristic arguments.  At the heart of the formalism is the evolution of the primordial density fluctuation field under gravitational amplification. The formalism treats the gravitational collapse of regions of space smoothed at some particular smoothing scale, under the assumption that the collapse occurs in a spherically symmetric manner. The collapse on some scale $R_f$ occurs when the spherical over-density on that scale exceeds a critical density $\delta_c$. The mass of the collapsed object is related to the smoothing scale $R_f \propto M^{1/3}$, and a function of the smoothing window. In the simplest setting, the smoothing window is a spherical top-hat function. In fact, the top-hat smoothing window is central to subsequent more accurate developments in the excursion set formalism in different settings. This is chiefly motivated by the fact that this choice of window renders the density fluctuations at different scales independent of each other, in turn substantially simplifying calculations and yielding closed-form expressions. The standard model assumes the initial fluctuation field to be a Gaussian random field. For field smoothed at a scale $R$, and with variance at that scale $\sigma(R_f)$, the probablility of density between $\delta$ and $\delta+\partial\delta$ is given by 

\begin{equation}
P (\delta; R) = \frac{1}{\sqrt{2\pi\sigma(R)}} e^{\frac{-\delta^2}{2\sigma^2 (R)}}\,d\delta.
\end{equation}

\noindent Integrating this equation, with the lower limit as $\delta_c$ gives the fractional volume of space occupied by virialized objects larger than the smoothing scale F(M), so that 

\begin{equation}
F(M) = \frac{1}{2}\text{erfc}\left(\frac{\delta_c}{\sqrt{2}\sigma(M)}\right).
\end{equation}

The heuristic excursion set formalism suffers from the discrepancy that for standard model Harrison-Zeldovich spectrum, the variance becomes arbitrarily large for small smoothing scales, and as such all the mass of the Universe should be contained in virialized objects for $F(0)$. However, the mass function of the Press-Schechter formalism accounts for only half of the matter in the Universe in virialized objects. This discrepancy is referred to as the \emph{cloud-in-cloud} problem, which has to do with the possibility that the over-density may not have attained the critical value $\delta_c$ at some smoothing scale $R_1$, while it might have for a different scale $R_2 > R_1$. The formalism introduces an arbitrary factor of $2$ to solve the cloud-in-cloud problem, in order to account for all the mass in the Universe. 

 \cite{bond1991} treated the excursion set formalism in its full rigor, to solve the cloud-in-cloud problem exactly, without needing to resort to arbitrary heuristic factors. The problem is solved by computing the largest scale at which the barrier threshold crosses the critical density for for collapsed objects. The central idea is to consider the hierarchy of scales, and examine the trajectory  of the fluctuations in scale-space. When the window function is a spherical top-hat, the steps in the trajectory are independent and the problem reduces to analyzing a random walk in scale space. Starting from a large scale, the idea is to compute the probability of the first up-crossing of the random trajectory. For this case, the equations reduce to the classical heat equation, which gives the correct analytic solution without resorting to  introducing arbitrary factors through heuristic arguments. 
 
 Since the rigorous treatment by \cite{bond1991}, the excursion set formalism in cosmology has developed significantly, with various works extending the usage in increasingly complex settings. \cite{sheth2001} develop the setting further to treat ellipsoidal collapse, which may represent a more realistic picture of halo formation in the cosmological context, compared to the spherical collapse model, which assumes perfect symmetry. Also noteworthy is the seminal work of \cite{shethwey2004} in this context, which extends the excursion set formalism to model the hierarchical distribution of cosmological voids.

\subsubsection{Excursion set formalism and approximations}
\label{sec:excursion_approx}

As discussed in the previous sections, the excursion set formalism has yielded extremely insightful results into the structure iof random fields. This has most often been accomplished by studying the topo-geometrical characteristics of manifolds induced by random field that it acts as the support of. In many cases, the study of excursion sets has yielded close-form analytical expressions, most famously concerning the geometrical characteristics. Achieved by studying the \LKCs, or equivalently the Minkowski functionals, this has culminated in the formulation of the Gaussian Kinematic formula, that gives the exact formulae for the Minkowski functionals of Gaussian and Gaussian related random fields in a wide variety of settings \citep{taylor2006,adl10,pranav2019a}.The GKF gives exact formulae for predominantly geometric quantities, the exception being the Euler characteristic, which while being a purely topological quantity, has geometric interpretation via Gauss' \emph{Theorema Egrerium}. 

With the exception of Euler characteristic, the excursion set formalism has failed to provide analytical insights in to the topological characteristics of random fields, even in the case of the basic Gaussian random fields. This is equally true for topological quantities that are localized in nature, such as the critical point distribution, or non-local, such as the homology and persistent homology characteristics. In the case of critical points, it is noteworthy to mention that the distribution for Euler characteristic is often used as an approximation for the distribution of maxima asymptotically \citep{adler1981,adl10,adler2017}. Beyond this, there are no known results for the height distribution of critical points, let alone their spatial distribution, the latter of which has been completely intractable of yet. The latter is also intimately connected to the homology characteristics, as the characteristics of the topological holes will depend on the spatial properties of the critical points. Recently however, \cite{cheng2015} have developed methods that give integral expressions for critical points of al indices that are computationally solvable.

Understanding the height and the spatial distribution of the critical points is relevant for cosmological density fields, as the critical points, together with the gradient and shear field, determine the kinematics of the matter flow in the universe \cite{bond1996,weygaertlecturenotesi}. As such, the true characterization of the critical point structure of the field is an essential pre-requisite to understanding the formation and evolution of structure in the cosmos. An extremely important and relevant example in this context is the structure and properties of dark matter halos in cosmological density fields \citep{matarrese1991,hahn2007,abel2012}. Halos are assumed to form at the seat of maxima, with additional physical considerations imposed for realistic scenarios. While considering the scale-dependent size and shape of the halos, it is evident that while the center of halos will typically be located at the seat of maxima, their shape and geometry will be influenced by the critical point structure in its neighborhood on the examined scale. While a simplistic treatment will assume perfect symmetry, resulting in spherical collapse models, the realistic situation may be arbitrarily complicated, and intractable  in the absence of a theoretical understanding of the spatial distribution of critical points. 

In view of this, observable properties of the mass distribution in the Universe, as for example the number distribution of halos, have been computed under certain simplifying approximations such as the spherical collapse model \cite{ps74,bond1991,bond1996}. While being an approximation, this has still yielded extremely insightful results nevertheless, both about number distribution and sizes and mass of the halos. The chief reason is that the spherical collapse model, implemented via the top-hat filter, which while rendering the fluctuations at different scales independent of each other, also ascribes meaningful properties such as size and mass to the halos, constrained by the scale of observation and the a-priori knowledge of density fluctuation at that scale, typically constrained by power spectral density characteristics.

 \begin{figure*}
 	\centering
 	\subfloat[]{\includegraphics[width=0.25\textwidth]{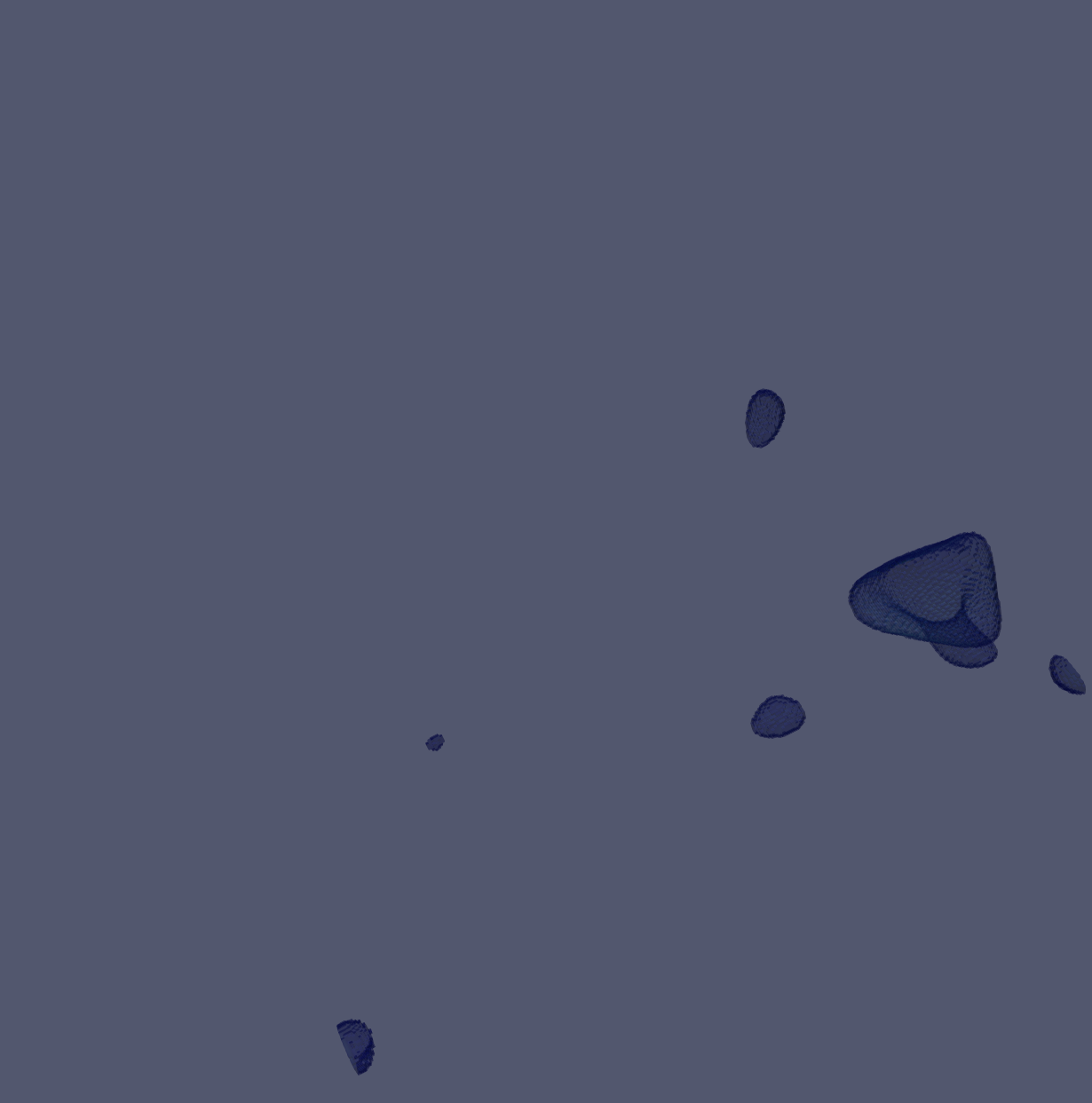}}
 	\hspace{0.01\textwidth}
 	\subfloat[]{\includegraphics[width=0.25\textwidth]{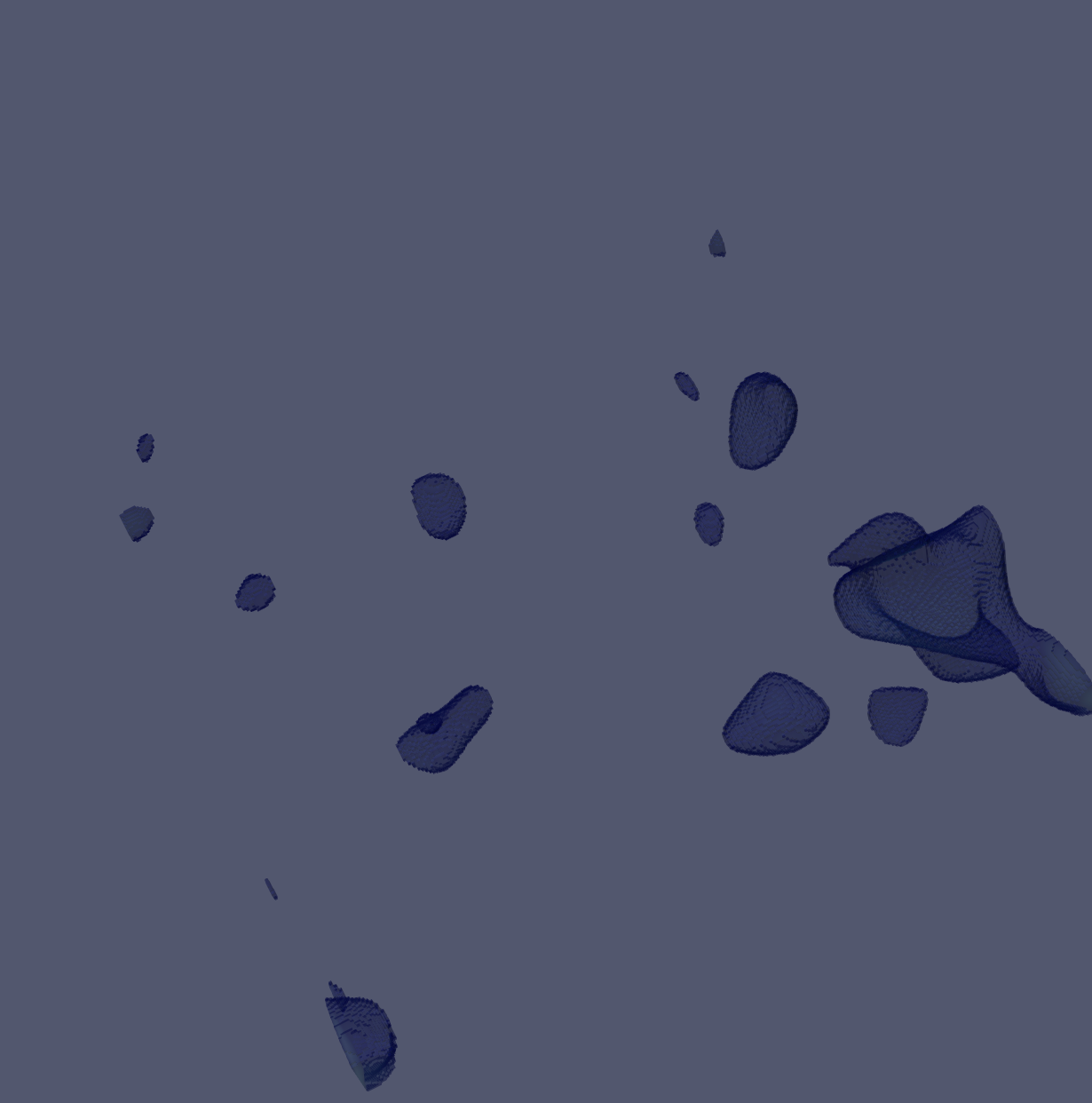}}
 	\hspace{0.01\textwidth}
 	\subfloat[]{\includegraphics[width=0.25\textwidth]{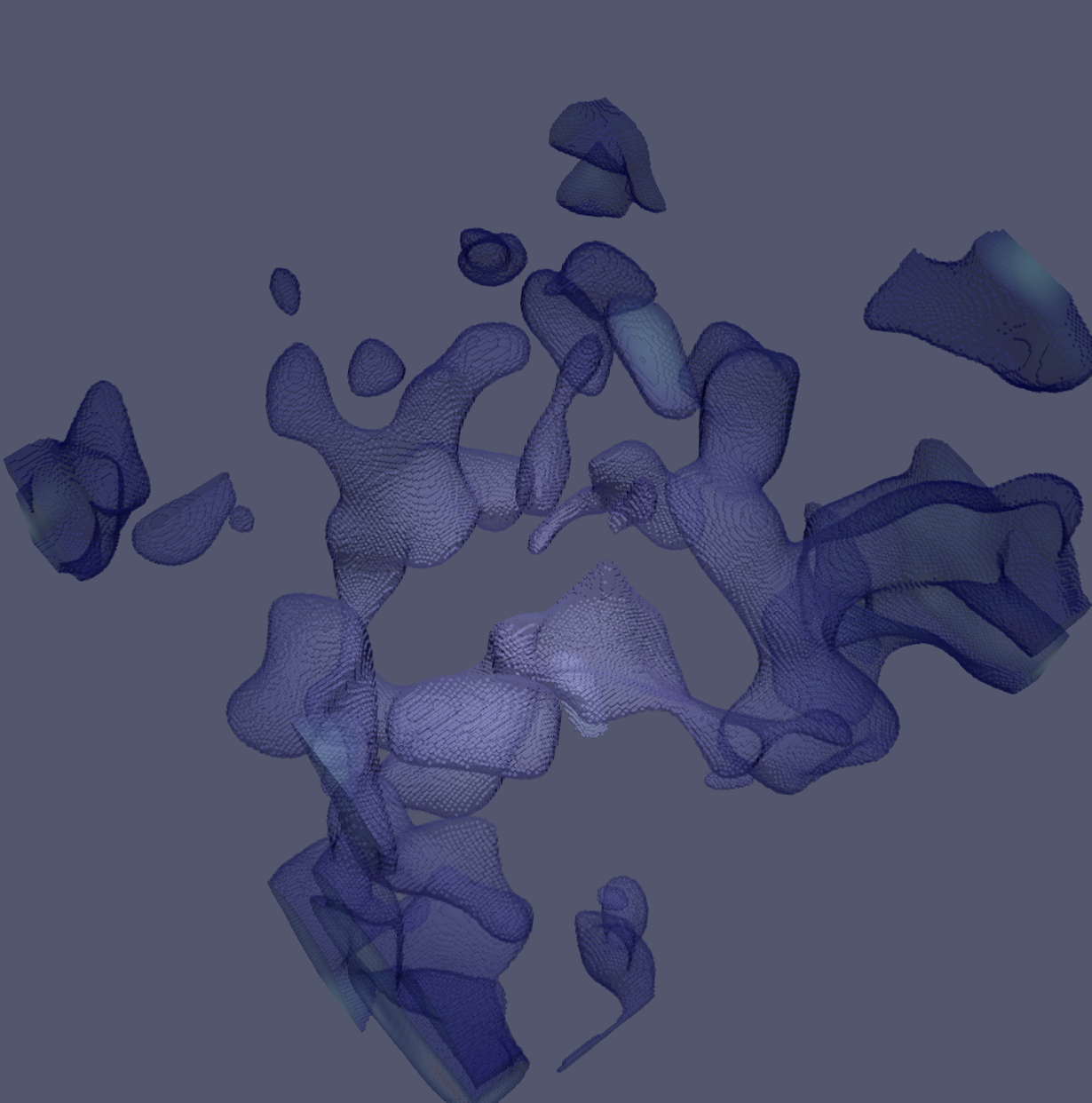}}\\
 	
 	\subfloat[]{\includegraphics[width=0.25\textwidth]{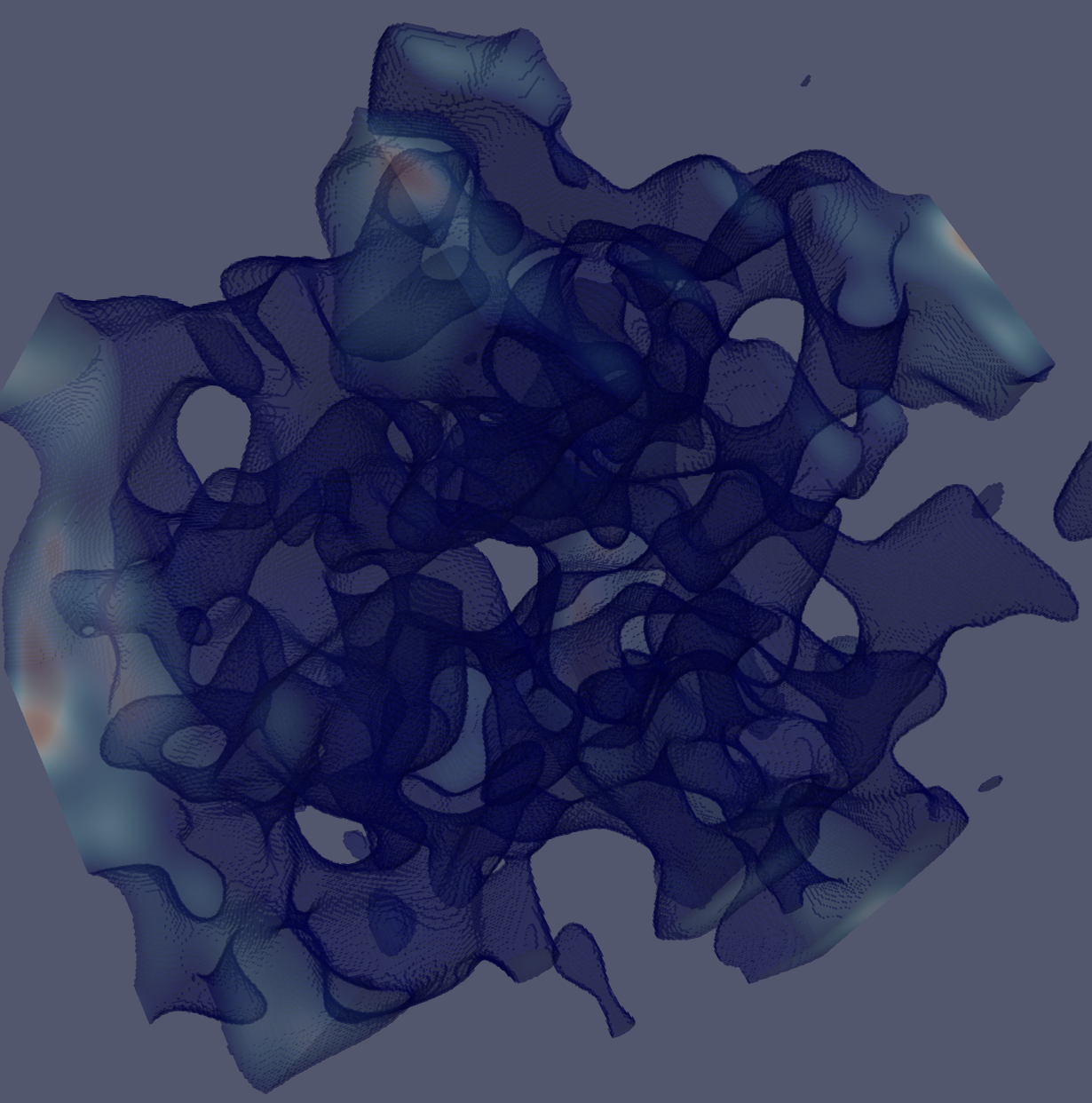}}
 	\hspace{0.01\textwidth}
 	\subfloat[]{\includegraphics[width=0.25\textwidth]{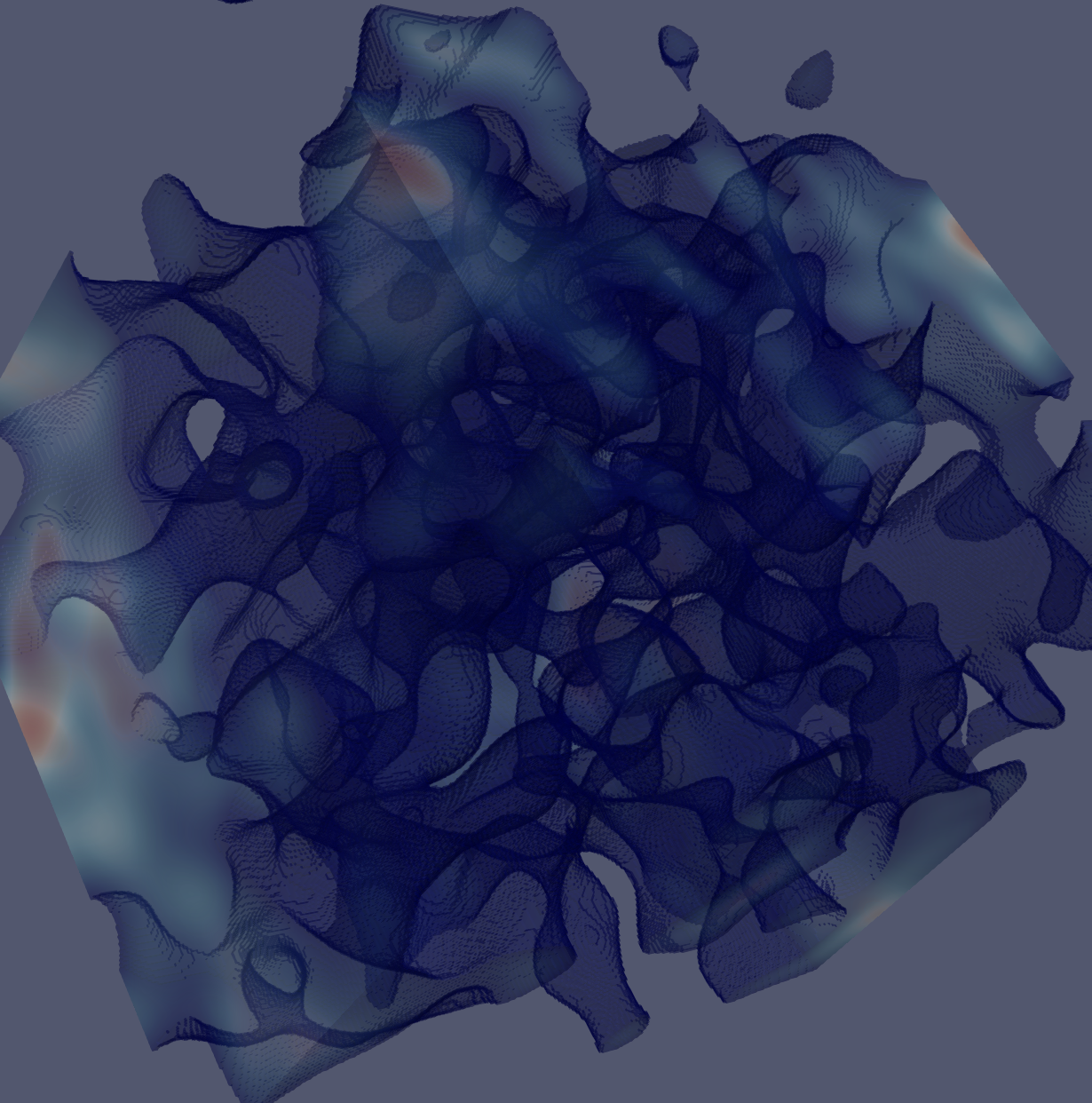}}
 	\hspace{0.01\textwidth}
 	\subfloat[]{\includegraphics[width=0.25\textwidth]{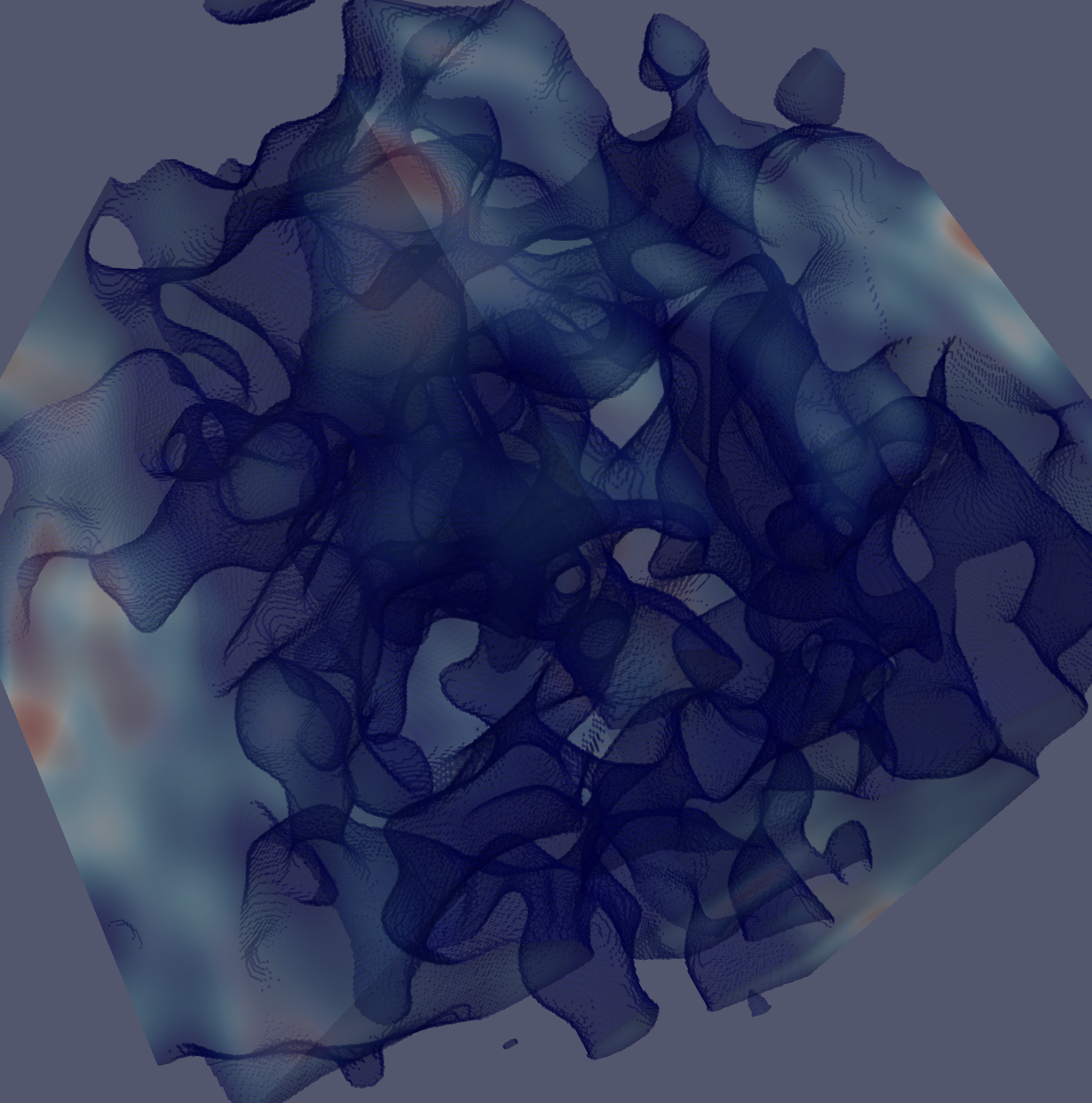}}\\
 	
 	\subfloat[]{\includegraphics[width=0.25\textwidth]{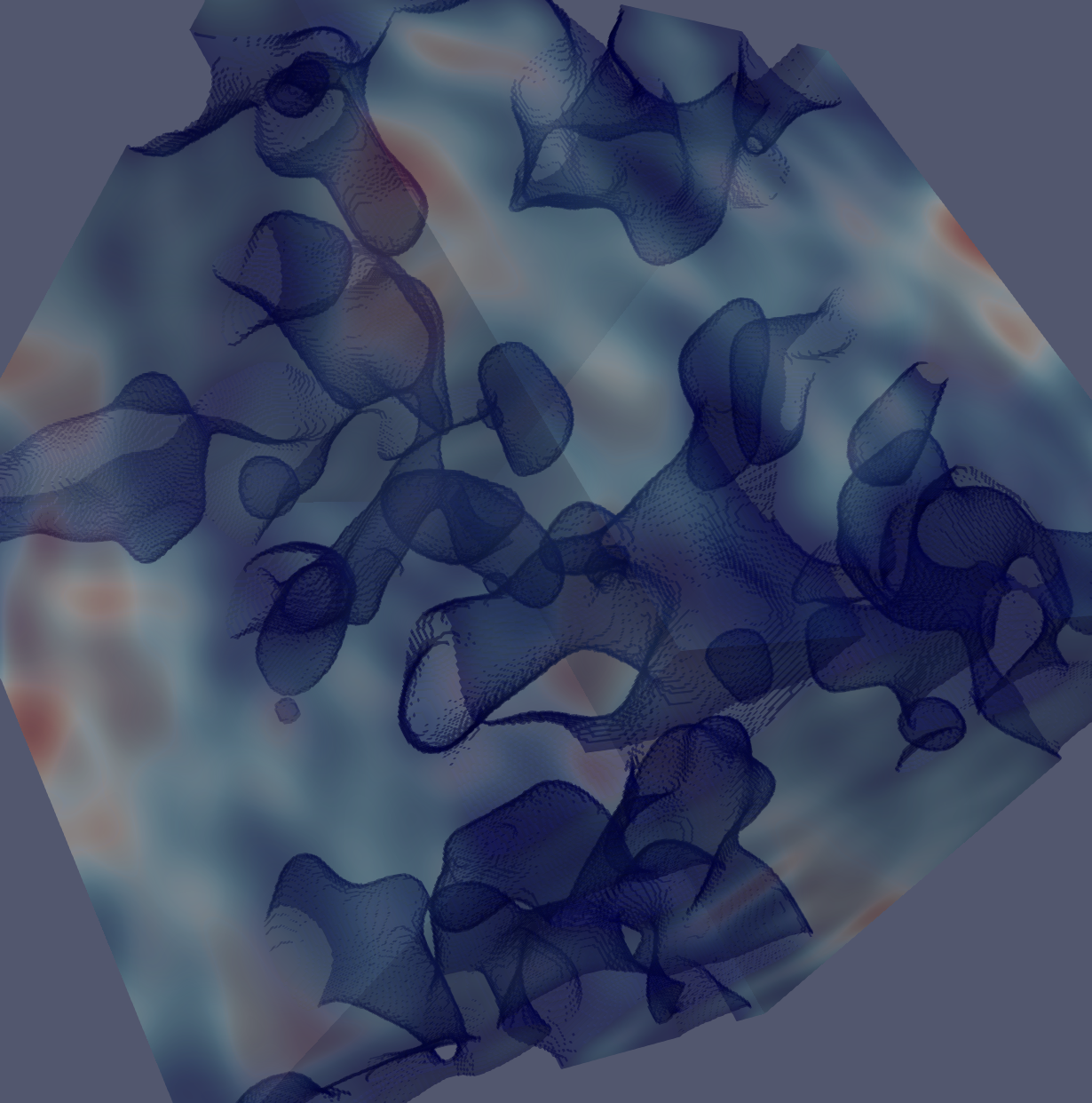}}
 	\hspace{0.01\textwidth}
 	\subfloat[]{\includegraphics[width=0.25\textwidth]{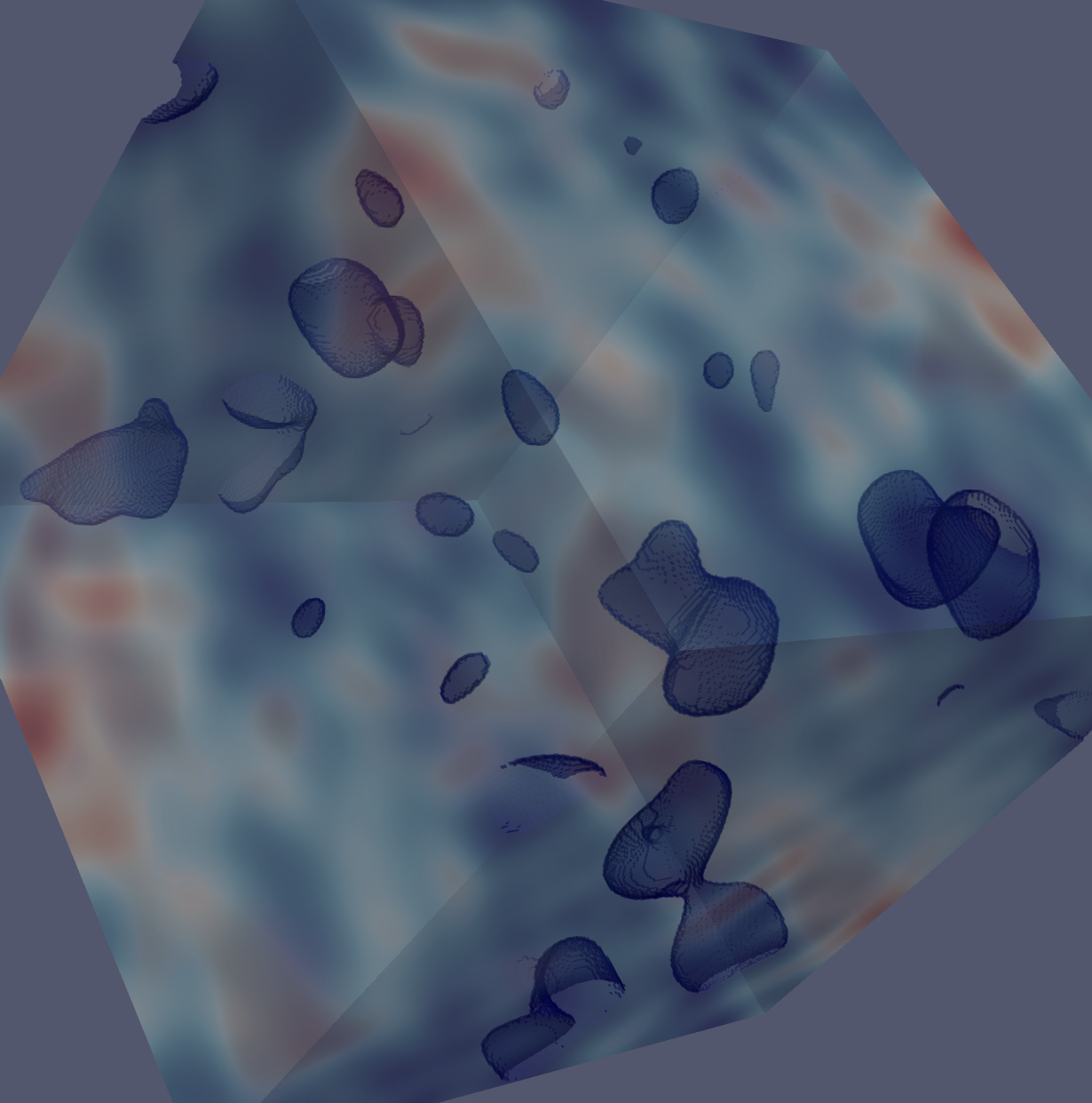}}
 	\hspace{0.01\textwidth}
 	\subfloat[]{\includegraphics[width=0.25\textwidth]{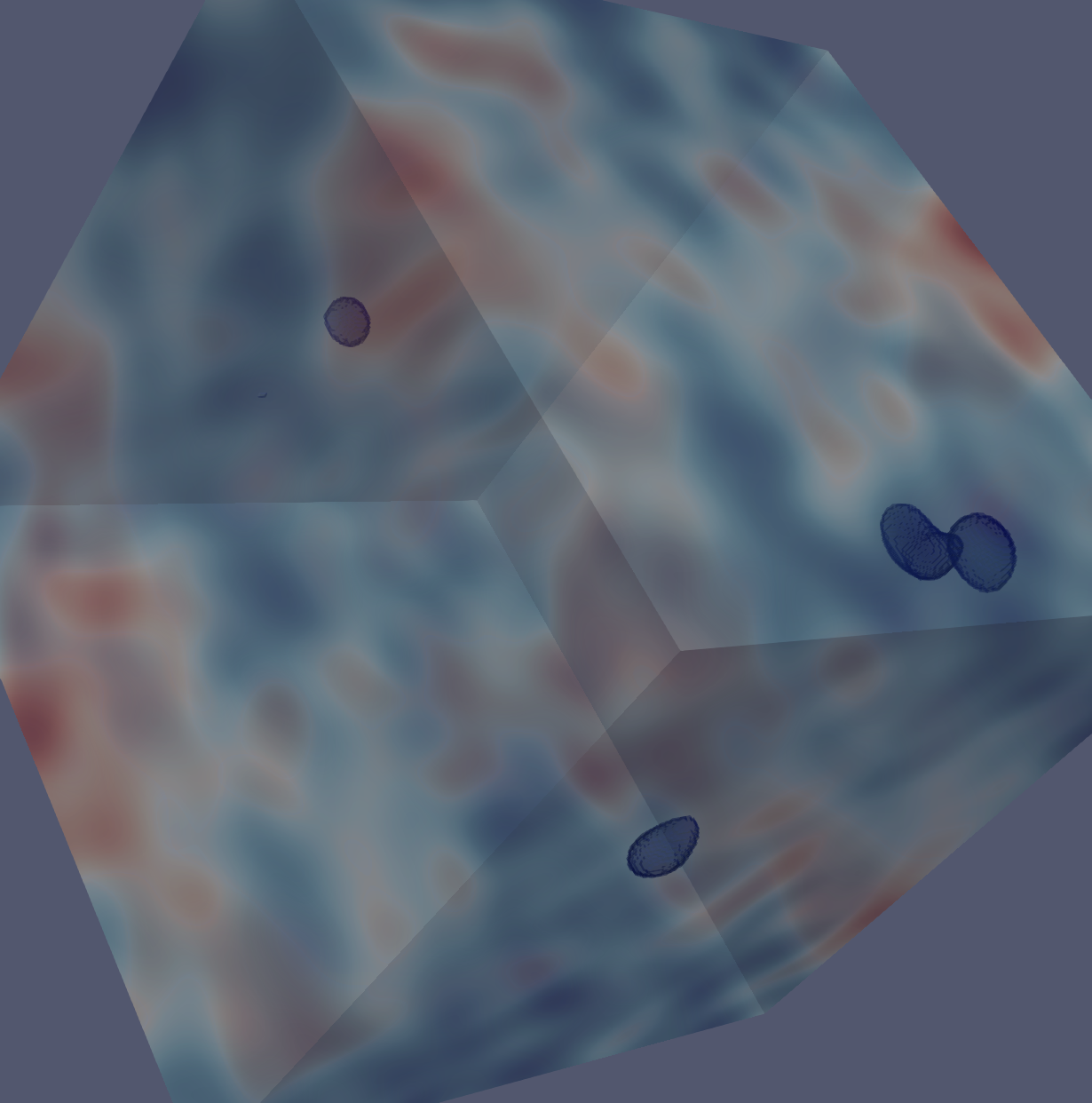}}\\	
 	\caption{ Visualization of manifold consisting of the superlevel set for various threshold values for Gaussian random field smoothed at $R_g =5 \Mpch$. The volume and surfaces are rendered in translucent form, in order to facilitate a visualization of the interior. From top-left (panel (a)) to bottom-right (panel (i)), we present the manifolds corresponding to decreasing super-level sets. For high levelsets, in the first row, the manifold largely consists of disconnected pieces. The topology resembles that of an arrange ment of disconnected meat-ball like objects. For median thresholds, presented in the middle row, the manifold characteristic gradually transforms to one with high number of loops and tunnels, accompanied by a merger of isolated objects, such that it attains a more sponge-like appearance. For low thresholds, depicted in the bottom row, the numerous tunnels split up gradually, shrink, and finally close up to form numerous isolated voids or cavities.  The topology resembles that of swiss-cheese -- a connected piece with air bubbles trapped inside, giving rise to fully enclosed cavities. Finally, for sufficiently low thresholds, all the cavities fill up, and the manifold is a single connected component, with no tunnels or cavities.}
 	\label{fig:filtration}
 \end{figure*}

\subsection{Hierarchical topology: Persistence}
\label{sec:pers_ch2}

%

In the previous sections, we presented a brief  account of the excursion set formalism in mathematical literature, as well as their derivative variants in the cosmological literature, that have developed propelled by specific cosmological requirements. Central to all of them is the computation of either the exact, or approximate (height) distribution of the critical point structure of the scalar field. It is only in the 1D case that there are exact formulae, while higher dimensions rely either on qualitative descriptions, or approximations in the cosmological case, with the exception being the alternating sum of the critical points, which has exact computations due to the link with the Euler characteristic. From the cosmological perspective,  the importance, as well as the intractability, of the height as well as the spatial distribution properties of critical points is of paramount importance, as they are crucially linked to the exact description of properties of cosmological objects, such as the dark matter halos. While the height distribution would be linked to the formation, collapse and merger of cosmological objects, their spacial distribution is linked to accurate description of the geometry, such as shape, and associated derived properties such as number densities and mass, of these cosmological objects. As we have noted in the previous section, the height distribution is tractable only in certain specific cases, while the spatial distribution properties have no known exact analytical results at all as of yet, giving rise to the approximations arising in cosmology. The intractability of spatial distributions is in turn linked to the absence of close-form expressions of non-local topological characteristics, such as those represented by the topological notion of homology \citep{edelsbrunner2010,pranav2017,pranav2019a}.

Theoretical limitations aside, there have been developments in the recent past on the computational side in topology, that have paved ways for exact computations of topological quantities, both local and non-local in nature, such as the information on critical points, as well as the information on homology properties, in a hierarchical setting, which is of crucial relevance  in the cosmological scenario, given the hierarchical nature of structure formation in the cosmos. Finding its root in homology and morse theory, this hierarchical computation formalism, known as \emph{persistent homology}, is the main theme of the rest of the paper. We briefly describe it next, referring the reader to the canonical textbook \citep{edelsbrunner2010}, as well as more user-friendly exposition in the cosmological context \citep{pranavthesis,pranav2017}.

In Section~\ref{sec:morse_ch2}, we noted the relation between the 
critical points of a function with the topological changes it induces in 
a manifold. In this section, we use the notions described above to 
sketch an intuitive understanding of persistence homology 
\citep{elz02,edelsbrunner2010,pranav2017}. 
Having deep connections 
with Morse theory \citep{mil63}, at the heart of the formalism of \emph{persistence} 
is the key observation is that the topology of the manifold 
changes only when passing through a critical point. More specifically, 
the addition of an index $p$ critical point can result in either the 
\emph{birth} 
of a $p$-dimensional hole or the \emph{death} of a ($p-1$)-dimensional 
hole \cite{elz02,zom05,edelsbrunner2010}. 
In addition, 
each topological hole is associated with two unique function value: 
$f(u_{birth})$ associated with the critical point $u_{birth}$ that gives birth to the 
hole, and $f(u_{death})$ associated with the critical point $u_{death}$ that is 
responsible for filling the hole. The 
\emph{life-time} or \emph{persistence} $\Pi$ \citep{elz02,edelsbrunner2010}, of the hole is then given by the 
absolute difference between the death and the birth values associated 
with the hole 
\begin{equation}
\Pi = |u_{birth}-u_{death}|.
\end{equation}

\noindent The hierarchical nature of persistence is reflected in the fact that it is computed from a nested sequence of sub- or super-level sets, resulting in a \emph{filtration} \cite{pranav2017}. The formulation of persistence allows tracking of creation and destruction of topological entities as the super-level set (in the case of this paper) grows. Figure~\ref{fig:filtration} presents a visualization of a few representative superlevel sets of a simulation of Gaussian random field. The levelset threshold decreases from the top-left to the bottom-right panel. The top row comprises of high levels that are dominant in isolated components. The level sets in the middle row belong to the median range of thresholds, and are dominant in tunnels. The bottom row presents the visualization for low threshold values, dominated chiefly by cavities fully enclosed by surfaces. The  sequence of images is a visual representative the filtration (with only a few levels shown), where the manifold at a a higher threshold is nested inside the one at a lower threshold.

\begin{figure*}
	\centering
	\includegraphics[width=0.8\textwidth]{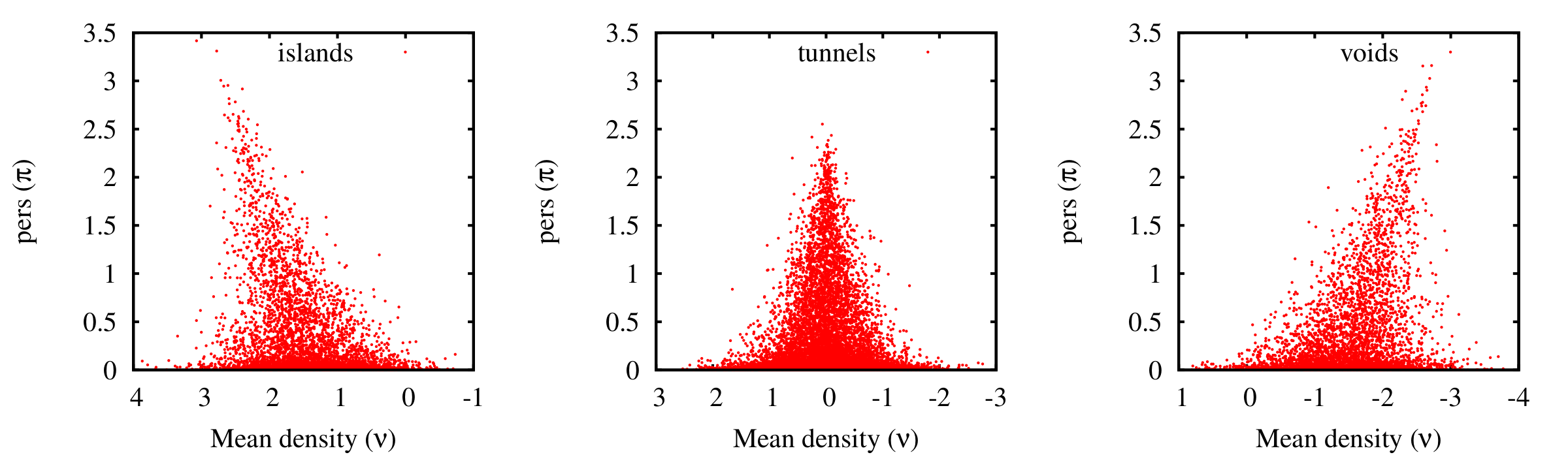}\\
	\caption{  Typical persistence diagram for islands, tunnels and voids 
		for a white noise Gaussian random field in 3D.  Each dot in the diagram represents a unique topological change in the filtration, that keeps track of creation and destruction of topological holes with decreasing levelset. Diagrams for all dimensions are triangular in shape, and exhibit a symmetry. Diagram for islands is symmetric to the diagram for voids, and the diagram for loops exhibits internal symmetry with respect to the vertical axis ($y = 0$).}
	\label{fig:dgm_grf_n0}
	
\end{figure*}

\subsubsection{Persistence diagrams}

Persistence homology is represented in terms of 
\emph{persistence diagrams} \citep{elz02,edelsbrunner2010,pranav2017}. Given a function $f(x)$, and the superlevel set  $f_\nu (x) =\{x: f(x) \geq \nu\}$, the p-th
persistence diagram $\texttt{PD}_p $ is the collection of points in the extended plane $\Rspace^2$ such that each point $(x,y)$ in the diagram represents a distinct $p$-dimensional topological feature created when $\nu = x$ and destroyed when $\nu = y$. Each persistence diagram can be represented as a collection of points 
\begin{equation}
\texttt{PD}_p = \{(b_j, d_j):b_j> d_j; j=0,\ldots,N; p=0,\ldots,d\},
\end{equation}
such that $N$ is the number of off-diagonal elements in the diagram, and $d$ is the dimension of the manifold. In the above equation, the condition $:b_i> d_i$ implies that we are analyzing superlevel sets. Each $(b_j,d_j)$ denotes the birth and death levels respectively, of a unique feature. A persistence diagram has infinitely many points with equal birth and death values along the diagonal. Such features are transient, and are not usually considered. However, they are useful when computing the bottleneck or Wasserstein distance between two diagrams \citep{cohensteiner2007,wasserman2016}. 
There is a 
diagram for each ambient dimension of the manifold. In 3D, $0$-dimensional 
diagrams record the merger events of two isolated objects. 
$1$-dimensional diagrams record the formation and destruction of loops, 
while $2$-dimensional diagrams record the birth and death of 
topological voids. 

We have introduced a representation of the persistence 
diagram \citep{pranav2017} that involves a rotation according to
\begin{equation}
b:d \to \frac{d+b}{2}:d-b.
\end{equation}
\noindent In this representation, 
the horizontal axis $\left(\frac{d+b}{2}\right)$ is the mean-density of 
the feature. The vertical axis $(d-b)$ is the \emph{persistence}, or 
\emph{life-span} 
of the feature. Figure~\ref{fig:dgm_grf_n0}, presents the typical dot diagrams for islands, 
tunnels and voids for a random realization of a 3D white noise 
Gaussian random field. Note that our representation is based on the transformation of the more familiar birth-death diagrams (c.f. \cite{cole2018,xu2019,feldbrugge2019,biagetti2020}), where we present the mean density on the horizontal axis, and persistence on the vertical axis. We choose this representation specifically in the case of Gaussian fields to appreciate the symmetry in the structure of the maps, which is otherwise not discernible in the birth-death maps. Note that this symmetry is not a general feature of diagrams, and vanishes in general for non-Gaussian cases, as pointed out in \cite{feldbrugge2019} and \cite{biagetti2020}.

\subsection{Statistics of persistence diagrams}
\label{sec:intensitymap}

\subsubsection{Intensity maps}
 
First introduced in \cite{edelsbrunner2012} (see  their Figure 1), and later treated in detail in \cite{pranavthesis} and \cite{pranav2017}, intensity maps are a useful way of visualizing the inherent structures of the persistence diagrams. Constructing the intensity maps from the persistence diagrams endows them with classical statistical properties such at mean and variance. This is a powerful intermediate step towards obtaining statistically meaningful and robust results. 

We are interested in the statistics of persistence diagrams that arise from stochastic processes, of which Gaussian random fields are a prime example. To facilitate this, we construct the intensity function $p: \Rspace^2 \to \Rspace$, the visualization of which results in the intensity maps. In the simplest form, the intensity maps can be computed as a bin-wise histogram of the persistence diagrams. Dividing the plane of persistence diagram into $n \times n$ cells, such that the number of points in the cell (i,j) is $N_{ij}$,  the \emph{absolute intensity} is given by:
\begin{equation}
I_{ij} = N_{ij}.
\label{eqn:intensity}
\end{equation}

\noindent The \emph{total intensity} $I_{t}$ is given by the sum of intensities in each bin
\begin{equation}
I_t =  \sum\limits_{i, j} N_{ij},
\label{eqn:t_intensity}
\end{equation}
\noindent and the \emph{normalized intensity} is given by 
\begin{align}
I_{nrm} = \frac{I_{ij}}{I_{t}} = \frac{N_{ij}}{ \sum\limits_{i, j} N_{ij}}.
\label{eqn:nrm_intensity}
\end{align}
The total intensity of the map is proportional to the total number of dots in the persistence diagrams. Since the persistence diagrams in turn summarize the topological characteristics of the field, the intensity maps become an excellent tool to probe the topological structure of the corresponding stochastic process. Finally, we note that \cite{chenintensitypd} and \cite{adams2017} propose to construct intensity maps and images by replacing the  binwise histogram construction with kernel density estimation methods. Such methods take a point set as the input, and filter it with an appropriate kernel (Gaussian usually) to arrive at a continuous density function estimate at each point in space.

\subsubsection{Difference maps}

When comparison between models is the primary objective, a bin-wise difference of the intensity maps between two models can be used to construct the \emph{difference maps}, which we introduced in \cite{pranavthesis}.
Suppose, two functions $f$ and $g$ produce the binwise intensity functions $I^f_{i,j}$ and $I^g_{i,j}$. The binwise difference function is given by
\begin{equation}
d^{f,g}_{i,j} = I^{f}_{i,j} - I^{g}_{i,j}.
\label{eqn:ratio_map}
\end{equation}
Henceforth, we drop the indices (i,j) as well as the subscripts $f$ and $g$ from the equations, and assume them implicitly.
Given $n$ independent realizations of the models, the binwise mean of the difference is given by
\begin{equation}
\bar{d} = \frac{1}{n} \sum_{k = 0}^{n} d_k,
\end{equation}
and the standard deviation is given by
\begin{equation}
\sigma_d = \sqrt{\frac{1}{n}\sum_{k = 0}^n (d_k - \bar{d})^2},
\end{equation}
such that, the standardized difference function is the ratio of the mean to the standard deviation of the difference function, computed bin-wise , and given by
\begin{equation}
\Delta  = \frac{\bar{d}}{\sigma_d}.
\label{eqn:std_diff}
\end{equation}

\section{Gaussian random fields: Definitions and Models}
\label{sec:grf}

In this section, we briefly review the definitions and properties of Gaussian random fields, followed by a description of the models analyzed in this paper. Standard references for this section are \cite{adler1981} and \cite{bbks};  also see \cite{pranav2019a}. 

\subsection{Definitions and properties}

At the simplest level, a \emph{random field} is nothing but a collection of random variables $f(x)$, where $x$ runs over a parameter space $\mathcal{X}$. In this paper, the parameter space $\mathcal X$ is the 3-dimensional Euclidean space. The probabilistic properties of  random fields are determined by their   $m$-point, 
joint, distribution functions, 
\begin{equation}
P[f_i]\,df_i, \, i = 1,\dots,m
\end{equation}
where, $f_i$ are the values of the random field at $m$ points $x_i$.

A random field is \emph{Gaussian} if the $m$-point distributions are all multivariate Gaussian. Assuming \emph{zero mean}, it is given by:

\begin{equation}
\begin{aligned}
&P\left[f_1, \ldots , f_m\right]df_1 \ldots df_m  \\
&= \frac{1}{(2 \pi)^N (\text{det} M)^{1/2}}
\times \exp\left(-\frac12
\sum f_i(M^{-1})_{ij}  f_j
\right)\,df_1 \ldots df_m. 
\label{eqn:prob_grf_ch2}
\end{aligned}
\end{equation}
\noindent For a Gaussian random field, the \emph{covariance} or \emph{autocovariance function}, given by
\begin{equation}\xi(x_1,x_2) \, =\,  \left\langle f(x_1) f(x_2)\right\rangle
\label{eq:rob:xi}
\end{equation}
\noindent fully determines all its properties. The matrix $M_{ij}$ in Equation~\ref{eqn:prob_grf_ch2} is the discretized equivalent of the continuous autocovariance function. The angle brackets denote ensemble averaging.

For random fields defined in $\mathbb R^D$, where $D\geq 1$, the points in the parameter space are vectors, and we introduce the concepts of \emph{isotropy} and \emph{homogeneity}. A Gaussian random field is homogeneous if  the autocovariance function $\xi (\vec x, \vec y)$ depends only on the difference $ \vec x - \vec y$. It is isotropic if it is also a function only of the (absolute)  distance $\| \vec x - \vec y\|$.  In the homogeneous, isotropic, case the autocovariance function is specified by:

\begin{equation}
\xi(r)\,=\,\xi(\|{{\vec r}}\|)\,\equiv\,\langle f({\vec x}) f({\vec x}+{{\vec r}}) \rangle \,.
\label{eq:xi_ch2}
\end{equation}

If a Gaussian random field is homogeneous, its variance defined by
\begin{equation}
\label{eq:rob:sigma2}
\sigma^2=\xi(0)=\langle f^2(\vec x)\rangle
\end{equation}
is constant. Normalizing the autocovariance function by $\sigma^2$ gives  the \emph{auto-correlation function}.

The Fourier transform of $f$ is given given by:
\begin{eqnarray}
\label{eq:rob:fhat}
\hat f(\vec k) &\, =  \,& \int_{\mathbb R^D} d^D\vec x\, \,f(\vec x)\,\exp(i\vec k\cdot \vec x) ,\nonumber\\
\ \\
f(\vec x) &\, = \,& \int_{\mathbb R^D} \frac{d^D\vec k}{(2\pi)^D}\, \,{\hat f}(\vec k) \exp(-i\vec k\cdot \vec x) \,\nonumber
\end{eqnarray}
An important quantity defined through the route of Fourier transforms is the \emph{power spectrum} $P({\vec k})$, which is the  transform of the autocovariance function $\xi (\vec{r})$. If a random field is strictly homogeneous and Gaussian, its Fourier modes ${\hat f}(\vec k)$ are mutually independent. Additionally 
the real and imaginary parts ${\hat f}_r(\vec k)$ and ${\hat f}_i(\vec k)$, where
\begin{equation}
{\hat f}(\vec k)\ = \ {\hat f}_r(\vec k) \,+\,i {\hat f}_i(\vec k)\,,
\label{eqn:real_imag}
\end{equation}
are both individually Gaussian distributed. The dispersion of the Fourier modes is  determined by the value of the power spectrum for the corresponding wavenumber ${\vec k}$,
\begin{eqnarray}
P({\hat f}_r(\vec k))\, &=& \, \frac{1}{\sqrt{2\pi\,P(k)}}\ \exp{\left(-\frac{{\hat f}_r^2(\vec k)}{2P(k)} \right)}\,,\nonumber\\
P({\hat f}_i(\vec k))\, &=& \, \frac{1}{\sqrt{2\pi\,P(k)}}\ \exp{\left(-\frac{{\hat f}_i^2(\vec k)}{2P(k)} \right)}\,. 
\end{eqnarray}

\noindent Equation~\ref{eqn:real_imag} can be rewritten as 
\begin{equation}
{\hat f}(\vec k)\,=\,\|{\hat f}(\vec k)\|\,e^{i \phi(\vec k)}\,,
\end{equation} 
\noindent which defines the amplitude (or moduli) $|{\hat f}(\vec k)|$ and the phase $\hat \phi(\vec k)$ of the Fourier modes. The phases of a  homogeneous Gaussian random field are uniformly distributed, $U[0,2\pi]$.The amplitudes have a  
Rayleigh distribution \citep{bbks}. Assuming ergodicity the power spectrum is continuous. Restricting to $\mathbb R^D$, it is given by
\begin{equation}
\label{eq:rob:PF}
(2\pi)^D\,P(\vec k)\,\delta_D({\vec k}-{\vec k}')\, = \, 
\left\langle {\hat f}({\vec k}) {\hat f}^*({\vec k}')\right\rangle,
\end{equation}
\noindent where $\delta_D$ is the Dirac delta function. 

When $f$ is isotropic, $P$ is spherically symmetric, such that
\begin{equation}
P(\vec k)\, = \,P(\|\vec k\|) \, =\, P(k)  \,.
\end{equation}
The power spectrum specifies the contribution of the different frequencies to the total variance of $f$, so that 
\begin{eqnarray}
\label{eq:rob:sigmaP}
\sigma^2 \ =\ \int_{\mathbb R^D}  \frac{d^D\vec k}{(2\pi)^D}\, P(\vec k)  &=& \frac{2}{(4\pi)^{D/2} \Gamma(D/2)} \int_0^{\infty}dk\,k^{D-1}P(k)\nonumber\\
&=& \frac{2}{(4\pi)^{D/2} \Gamma(D/2)} \int_0^{\infty}d(\mathrm{ln}\,k)\,k^DP(k).\nonumber\\
\end{eqnarray}
\noindent where $\Gamma(x)$ is the Gamma function. Thus, from the above equations $k^D P(k)$ can be interpreted as the contribution of the power spectrum, on a logarithmic scale,  to the total variance of the density field.
The recurrence relation $\Gamma(1+x)=x\Gamma(x)$, and the values
$\Gamma(1)=1$ and $\Gamma(1/2)=\sqrt{\pi}$ determine the numerical prefactors. In 3D,
the variance $\sigma^2$ is given by
\begin{equation}
\sigma^2 \ = \ \frac{1}{2\pi^2} \int_0^{\infty}d(\mathrm{ln}\,k)\,k^3 P(k)\,.
\end{equation}

\begin{table*}
	\centering
	\begin{tabular}{llrclll} 
		\hline
		\hline
		\ \\
		Number & power & index $n$ & $\#$ grid & $\#$ field & normal. & $k_c $ \\
		& spectrum & & point & realizations & & ($\Mpchk$) \\
		\ \\
		\hline
		\ \\
		1 & LCDM & & 128$^3$ & 100 & $\sigma_8=1.0$ & \\
		\ \\
		2 & power law & -2.0 & 128$^3$ & 100 & $\sigma_8=1.0$ & 0.785 \\
		3 & power law & -1.0 & 128$^3$ & 100 & $\sigma_8=1.0$ & 0.785 \\
		4 & power law & -0.0 & 128$^3$ & 100 & $\sigma_8=1.0$ & 0.785 \\
		5 & power law & 1.0 & 128$^3$ & 100 & $\sigma_8=1.0$ & 0.785 \\
		\ \\
		\hline
		\hline
	\end{tabular}
	\caption{Parameters Gaussian field realization dataset. \\
		The columns specify: (1) class number, (2) name power spectrum, (3) index power spectrum, 
		(4) number of grid-points, (5) number of field realizations, (6) normalization power spectrum and (7) normalization wavenumber $k_c$. Table reproduced from \citep{pranav2019a}.}
	\label{tab:gaussdata}
\end{table*}

\begin{figure*}
	\begin{center}
		\rotatebox{-90}{\includegraphics[height=0.9\textwidth]{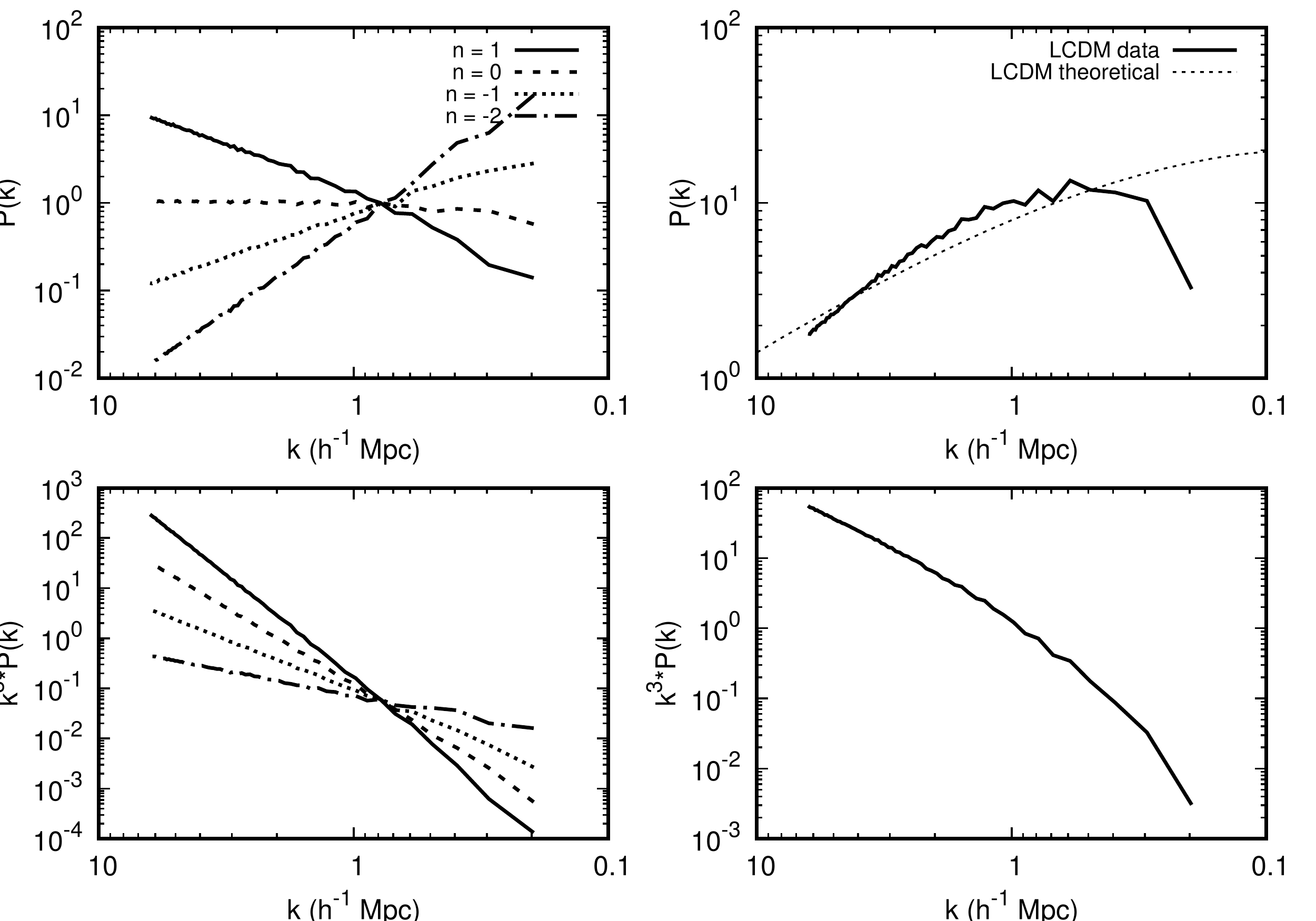}}\\
		\caption{The power spectrum $P(k)$, as well as power spectrum per unit logarithmic bin $k^3\,P(k)$. Graphs are presented for the different spectral indices of the power law model (left column), as well as the LCDM model (right column). The power law
			spectra are scaled such that different models have the same variance of the density fluctuations, when  
			filtered with a top hat filter of radius $8\Mpch$. Figure reproduced from  \citep{pranav2019a}.}
		\label{fig:plaw_spectrum}
	\end{center}
\end{figure*}

\subsection{Models}

In this section, we briefly describe the models analyzed in this paper, noting that these are the same as analyzed in the accompanying first part of the series \citep{pranav2019a}, where we also describe them in a far greater detail. The two main classes of models that we investigate are the cosmologically relevant ones, where the power spectra are defined by the LCDM spectrum and the power-law power spectrum.

\subsubsection{Power law spectra}

The power law power spectra are  a generic class of spectra, specified 
by the spectral index $n$, and given by \citep{pranav2019a}
\begin{equation}
P(k) = A_n\,k^n.
\label{eqn:powerlaw_spectrum_ch2}
\end{equation}
In this paper, we look at cases with $n = 1, 0, -1$ and $-2$.The  $n=1$ case is  the Harrison-Zel'dovich spectrum, which is the
conventionally expected spectrum for the primordial density perturbations \cite{harrison1970,peebles1970,zeldovich1972}. 
The measured spectrum of the primordial perturbations is very close to this,  with $n \sim 0.96$ 
\cite{dunkley2009,komatsu2011,planckcollaboration2016a}. ITo facilitate comparison between the field realizations we have normalized our spectra by equating the spectral
amplitude at one particular scale of $8 \Mpch$, corresponding to a frequency of $k_c \approx 0.785 \Mpchk$. Hence,
all spectra are set such that all power law spectra realizations have 
\begin{equation}
P(k_c)\,=\,A_n k_c^n\,=\,A_0\,=\,1.
\end{equation}

\noindent The left column in in Figure~\ref{fig:plaw_spectrum} shows  thew power spectra for the power law models, measured directly from the field realizations. For these models, there is relatively more power at the small scales 
for a higher spectral index, in comparison to a lower spectral index.  
As a result, the field fluctuates rapidly for high spectral indices. As the spectral index decreases, the power shifts 
towards larger scales. This results in a smoother  field with structures at larger scales.

\subsubsection{LCDM spectrum}
\label{sec:model_lcdm}

The LCDM power spectrum stems from the standard concordance model of cosmology. It fits the 
measured power spectrum of the cosmic microwave background as well as the power 
spectrum measured in the nearby large scale Universe to high accuracy. The shape of the
power spectrum can be inferred by evaluating the evolving processes through the
epoch of recombination, through the Boltzmann equation \citep{CMBfast}. A good numerical
fit is given by \cite{efstathiou1999,heavens1999}:
\begin{align}
&P_{CDM}(k) \propto \nonumber \\ 
&\frac{k^n}{\left[1+3.89q+(16.1q)^2+(5.46q)^3+(6.71q)^4\right]^{1/2}}
\times\frac{\left[\ln(1+2.34q)\right]^2}{(2.34q)^2}, \nonumber \\
\end{align}
\noindent where
\begin{align}
&q = k/\Gamma,\,&\Gamma = \Omega_m h \exp\left\{-\Omega_b-\frac{\Omega_b}{\Omega_m}\right\},
\end{align}
$\Omega_m$ and $\Omega_b$ are the total matter density 
and baryonic matter density respectively and $\Gamma$ is referred to as the shape 
parameter. We have used the value $\Gamma \sim 0.21$, which forms a reasonable approximation
for the currently best estimates for $\Omega_b$ and $\Omega_m$ as obtained from
the Planck CMB observations \citep{planck2016}. 
In our study, the power spectrum of the LCDM Gaussian field realizations is normalized by
means of $\sigma_8=1.0$. 

Locally, the spectrum resembles a power law, with  spectral index 
$n_{eff}(k)$,  showing a dependence on the scale $k$, through the relation
\begin{equation}
n_{eff}(k) = \frac{\text{d\,ln} P(k)}{\text{d ln}k}.
\end{equation}
\noindent In the asymptotic limit of  
small and large $k$, the limits of $n_{eff}(k)$ are well defined. At very large scales, 
its behavior tends towards a power law with index $n = 1$. At small scales, the LCDM power spectrum behaves like a 
power law power spectrum with index $n = -3$. The effective index 
of the model varies steeply between $n_{eff} \sim -0.5$ to $n_{eff} 
\sim -2.5$ for our models, and it resembles $n_{eff} = -1$ the most. At the lower limit, the Nyquist mode of the box corresponds to the scale of  galaxies of the size of the Milky Way. At the other end the fundamental mode of the box corresponds to wavelengths well beyond the scales at which the Universe appears homogeneous. The right column of Figure~\ref{fig:plaw_spectrum} presents the power spectrum $P(k)$ computed from simulations in solid black line. For comparison,  we also present the theoretical curve in dashed lines. The bottom panel presents the power spectrum per unit logarithmic bin $k^3\,P(k)$.

\subsubsection{Model realizations and Data sets}
\label{sec:model_real}
\bigskip
The samples of Gaussian field realizations are generated in a cubic volume on a finite grid, with periodic boundaries, there by transforming  a finite domain in $\Rspace^3$ to a periodic domain $\Tspace^3$. The 
field realizations are produced on a grid with $N=128^3$ grid-points, generated by (constrained) initial conditions
code \cite{vandeweygaert1996}. Briefly, we generate $128^3$ independent Gaussian distributed Fourier field
components ${\hat f}(\vec k_i)$, and the subsequent inverse FFT transform to yield the corresponding
density field. 
Table~\ref{tab:gaussdata} lists the relevant parameters of the sample of Gaussian field realizations used in our study. 

The field realizations are 
between the fundamental mode and Nyquist mode of the grid, corresponding to a fundamental mode of $k_{fund} =2\pi/128 \Mpchk \approx 0.049 \Mpchk$ and a Nyquist frequency
$k_{Nyq} = 2\pi /2 \Mpchk \approx 3.14 \Mpchk$. They are zero for lower and higher frequencies, effectively resulting in a a block spectrum. For the power law spectra this circumvents the
divergences that beset pure power law spectra. The statistical results are based on
100 different field realizations for each tested power spectrum.

\begin{figure*}
	\centering
	\includegraphics[width=\textwidth]{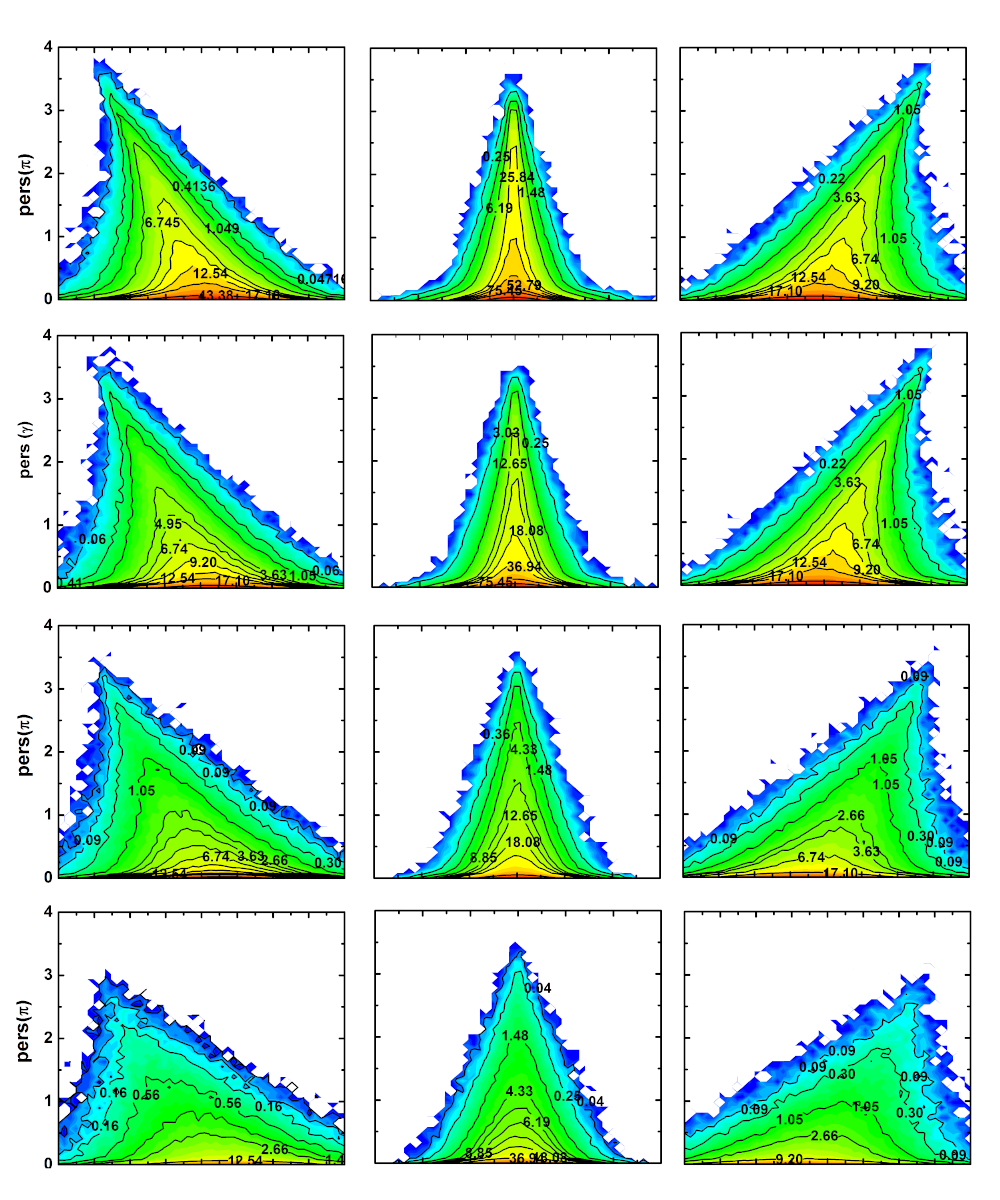}\\
	\caption{  Absolute intensity maps of islands, tunnels and voids (left to right) 
		of the 3D Gaussian random field models. The intensity map, in the simplest form, is the bin-wise histogram of persistence diagrams. The intensity function is defined as the number of dots
		in a grid cell. The maps show a characteristic dependence 
		on the choice of the power spectrum., with intensity concentrating more towards the horizontal axis, as the spectral index decreases.}
	\label{fig:abs_PLAW_intensity_map}  
\end{figure*}

\begin{figure*}
	\centering
	\includegraphics[width=\textwidth]{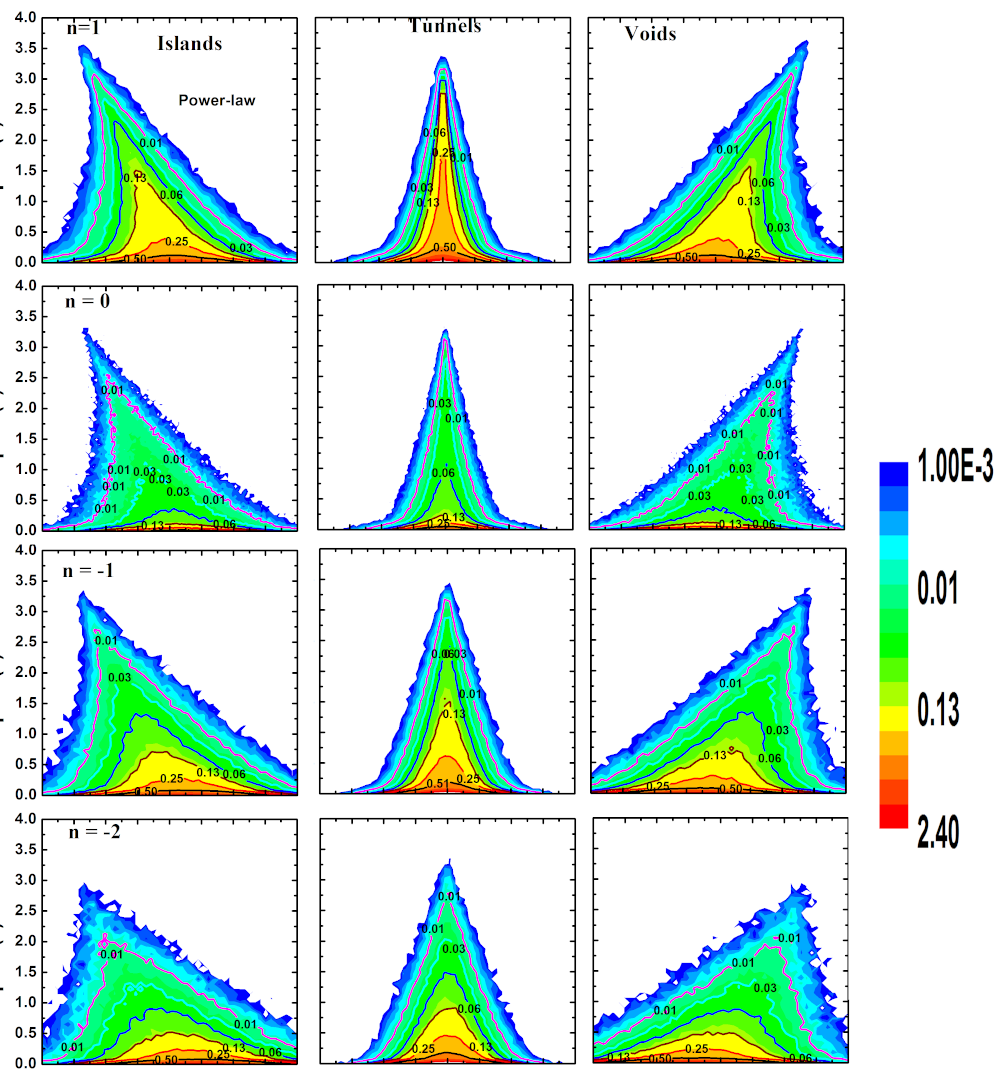}\\
	\caption{ Normalized intensity maps of islands, tunnels and voids (left to right) 
		of the 3D Gaussian random field models. Normalized intensity $I_{nrm}$ is defined as the ratio of intensity in a grid cell to the total intensity of the map, where intensity is defined as the number of dots in a bin. In this representation, the $n = 0$ model, stands out as the watershed between models with positive and negative spectral indices. The $n = 0$ model is characterized by white noise, which has maximal fraction of critical pairs with lowest persistence. This is reflected in the intensity concentrated mostly around horizontal axis. On either side of the $n = 0$ model, the intensity spreads upwards, signaling an increase in the number of high persistence pairs. For the positive indices, this is due to the increase in small scale features with large persistence. For the negative indices, this is due to the increase in large scale features with high persistence. Within the negative indices, the fraction of high persistence pairs starts decreasing with decreasing index. This is due to power shifting towards larger scales, resulting in prominent large scale features where there are lesser numbers of feature possible to pack in a given fixed volume.}
	\label{fig:PLAW_intensity_map}  
\end{figure*}

\section{Persistence characteristics of 3D Gaussian random fields}
\label{sec:intensity_map_result_plaw}

In this section, we 
present the persistence characterization of the models described in 
Section~\ref{sec:grf}. We begin by investigating the general characteristics of the intensity maps, by examining maps arising from Gaussian random field simulations characterized by the power-law  power spectra with varying spectral indices, in order to ascribe the spectral dependence to their topological properties.  Following a description of the characteristics of the power-law models, we investigate the maps arising from Gaussian field models characterized by the LCDM power spectrum. All the results in this section, as well as subsequent sections are based on $100$ realizations.

\subsection{Power-law models}

\subsubsection{Absolute intensity maps}

Figure~\ref{fig:abs_PLAW_intensity_map} presents the absolute intensity 
maps of 3D Gaussian random field models with power-law power spectra, for varying spectral index $n = 1, 0, -1, -2$. The contour labels may be utilized to track matching iso-levels of intensity in the different maps.
The shape of the
maps shows various degrees of concavity in the vertical arms of the triangle. The level of 
concavity depends
on the spectral index. For the 
$0$- and $2$-dimensional maps, the concavity is more pronounced in the arm towards 
which the maps tilt. The quantification of the model dependent 
concavity appears non-trivial, and may be an interesting exercise for 
the future. The overall total intensity decreases for decreasing spectral index. This is because lowering the spectral index results in  larger features becoming more prominent, which has inverse relationship with the packing number for a fixed volume (c.f. \cite{adler1981,bbks}). In addition, the width in the mean density distribution, as well as maximum persistence 
are important markers of difference in the map characteristics. They 
are a reflection of the features in the density landscape of the field. We study this in Section~\ref{sec:intensity_map_stats}, where we characterize the distribution of mean density and persistence.

\begin{figure*}
	\centering
	{\includegraphics[width=0.85\textwidth]{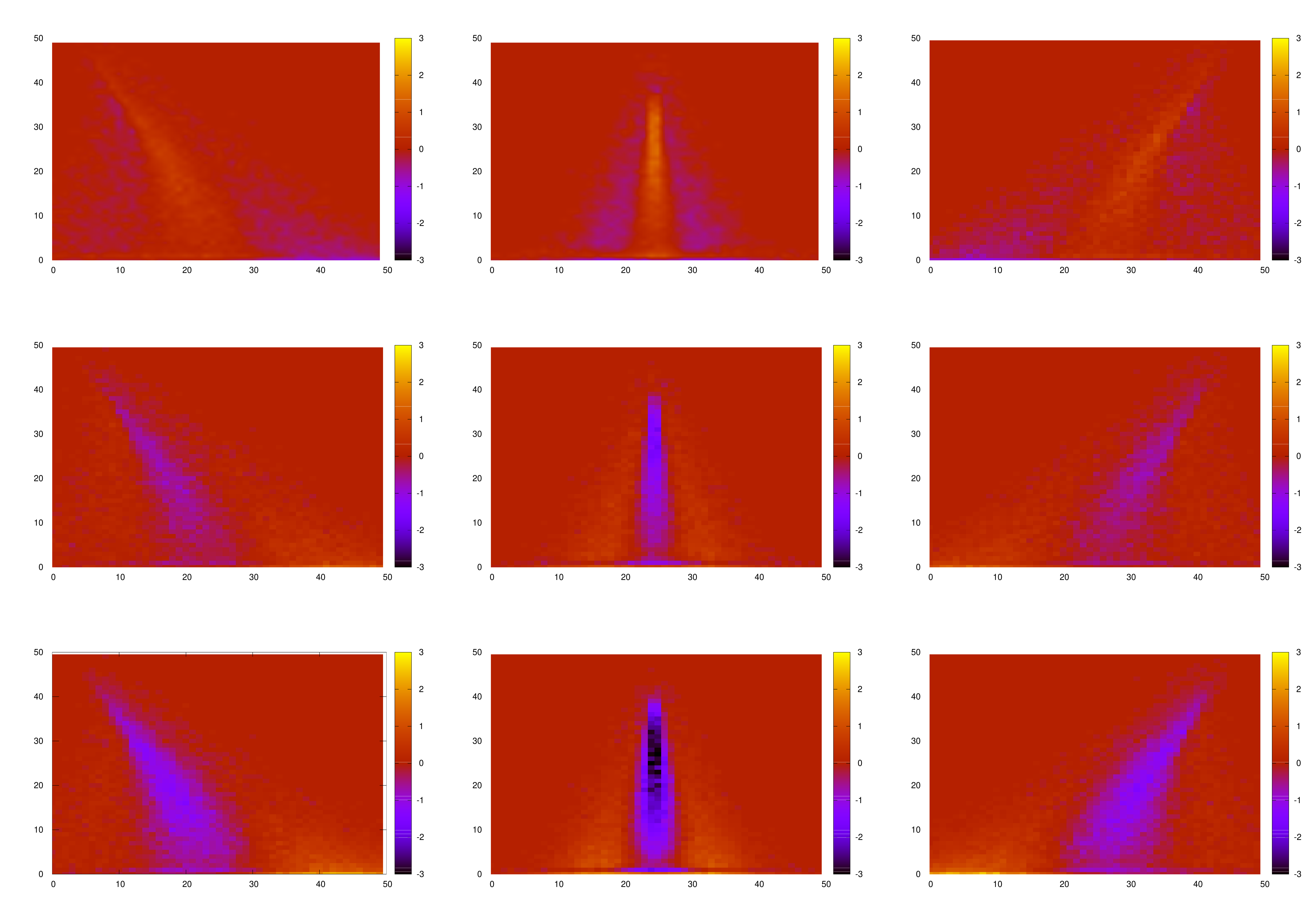}}\\
	\caption{  Difference maps of islands, tunnels and voids (from 
		left to right) for Gaussian random fields for a power-law power 
		spectrum, for spectral indices $n=1, -1, -2$ (from top to bottom). The model of reference for comparison is the white 
		noise model with spectral index $n=0$. The bin-wise difference values are in terms of the dimensionless standardized difference as defined in Equation~\ref{eqn:std_diff}. 
		The difference maps, like the intensity maps, also exhibit a symmetry. The map for
		islands is symmetric with respect 
		to the map for voids about $\nu = 0$. The map for tunnels exhibits internal symmetry under 
		reflection about $\nu = 0$. For all the models, the overall shape 
		of the maps is triangular. The details of 
		the features present in the difference maps show a 
		systematic dependence on the value of the spectral index. }
	\label{fig:PLAW_diff_map}
	
\end{figure*}

\subsubsection{Normalized intensity maps }
There is more to be learned from the intensity maps regarding the inherent structure of the field if we normalize them by the total intensity. This normalization recasts the intensity function into an empirical probability distribution function. 
By definition, 
the integral of the normalized intensity function over the plane defined by the mean 
density and persistence is unity (cf. Equation~\ref{eqn:nrm_intensity}). The effect of the 
choice of the model is reflected in the
redistribution of intensity in the plane. This has repercussions on the characteristics of 
the intensity function, which can be observed in Figure~\ref{fig:PLAW_intensity_map}. 

For the white-noise case (n = 0), almost all the intensity is concentrated near the horizontal axis. This can be attributed to the flat power spectrum, which has equal power at all scales \footnote{Note that this is different from the $n = -3$ case which has equal power per \emph{logarithmic} interval.}. This is reflected in low persistence for majority of  the features, and results in the concentration of intensity near horizontal axis. On either side of the $n = 0$ model, the intensity spreads towards higher persistence values. For the positive spectral indices, this is due to significant power at smaller scales, resulting in high persistence small scale features. For the negative spectral indices, this is due to significant power at larger scales, which result in prominent large scale features that persist for long intervals. Yet still, for the negative spectra, the concentration of intensity starts shifting towards the horizontal axis, as the spectral index decreases. This is again connected to the nature of the power spectrum and its effect on the structure of the density field. As the index decreases, the power shifts to larger scales, resulting in the prominent features corresponding to these scales. However, for a constant box-size the packing number has an inverse relationship with the size of the objects, and hence progressively smaller number of objects can be accommodated in the box as their sizes increase. On the other hand, decreasing spectral index also results in an increase in population for  low persistence small scale features. The shift of concentration of intensity towards horizontal axis is a direct reflection of these intertwined phenomena.

\subsubsection{Difference maps }
\label{sec:ratio_map_result_plaw}

To compare two models, we construct their standardized difference as 
defined in Equations~\ref{eqn:ratio_map} and~\ref{eqn:std_diff}. Quantification of the difference function 
$d_{f,g}$ between two models $f$ and $g$ results in 
maps which are at various factors of elevation or depression 
with respect to each other, in the local neighborhood defined by a given mean density and persistence value. These elevations and depressions indicate
an excess or deficit in 
the number of topological features for one model compared to the other.




Figure~\ref{fig:PLAW_diff_map} 
presents the difference maps for Gaussian random field models. The model of reference for comparison is the white 
noise model with spectral index $n=0$. The bin-wise difference values are in terms of the dimensionless standardized difference as defined in Equation~\ref{eqn:std_diff}. 
The difference maps, like the intensity maps, also exhibit a symmetry. The map for
islands is symmetric with respect 
to the map for voids about $\nu = 0$. The map for tunnels exhibits internal symmetry under 
reflection about $\nu = 0$. For all the models, the overall shape 
of the maps is triangular. The details of 
the features present in the difference maps show a 
systematic dependence on the value of the spectral index. A visual inspection of 
Figure~\ref{fig:PLAW_diff_map} reveals this. The differences in the maps for tunnels are consistently higher than those for islands and voids across all the models. 

For the maps corresponding to tunnels, the $n = 1$ model exhibits an elevated ridge
along and around $\nu = 0$ (top-row, middle panel). This means that the $n = 1$ 
model has a larger number of topological features with zero or near-zero 
mean density with different persistence values, as compared to 
the white noise case. On either side of this elevated ridge, there are regions of depression, shaded in blue,  where the $n = 1$ model has lesser intensity than the $n = 0$ model. In contrast, the trend inverts for the maps for 
$n=-1,  \text{ and } -2$. These are depressed around $\nu = 0$, forming a canyon like shape, accompanied by an elevation on either sides. The extent of depression of the canyon, as well as the elevated regions surrounding it, increase monotonically with decreasing spectral index. This indicates that as the spectral index decreases, the 
number of features with the mean density equal to the mean of the field 
decreases monotonically. It is simultaneously accompanied by an 
increase in the features with a non-zero mean density. 

Assuming the null hypothesis that the maps emerge from the same distribution, we examine the significance of their difference. The number of islands and voids in the $n = 1$ model exhibits difference in the range of $2.5\sigma$, and in the range of $3.5\sigma$ for tunnels. The number of islands and voids in the persistence interval of $1.5\Pi$ to $3.5\Pi$ exhibit differences which are $1\sigma$ significant. For the tunnels, the significant differences arise for features with zero mean density. For such features with persistence value in the range of $\sim 1.5\Pi - 2.5\Pi$ the difference is $2\sigma$ significant.
For the negative spectral indices, the significance of difference increases as the spectral index decreases. For the tunnels, the most significant differences arise for features with zero mean density, and decreases as the spectral index decreases. As an example, for the $n = -1$ model, the number of independent tunnels with a persistence between $1.5\Pi$ and $3.5\Pi$, is different from those of the $n = 0$ model with $2\sigma$ significance. The same value drops to approximately $3\sigma - 4\sigma$ for the $n = -2$ model. A similar trend is also observed in the difference maps for islands and voids.
The maps show an opposite trend along the edges of the triangular 
arms, compared to the neighborhood defined by $\nu = 0$, where there are more features compared to the $n = 0$ model, and the significance of difference increases as the spectral index decreases.

\begin{figure*}
	\centering
	\includegraphics[width=\linewidth]{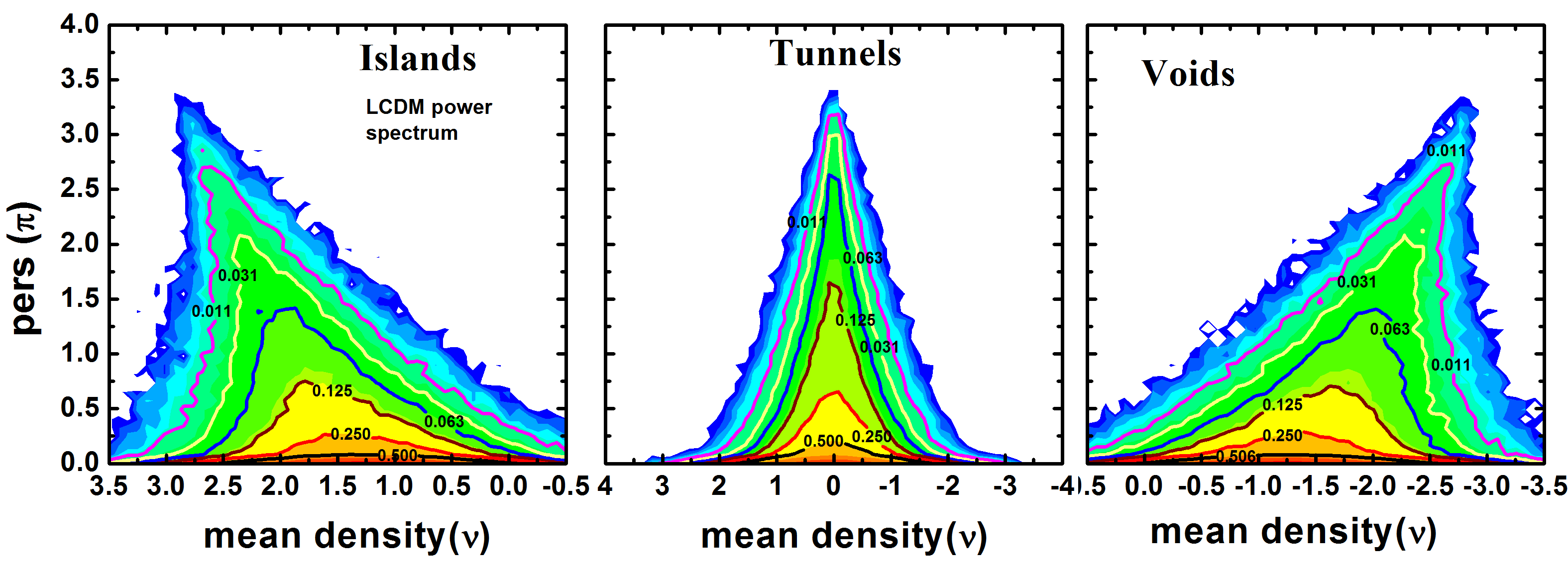}
	\caption{Intensity maps of Gaussian fields based on LCDM power spectrum. The maps are based on a $100$ realizations. The maps for all dimensions are triangular in structure, and exhibit a symmetry. Iso-level contours are labeled and maybe used to trace the structure of the maps. The maps for islands are symmetric with respect to the maps of voids, with reflection about $\nu = 0$. The map for tunnels exhibits an internal symmetry, also with respect to $\nu = 0$. The symmetry in the maps refect the underlying symmetry of the Gaussian fields, where half the expected volume is overdense, and half underdense. The map for tunnels indicates a higher spread along vertical axis, compared to the islands (and voids), indicating an independent behavior.}
	\label{fig:abs_LCDM_intensity_map}
\end{figure*}

\begin{figure*}
\centering
{\includegraphics[height=\textwidth]{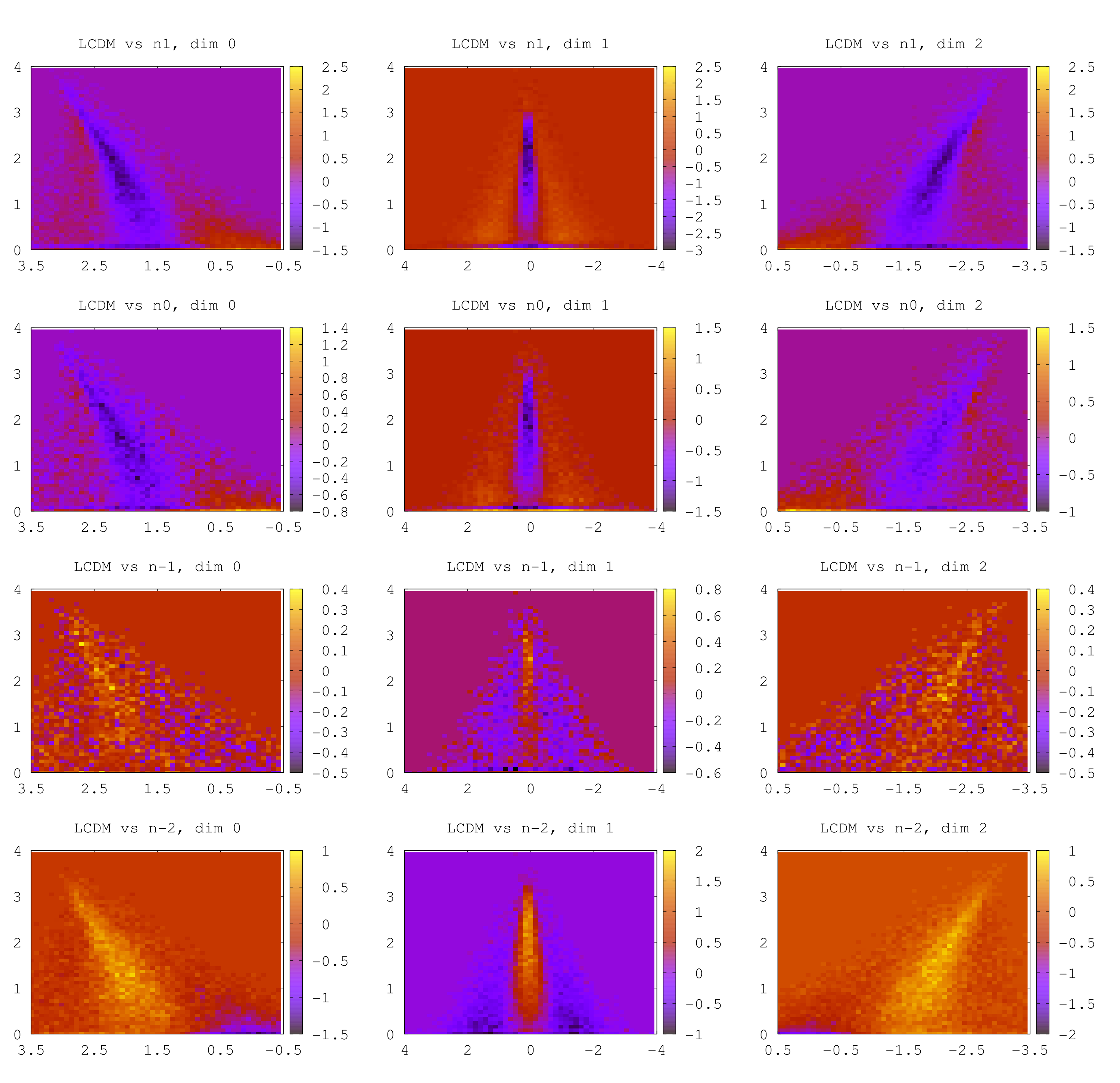}}\\
 \caption{ Difference maps of the LCDM model with respect to the various power-law models, with $n = 1, 0, -1, -2$. The bin-wise difference values are in terms of the dimensionless standardized difference as defined in Equation~\ref{eqn:std_diff}. The difference maps are the noisiest for the LCDM and $n = -1$ pair, and demonstrate their proximity clearly. Differences in the tunnel maps are consistently starker than the differences in the island and void maps.}
 \label{fig:LCDM_diff_map}
\end{figure*}


\subsection{The LCDM model}
\label{sec:intensity_map_result_lcdm}

\subsubsection{Intensity maps}

Figure~\ref{fig:abs_LCDM_intensity_map} presents the normalized intensity 
maps for the 3D Gaussian random field models characterized by the LCDM spectrum. The left column presents the 
intensity maps for islands, the middle column for the
tunnels, and the right column for the voids. We present equivalent contours of the functions for all the three maps for ease of comparison. Successive contours decrease in value by a factor of $2$ upwards from the horizontal axis. 
The intensity maps are triangular in shape for all the topological entities, i.e., for 
the islands, tunnels and voids. The 
maps also exhibit a symmetry under reflection along the vertical axis
The map
for islands is a mirror image of the map for voids. The map for tunnels exhibits internal symmetry, the axis of symmetry being $\nu 
= 0$. This reflects the symmetry of the 
Gaussian field itself, which has on an average half the volume underdense, and the other half overdense with respect to the mean field value. The maps also exhibit a tilt for the islands and voids towards high and low thresholds respectively.  Noticeable is also the different behavior of loops compared to the islands and tunnels. This can be ascertained from tracking the matching contours in the maps corresponding to islands and tunnels, as observed from the shape of the contours. As we have noted, topological entities of different dimensions are independent. However, the islands and voids exhibit a symmetry, specific to the Gaussian case.   In this scenario, the different behavior of tunnels points to a truly independent quantity. 

\subsubsection{Difference maps }
\label{sec:diff_map_result_lcdm}

Figure~\ref{fig:LCDM_diff_map} 
presents the difference maps of the LCDM model 
with respect to the power-law models. 
The left column in the figure presents the maps for islands, the 
middle column presents the maps for tunnels, and the right column presents 
the maps for voids. We are motivated to compare the LCDM model with all the power law 
models. This is because the LCDM model has a running, 
scale-dependent spectral index, and its slope locally resembles a power 
law, with the spectral index $n_{eff} = d\,ln\,P(k)/d\,ln\,k$.

The LCDM model shows evident 
differences with respect to the power-law models. As seen in the case of intensity maps, the differences in tunnels is consistently higher than that of islands and voids. The difference maps show a trend as the spectral index decreases from $1$ to $-3$. To elucidate this, we concentrate on the maps for tunnels, noting that the maps for islands and voids show similar trends. The most significant differences arise for the features with near-zero mean density. For the $n = 1$ model, the number of features with $1 -- 3.5\sigma$ is more than the LCDM model with a significance of $2 - 3\sigma$. For the same persistence range, the $n = 0$ model exhibits difference which are significant to the tune of $1 - 1.5\sigma$. The $n = -1$ model 
appears closest to the 
LCDM model. As can be verified from the values  in 
Figure~\ref{fig:LCDM_diff_map}, the bin-wise difference function  
reaches a minimum for the $n = -1$ model and the LCDM model, with the range of significance spanning $1\sigma$ for islands and voids, and $1.4\sigma$ for tunnels. the binwise difference is $0.4\sigma$ on the higher side and $0.5\sigma$ on the lower side for islands and voids, and $0.8\sigma$ on the higher side and $0.6\sigma$ on the lower side for the tunnels. As the spectral index decreases still lower, the LCDM model has more number of features compared to the base model. In the case of $n = -2$ model the difference is approximately $2\sigma$ different for the features with persistence in the range of $2 - 3.5\sigma$. In terms of differences, we also notice that the difference maps between the LCDM and the power-law spectra appear to be the noisiest, with least pronounced structures. The noisiness of the maps is a direct a reflection of the proximity between two models. In future, we will devote attention to developing tools for quantifying the noisiness of the difference maps, which will serve as an important tool for model discrimination.

\section{Statistical properties of persistence diagrams}
\label{sec:intensity_map_stats}
\begin{figure*}
	\centering
	\rotatebox{-90}{\includegraphics[width=0.8\textwidth]{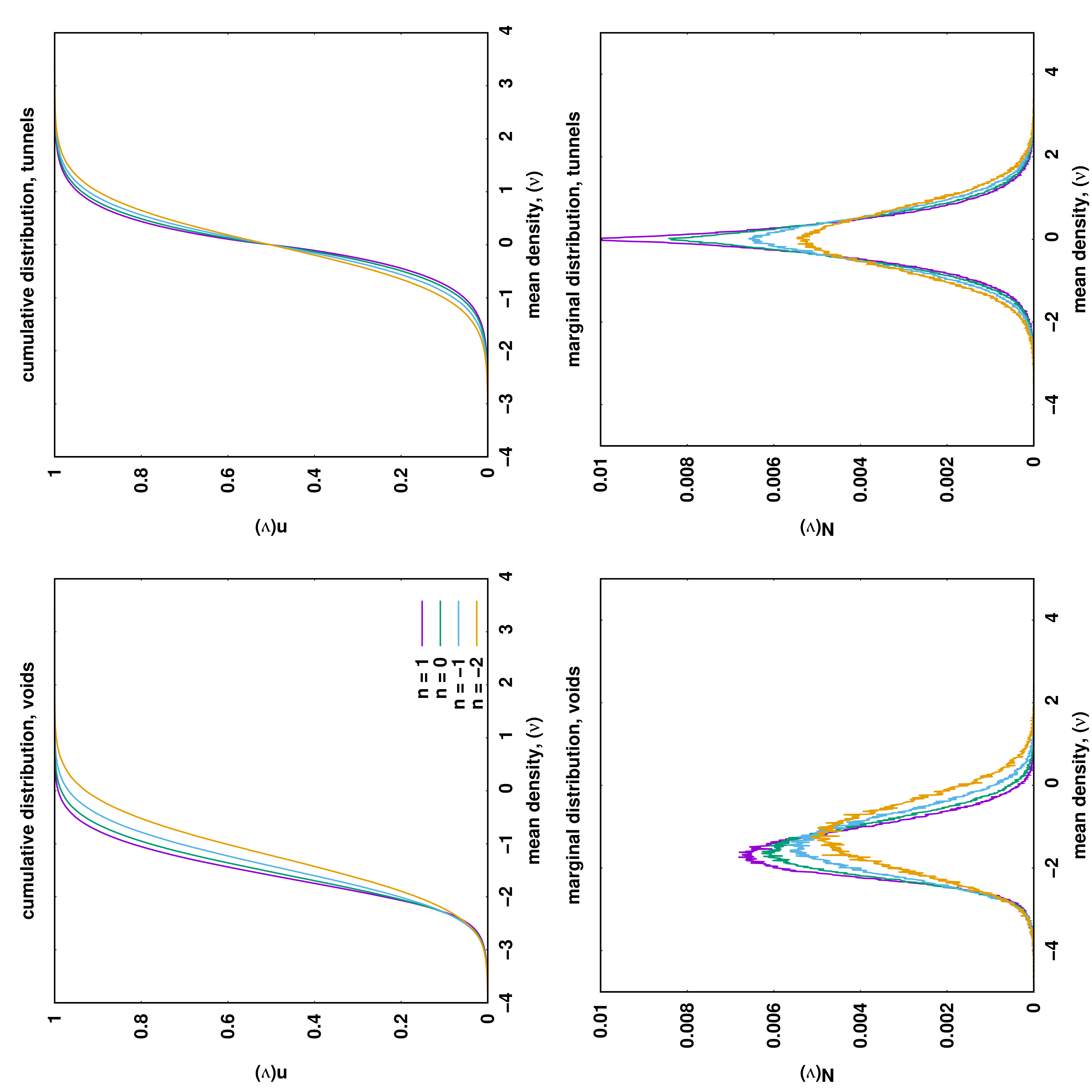}}\\
	\caption{  Normalized marginal and cumulative distribution of mean 
		density of the features in the models of 3D Gaussian fields. The graphs are 
		drawn for varying spectral index. The left column plots the cumulative and marginal distributions of topological voids in the top and bottom panel respectively. Curves 
		for the topological islands are identical to the curves for the 
		voids, under reflection about $\nu = 0$. The cumulative and marginal distribution of tunnels is plotted in the right column.}
	\label{fig:mean_density_dist}
	
\end{figure*}

\begin{figure}
\centering
   \includegraphics[width=0.4\textwidth]{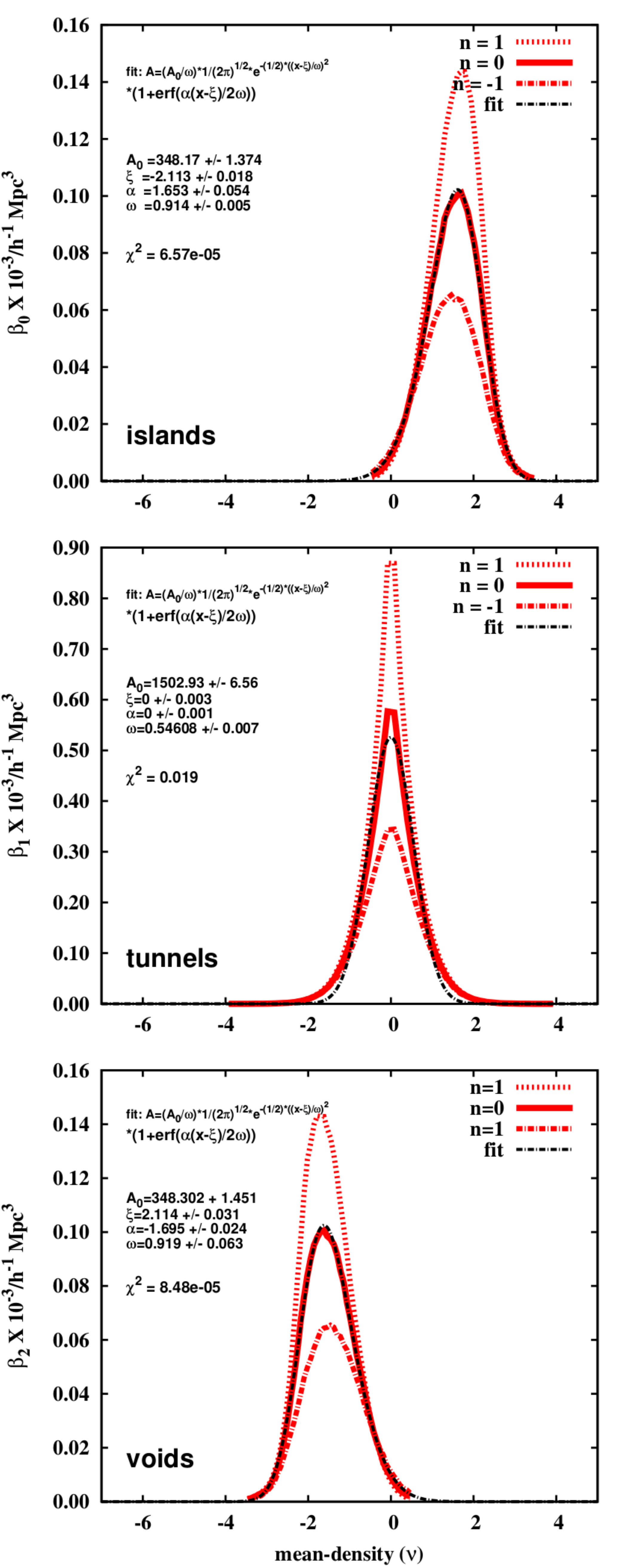}\\
  \caption{  Marginal distribution of the mean density of islands,
   tunnels and voids for $n = 1, 0 \text{ and } -1$. The 
   solid curve in red is the average computed from multiple 
   realizations. The curve in 
   black(dot-dashed) plots the best fit skew-normal distribution
   for each of islands, tunnels and voids. The values of parameters of
   distribution $A$(amplitude), $\zeta$ (location), $\alpha$ 
   (skewness-parameter) and $\omega$(scale/width) are depicted in 
   Table~\ref{tab:comp_param_ch2}.}
   \label{fig:mean_age_fit}
  
\end{figure}


\begin{figure*}
	\centering
	\rotatebox{-90}{\includegraphics[width=0.8\textwidth]{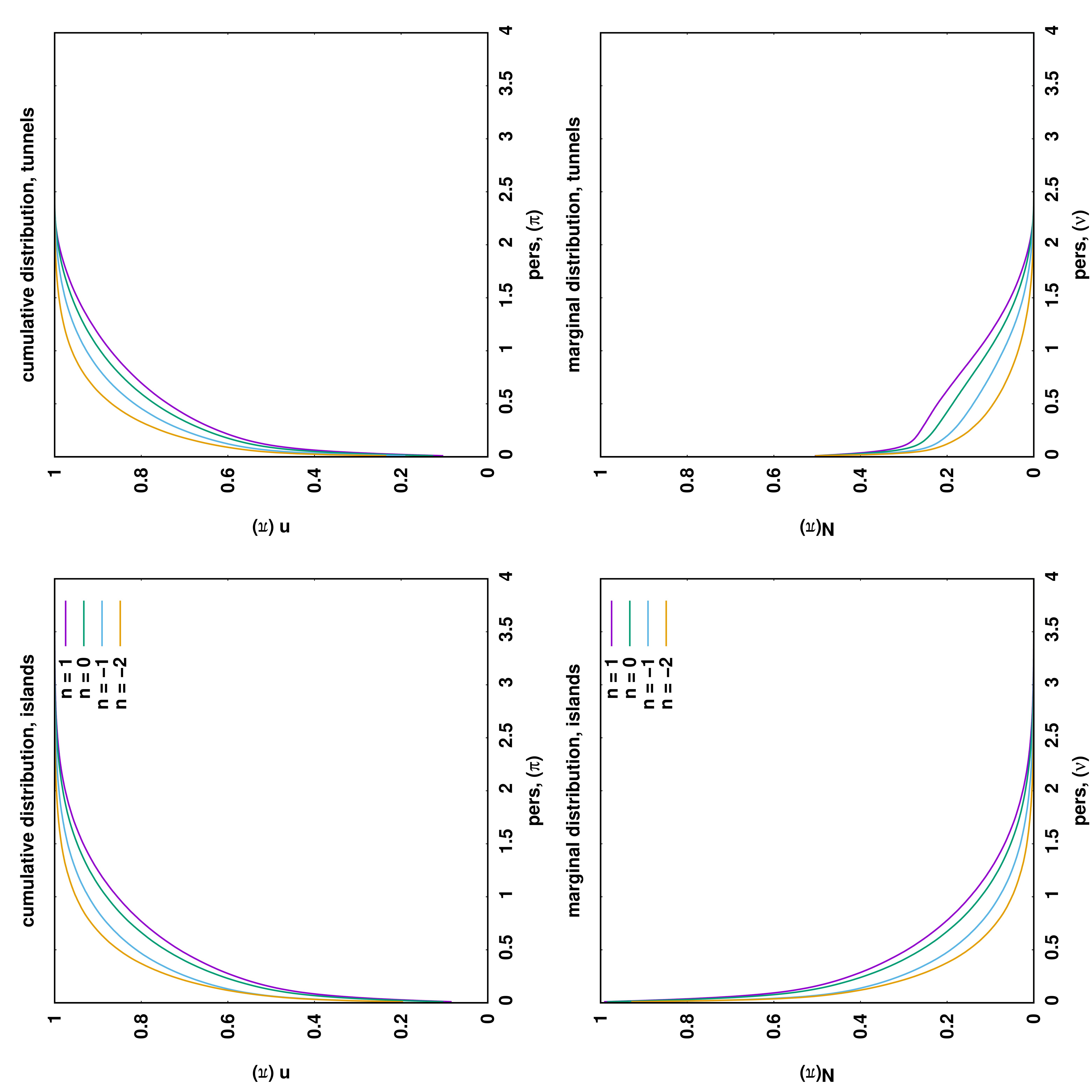}}\\
	\caption{  Normalized marginal and cumulative 
		distribution of persistence of the features. The left column plots the cumulative and marginal distributions of topological islands in the top and bottom panel respectively. Curves 
		for the topological voids are identical to the curves for the 
		islands, under reflection about $\nu = 0$. The cumulative and marginal distribution of tunnels is plotted in the right column.}
	\label{fig:pers_dist}
	
\end{figure*}

\begin{figure} 
	\centering
	\rotatebox{-90}{\includegraphics[height=0.4\textwidth]{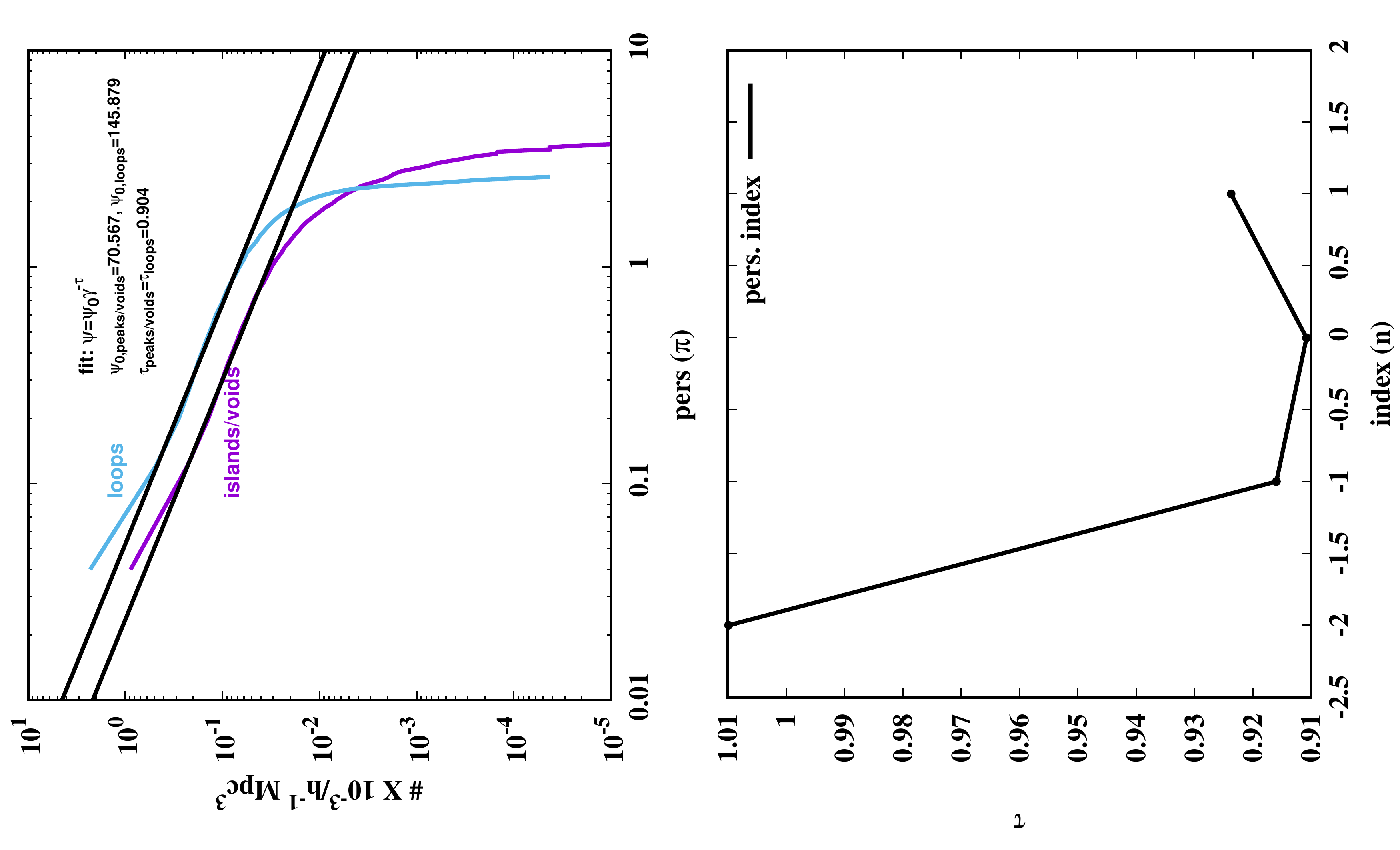}}\\
	\caption{  Left: Marginal distribution of persistence of the toplogical 
		holes. The rate of decrease is the 
		same for islands tunnels and voids. For small values of persistence,
		the rate of fall fits a power law. Right : the trend in 
		the dependence of the value of the persistent index on the index of 
		the power spectrum. Persistence index
		decreases while going from spectral index $1$ to $0$, and increases 
		monotonically while going from $0$ to $-2$.}
	\label{fig:pers_marg_distr_and_index_trend}
	
\end{figure}

\begin{table*}
	\centering
	\scriptsize
	\begin{tabular}{r|rrrrrrrrrr}
		&      &Relative          &            &Excess      & \emph{Mean-density} & \emph{Mean-density} & \emph{Mean-density}       & \emph{Mean-density} & \emph{pers.($\gamma$)} & \emph{pers.($\gamma$)}\\
		&index &intensity              &Skewness    &Kurtosis    & Amplitude       & Location        & Skewness              &  scale/width    & Amplitude              & Dimension             \\
		&(n)   &($\Lambda_{f1,f2}$)&($\gamma_1$)&($\gamma_2$)&     ($A_0$)     & param.($\zeta$)   & param.($\alpha$)      &  ($\omega$)     & ($\psi_0$)             & ($\tau$)              \\ 
		\hline
		&1     &1.31           & 0.006    &-2.97    &   908.26       & 4.30         & -1.91               & 1.73          &  92.48               & 0.92               \\ 
		&0     &1.00           &-0.018    &-2.96    &   693.34       & 4.22         & -1.65              & 1.83         &  69.86               & 0.91              \\
		Islands     &-1    &0.72           &-0.064    &-2.95    &   502.95       & 4.01         & -1.38              & 1.94         &  48.53                & 0.91               \\
		&-2    &0.46           &-0.135    &-2.93    &   326.56       & 3.52         & -0.99              & 1.99         &  26.79               & 1.00               \\
		\hline
		&1     &1.31           &-0.00072  &-2.98    &   3845.54       & 0               &  0             & 0.95        &  186.51               & 0.92               \\ 
		&0     &1.00           & 0.00048  &-2.98    &   3005.86       & 0               &  0             & 1.09         &  145.87               & 0.90               \\
		Tunnels     &-1    &0.72           &-0.00106  &-2.97    &   2214.19       & 0               & 0              & 1.29          &  101.99               & 0.90               \\
		&-2    &0.47           &-0.00099  &-2.96    &   1468.61       & 0               & 0              & 1.52         &  56.60               & 1.00               \\
		\hline
		&1     &1.31           &-0.005    &-2.97    &   910.41      &-4.32         & 1.97                & 1.75          &  92.88               & 0.91               \\ 
		&0     &1              &0.02      &-2.97    &   696.41       &-4.22         & 1.69               & 1.83         &  70.56               & 0.90               \\
		Voids       &-1    &0.72           &0.067     &-2.96    &   502.84       &-4.03         & 1.41               & 1.95         &  48.28               & 0.91               \\
		&-2    &0.46           &0.13      &-2.93    &   327.71       &-3.56         & 1.04               & 2.01         &  26.47               & 1.01               \\
	\end{tabular}
	\caption{  Statistical parameters of the intensity maps as a 
		function of varying spectral index for the power-law models. The 
		parameters are shown for islands, tunnels and voids. Column 1 shows the 
		value of the spectral index, column 2 presents the 
		relative intensity, $\Lambda_{f1,f2}$ which is calculated with respect to 
		the shot-noise maps (power-law, n=0). Columns 3 \& 4 present the 
		shape description parameters for the mean-density histograms -- Column 3 
		presents the skewness factor $\gamma_1=\mu^3/\sigma^3$, which is an indicator for 
		asymmetry around mean and Column 4 presents the excess kurtosis, 
		$\gamma_2=\frac{\mu^4}{\sigma^4}-3$, with respect to the standard 
		normal distribution. Columns 5 through 10 present the parameters of 
		fit for the fitting functions of mean-density(column 5--8) and 
		persistence(columns 9--10) histograms.  The mean-density histograms for 
		all islands, tunnels and voids are fitted with a skewed-normal 
		distribution as given by equation~\ref{eqn:mean_age_fit}. Peaks and 
		voids are visibly skewed distributions with significant values of 
		skewness parameter $\alpha$. Loops have a symmetric distribution with 
		$\alpha \sim 0$. The persistence histograms are fitted with a 
		power-law distribution. The value of \emph{persistence-dimension} 
		($\tau$) is presented in column 10.  Indices for $n=1, 0 \& -1$ 
		show similar values of $\sim 0.9$, and start increasing significantly 
		thereafter for lower values of spectral index.}
	\label{tab:comp_param_ch2}
\end{table*}

In this section, we study the marginal and cumulative distribution of 
mean density and persistence of the topological 
features of the models. We are 
also interested in an empirical fitting formula for the distributions. 
The form of the fitting function is motivated 
by the observed curve characteristics.

\subsection{Distribution of mean density}

Figure~\ref{fig:mean_density_dist} presents the 
marginal and cumulative distribution of the mean density of the 
features for the various models.
The left column plots the cumulative and marginal distributions of topological voids in the top and bottom panel respectively. Curves 
for the topological islands are identical to the curves for the 
voids, under reflection about $\nu = 0$. The cumulative and marginal distribution of tunnels is plotted in the right column.

The rate of 
change in the cumulative distributions of voids and tunnels is similar, but both the quantities show a dependence on the choice of 
the power spectrum. This follows from the observation that 
the local slope of the cumulative distribution is the same for both 
voids and tunnels, which may be visually confirmed from the graphs. 
The cumulative distribution increases most steeply for the $n =  1$ 
model, and the slope
decreases with decreasing spectral index. The curves for the tunnels cross 
each other at $\nu = 0$. A reflection of this can also be seen in the 
curves for the marginal distribution of mean density. The curves for the 
tunnels are symmetric about $\nu = 0$. The peaks of the curves are 
located at $\nu = 0$, irrespective of the model. 

The cumulative and the marginal distribution curves
for the voids are shifted towards lower density thresholds for 
higher spectral indices. The location of the maximum in the marginal 
distribution curve of voids follows a similar trend. This indicates that the abundance of topological entities as a function of density threshold is different for different models, there by pointing to the inherent differences in the topological structure within the power law models.
As noted earlier, the 
curve for islands is a mirror image of the curve for voids. This means 
that for a smaller spectral index, the cumulative and the marginal 
distribution curves are shifted towards lower density thresholds.

\subsubsection{Mean density fit: the skew-normal distribution} 

In this section, we attempt an empirical fit 
to the distribution. To this end, we take into account that the 
distributions for islands and voids are skewed and a mirror image of each 
other about $\nu = 0$. We also note that the distribution of tunnels is 
symmetric to itself and exhibits no skewness. The
observation that the curves for islands and voids exhibit a skewed 
distribution, while the curve for tunnels has no skew motivates
to introduce a generic class of distribution
called the \emph{skew-normal distribution}, as a fit simultaneously for 
the marginal distribution of mean density of islands, tunnels and voids. 
The well known normal distribution is emergent from the skew-normal 
distribution.

The skew-normal distribution is given by \citep{ohagan1976,azzalini1985,pranavthesis,wilding2020}:

\begin{equation}
	f(\nu)=\frac{A}{\omega\pi}\mathrm{e}^{-\frac{(\nu-\xi)^2}{2\omega^2}} \int_{-\infty}^{\alpha\left(\frac{\nu-\xi}{\omega}\right)} \mathrm{e}^{-\frac{t^2}{2}}\,\mathrm{d}t,
	\label{eqn:mean_age_fit}
\end{equation}
\noindent where, $A$ is the amplitude, $\alpha$ is the \emph{skewness 
parameter}, $\xi$ is the location parameter and $\omega$ is the scale 
parameter. For more details on the skew-normal distribution, refer to the 
Appendix.

We present the fit for the marginal distribution curves for $n = 0$ 
model in Figure~\ref{fig:mean_age_fit}. The three different panels present the 
marginal distribution of islands, tunnels and voids respectively. The solid red curves present the average marginal distribution curves computed over many realizations. The black dot-dashed lines 
presents the skew-normal fit curves 
for the same. In addition, the curves for $n = 1$ and $n = -1$ are 
also presented in red dotted and dot-dashed lines 
respectively for comparison. The fitted curves for 
islands and voids match the actual distribution remarkably well. The 
curve of actual distribution for tunnels has slightly broader 
tails and higher peak than the fitted curve. The fit 
for tunnels has the skewness factor $\alpha=0$, indicating that the 
mean density of the tunnels may be approximated by the normal distribution.

Table~\ref{tab:comp_param_ch2} presents the values of the various statistical properties of the intensity maps and the 1D distribution functions computed from them. Column 1 presents the index of the power spectrum. Column 2 presents the relative intensity of the various models, with the $n = 0$ model being the base model. IT is defined as the ratio of the total intensity of the two models in comparison.
Columns 3 presents the skewness, defined as $\gamma_1=\mu^3/\sigma^3$. 
Column 4 presents the excess kurtosis, defined as $\gamma_2=\frac{\mu^4}{\sigma^4}-3$. By this 
definition, the excess kurtosis of the standard normal distribution is 
$0$. The absolute 
value of skewness for the curves corresponding to islands and voids 
increases as one progresses from $n = 1$ to $n = -2$. The distribution for tunnels 
exhibits a near-zero, negligible skewness for all values of the spectral index. The 
excess kurtosis for all islands, tunnels and voids is $\sim -2.9:-3.0$, 
with small variations that show a decreasing trend in the absolute 
magnitude as one lowers the spectral index.

Columns 5 through 8 of Table~\ref{tab:comp_param_ch2} also enumerate the best-fit values for the parameters of the skew-normal distribution. 
The absolute value of the location parameter decreases for both 
islands and voids as $n$ decreases.  The absolute value of skewness parameter 
decreases for decreasing $n$ in the case of islands and voids. On the 
other hand, it 
is $\sim 0$ for tunnels, as expected. The scale parameter $\omega$, 
which indicates the 
width of the curves, increases uniformly for decreasing $n$. This is 
true for all islands, tunnels and voids. 

\subsection{Distribution of persistence}
\label{sec:pers_dist}

Figure~\ref{fig:pers_dist} presents the 
cumulative and marginal number distribution as a function of  
persistence of the features for the various 
models. The left column plots the cumulative and marginal distributions of topological islands in the top and bottom panel respectively. Curves for the topological voids are identical to the curves for the 
islands, under reflection about $\nu = 0$. The cumulative and marginal distribution of tunnels is plotted in the right column. The marginal distribution of tunnels is different 
from the marginal distribution of islands and voids. However the 
cumulative distribution of islands, tunnels and voids are approximately 
coincident for all the models.

The top panel of 
Figure~\ref{fig:pers_marg_distr_and_index_trend} plots the marginal 
distributions of persistence for the 
islands, tunnels and voids for the 
$n = 0$ model on a logarithmic scale. For low values of 
persistence $\pi \sim (0.1-1\sigma)$, the distribution 
follows a power-law
\begin{equation}
	\psi=\psi_0 \pi^{-\tau}.
	\label{eqn:pers_fit}
\end{equation}
The index of the power-law $\tau$ 
indicates the rate of fall in the number of objects as a function of 
persistence. We call this the \emph{persistence index}. 

Columns 9 and 10 of 
Table~\ref{tab:comp_param_ch2} list the amplitude $\psi_0$ and the 
value of the persistence index $\tau$ for the power-law models. 
The persistence index $\tau$ is 
approximately the same for islands, tunnels and voids for a given 
spectral index $n$. For 
$n = 1, 0, -1$ , the persistence index 
is $\tau\sim 0.9$. It
shows an increasing trend for $n < -1$. For the $n=-2$ model, 
$\tau\sim 1$. 
The bottom panel of Figure~\ref{fig:pers_marg_distr_and_index_trend} 
presents the trend in the dependence of 
the persistence index $\tau$ on the power spectrum index 
$n$. 

\subsection{Number density per unit volume of the topological features}

\begin{figure}
	\centering
	\includegraphics[width=0.4\textwidth]{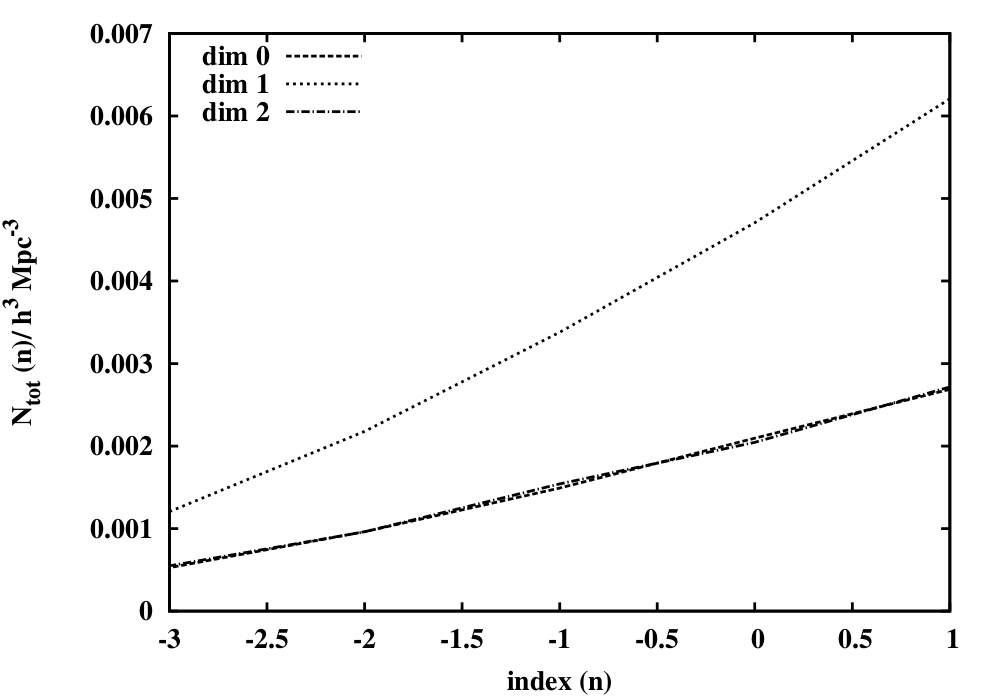}\\
	\caption{  The average total intensity as a function of the index of 
		the power spectrum. It indicates the average number of topological 
		objects per unit volume in the density field characterized by a given 
		power spectrum. The curve is presented as a function of the spectral 
		index $n$. The average total intensity decreases monotonically with 
		decreasing spectral index.}
	\label{fig:N_tot}
	
\end{figure}

The average total intensity $\langle I_{t} \rangle$ per unit volume maybe considered as
a measure of the topological structure of a field. It corresponds to
the average number of topological 
objects per unit volume in the density field. 
Figure~\ref{fig:N_tot} presents the average total intensity as a 
function of the index $n$ of 
the power spectrum. The average total intensity shows a 
characteristic dependence on the choice of the power spectrum. It decreases 
monotonically with decreasing spectral index. This means the number of 
topological objects per unit volume decreases monotonically with 
decreasing spectral index. This is expected on account of the fact that 
for lower spectral indices, the larger structures in the field become 
more prominent. As a consequence, lesser number of features can be 
packed in a given volume.

\section{Discussions and conclusions}
\label{sec:discussion_ch2}
The investigation of the structure and properties of stochastic random fields plays a central role in a multitude of theoretical as well as applied disciplines. Within this, the study of Gaussian random fields specifically occupies a key position, chiefly due to the availability of closed-form analytical results in certain cases, as well as the fact that Gaussian random fields serve as the baseline model for a variety of natural phenomena. Within the cosmological context, the relevance of Gaussian fields emerges from the fact that they serve as the null hypothesis model for the primordial stochastic matter fluctuation field, there by linked to the properties of the temperature fluctuations in the Cosmic Microwave Background, as well as the large scale matter distribution in the Universe, that emerges from the primordial fluctuations. At galactic scales, Gaussian fields are frequently used to model the ISM distribution. As such, our analysis of Gaussian fields is commensurate with the intention of establishing a baseline reference in terms of the properties of stochastic cosmological fields.

Given a stochastic random field, and its support manifold or the parameter space, both of which are smooth, there are possibly two  alternatives to study their properties. The first is to examine the image of the parameter space, or its subsets on which the field is defined, in which case it would be natural to express the stochastic structure as a combination of the topological structure of the parameter space, its ambient dimension, and the probabilistic structure of the stochastic field of interest. The second is to examine sets in the parameter space, where the field takes particular values of interest. While there are no known results in the former category, the latter has given rise to the framework of excursion set formalism, which has yielded illuminating results, both in the purely mathematical, as well as the cosmological context. These results are predominantly concerned with the topo-geometrical structure of stochastic fields. The study of excursion sets of stochastic fields has especially been fruitful with respect to their geometric properties in the integral geometric setting, resulting in closed-form analytic expressions for all the \LKCs, and so the Minkowski functionals, for Gaussian and Gaussian-related random fields, culminating in the Gaussian Kinematic Formula, in a single compact expression. As the 0-th \LKC~is related to the Euler characteristic, which is a purely topological measure, the GKF also informs on the topological properties. 

Beyond these results though, there are no known analytical closed form results on the topological properties, even for local measures such as the distribution of critical points, the exception being the Euler characteristic heuristic, which is used as a measure for the distribution of maxima asymptotically in the high threshold limit. More refined topological measures such as those emanating from homology theory are completely intractable from the theoretical view point, chiefly due to the fact that they are non-localized in nature, and depend on the overall structure and distribution of critical points, which themselves have no theoretical underpinning as of yet.

The formation and development of structure in the cosmos is crucially linked to the topo-geometrical properties of the stochastic fluctuation fields, such as the structure of critical points. For example, collapsed objects in the cosmological context, such as the dark matter halos necessarily form at the seat of local maxima, with their additional properties linked to the overall structure and distribution of saddles and minima around them. In the absence of a theoretical understanding of critical point distribution, a complete and accurate theoretical prediction of cosmological objects is intractable. These limitations aside, the excursion set formalism has developed in parallel in the cosmological context, motivated by specific problems in cosmology. Even though often under simplified assumption, this has resulted in extremely valuable insight into the properties of cosmological objects, specifically the properties of dark mater halos, in terms of their geometry as well as number distribution as a function of mass, providing a test-bed for comparing theoretical predictions with observations. 

Theoretical limitations, and the above considerations enumerated aside, recent developments in computational topology have paved way for methodologies and formalism that inform about the spatial as well as the height distribution of critical points, as well as the associated topological information emerging from homology theory, in a hierarchical setting, thereby providing additional and crucial tools and frameworks within which to analyze the hierarchical structure formation and evolution in the cosmos. These developments are also important from the the viewpoint that recent years have seen a massive proliferation in data acquisition in the cosmological context, propelled by a massive surge in the experiments and surveys in observational cosmology. This scenario demands increasingly more sophisticated methods of analysis in order to glean meaningful information out of raw datasets. 

This paper, second in the series and a follow-up to \cite{pranav2019a}, focuses on the hierarchical topological formalism known as \emph{persistent homology}, within the excursion set framework.  The development of persistent homology  has ushered in a new era in data analysis by means of topo-geometrical methodologies, such that Topological Data Analysis (TDA) has rapidly emerged as a discipline in its own right. The concepts and tools that form the backbone of TDA emanate chiefly from Morse theory and persistent homology. While providing a survey of the basic tenets of the excursion set formalism, as well as a description of persistent homology within the excursion set framework, this paper analyzes the topological structure of Gaussian random fields in a computational setting. In addition, we also present a brief survey of the mathematical aspects that form the foundations of the investigation into the structural properties of stochastic random fields. The numerical studies presented here and in \cite{pranav2019a} also accompany \cite{feldbrugge2019} in spirit, where we develop a semi-analytical formalism for persistence diagrams and Betti numbers, arriving at approximating formulae for the quantities, in the case of Gaussian and local-type non-Gaussian random fields. These investigations follow the broad program laid out in \cite{isvd10} and \cite{pranavthesis} aiming at the application of geometric and topological methods in the cosmological context.


 In \cite{pranav2019a}, the first part of the series, we focused on characterizing the topology of Gaussian random fields via homology, quantified by the Betti numbers. Simultaneously, we presented a study on the geometric and topological characteristics of Gaussian fields, exploring the differences and complementary nature of the topological and geometric measures. We focused on the investigation of the geometrical properties in terms of the Minkowski functionals, a program introduced in the cosmological context by Buchert and collaborators \citep{mecke94}. In the process, we introduced a theoretical exposition on the Gaussian Kinematic Formula (GKF) in the cosmological context \citep{adl10,pranav2019a}. Further, we established that homology and Betti numbers provide a more detailed topological description than the Euler characteristic, propelled by the observation that the shape of the Betti number curves depends on the spectral index, as opposed to the shape of the Minkowski functional curves that show no dependence on the choice of the power spectrum. This is an observation \cite{ppc13} had already established in the context of comparing Euler characteristic with Betti numbers. 



In the current paper, we extend the topological investigation of Gaussian fields to include their persistent homology characteristics. Persistent homology is the hierarchical extension of homology, introduced by Edelsbrunner and collaborators \cite{elz02,edelsbrunner2010}, and as such, presents a powerful tool for the analysis of hierarchical cosmic structures. We present evidence in support of this thesis via an analysis of the hierarchical structures for two classes of cosmologically relevant Gaussian fields -- the scale invariant power-law models with varying spectral index, as well as the LCDM model. We establish that the intensity maps capture the topology of the hierarchical structures, and present extra information that a homology analysis through Betti numbers does not.  Moving on to the analysis of power-law models,  we find that the normalized intensity maps reveal that the pure white noise case, acts as watershed between models with positive and negative spectral indices. This is because white noise has the maximum proportion of transient low persistent features among all the models. On either side of the spectrum - the positive and negative indices - the proportion of high persistent features starts increasing on account of relative shift of power to small and large scales respectively. This information is not available to Betti numbers, and establishes that the intensity maps constructed from the persistence diagrams reflect the distribution of power across the hierarchy of structures at different scales in the models. Further continuing the statistical analysis of persistence diagrams, we study the characteristics of the numerically computed distribution functions of mean density and persistence. In this context, we show that the distribution functions of mean density and persistence show a dependence on the spectral index. This behavior is similar to the the topological Betti numbers, but in contrast with the geometric Minkowski functionals \citep{pranav2019a}. In general, our observation is that topological measures show a sensitivity to the characteristics of the power spectrum, where as the integral geometric measures do not. Subsequently, employing the difference maps, we compare the LCDM model with power law models with a range of spectral indices.  For the given box size of the simulations, we demonstrate that the difference maps are able to establish the proximity of the LCDM model to the $n = -1$ model remarkably well. Finally, we introduce the skew-normal distribution in the cosmological context  (also see \cite{pranavthesis} and \cite{wilding2020}), as a fit for the distribution functions of mean-density and persistence. The observation that the curves for islands and voids exhibit a skewed distribution, while the curve for tunnels has no skew motivates to introduce this generic class of distribution for a simultaneous fit for all categories of topological holes. 

In concluding remarks, we have presented an account of novel topological methods aimed at providing tools for the analysis of hierarchical cosmic structures. In the process, we have focused on a  demonstration of the efficacy of the new developments in applied topology in revealing novel features in Gaussian random fields, and reveal aspects of the hierarchical properties of the structure and connectivity of the stochastic fields that are opaque to traditional measures. Given the recent thrust in cosmology towards data-driven investigations, these tools represent the state-of-the-art.  In view of these observations, we propose to utilize tools emanating from topological data analysis in addressing the challenges thrown by data emerging from ongoing and future experiments.

\section*{Acknowledgemnts}

I am extremely indebted to Robert Adler, Thomas Buchert, Herbert Edelsbrunner, Bernard Jones, Job Feldbrugge, Armin Schwartzman, Gert Vegter, and Rien van de Weygaert for insightful discussions and comments. This work has been partially supported by ERC advanced grant ARTHUS (Adavances in the research on Theories of the Dark Universe; PI: Thomas Buchert), number 740021.

\bibliographystyle{mn2e}


\bibliography{/Users/pratyuze/papers/mine/references/master_references.bib}

\appendix

\section{Skew-normal distribution}
\label{sec:skew_normal}

The skew-normal distribution with the \emph{skewness 
parameter} $\alpha$ is given by \cite{ohagan1976,azzalini1985,pranavthesis,wilding2020}:
\begin{equation}
	f(x) = 2\phi(\nu)\Phi(\alpha\nu).
\end{equation}

\noindent Here,
\begin{equation}
	\phi(\nu) = \frac{1}{\sqrt{2\pi}}\mathrm{e}^{-\frac{\nu^2}{2}}.
\end{equation}

\noindent is the standard normal distribution. The function
 
\begin{eqnarray}
 \Phi(\nu) &=& \int_{-\infty}^x \phi(t)\,\mathrm{d}t    \nonumber \\
         &=& \frac{1}{2}\left[1+\mathrm{erf}\left(\frac{\nu}{\sqrt{2}}\right)\right],
\end{eqnarray}

\noindent is the cumulative distribution function, in which \emph{erf(x)} is the error function. 
Note that one recovers the familiar normal distribution when the skewness parameter 
$\alpha=0$. he absolute value of skewness increases as the absolute 
value of $\alpha$ increases. Note that the skewness parameter $\alpha$ 
is different than the skewness, i.e. the 
third moment of the distribution
\begin{equation}
\gamma_1=\mu^3/\sigma^3. 
\end{equation}

\noindent By definition, a curve has positive skewness if it has a 
more prominent tail for increasing values of $\nu$, and a negative skewness 
when its balance is shifted towards decreasing values of $\nu$. 

To account for the location of the peak and the width of the curve, 
one usually makes the transformation 
\begin{equation}
\mathrm{x} \to \mathrm{x}-\xi/\omega, 
\end{equation}
\noindent where $\xi$ and $\omega$ are the location and scale parameters respectively. 

\noindent The probability distribution 
function with location $\xi$, scale $\omega$ and skewness parameter 
$\alpha$ becomes
\begin{equation}
	f(x) = \frac{2}{\omega} \phi\left(\frac{\nu-\xi}{\omega}\right)\Phi\left(\alpha\left(\frac{\nu-\xi}{\omega}\right)\right). 
\end{equation}
Introducing the amplitude parameter $A_0$, this takes the form
\begin{equation}
	f(\nu)=\frac{A_0}{\omega\pi}\mathrm{e}^{-\frac{(\nu-\xi)^2}{2\omega^2}} \int_{-\infty}^{\alpha\left(\frac{\nu-\xi}{\omega}\right)} \mathrm{e}^{-\frac{t^2}{2}}\,\mathrm{d}t.
\end{equation}

\end{document}